\documentclass{aa}
\usepackage{natbib,twoopt}
\usepackage{xcolor}
\usepackage{subcaption}
\usepackage[breaklinks=true, colorlinks=true, linkcolor=blue, citecolor=blue]{hyperref} %% to avoid \citeads line fills
\bibpunct{(}{)}{;}{a}{}{,}             %% natbib format for A&A and ApJ
\makeatletter
\newcommandtwoopt{\citeads}[3][][]{\href{http://adsabs.harvard.edu/abs/#3}%
    {\def\hyper@linkstart##1##2{}%
     \let\hyper@linkend\@empty\citealp[#1][#2]{#3}}}
  \newcommandtwoopt{\citepads}[3][][]{\href{http://adsabs.harvard.edu/abs/#3}%
    {\def\hyper@linkstart##1##2{}%
     \let\hyper@linkend\@empty\citep[#1][#2]{#3}}}
  \newcommandtwoopt{\citetads}[3][][]{\href{http://adsabs.harvard.edu/abs/#3}%
    {\def\hyper@linkstart##1##2{}%
     \let\hyper@linkend\@empty\citet[#1][#2]{#3}}}
  \newcommandtwoopt{\citeyearads}[3][][]%
    {\href{http://adsabs.harvard.edu/abs/#3}
    {\def\hyper@linkstart##1##2{}%
     \let\hyper@linkend\@empty\citeyear[#1][#2]{#3}}}
\makeatother
\usepackage{graphicx}
\usepackage{multirow}
\usepackage{siunitx}  %unit
\usepackage{txfonts}
\usepackage{graphicx}	% Including figure files
\usepackage{amsmath}	% Advanced maths commands
\usepackage{amssymb}	% Extra maths symbols

\usepackage{multirow}
\usepackage{color}

\definecolor{blue-violet}{rgb}{0.54, 0.17, 0.89}
\definecolor{bostonuniversityred}{rgb}{0.8, 0.0, 0.0}
\definecolor{brilliantrose}{rgb}{1.0, 0.33, 0.64}
\definecolor{calpolypomonagreen}{rgb}{0.12, 0.3, 0.17}
\definecolor{auburn}{rgb}{0.43, 0.21, 0.1}
\definecolor{brightpink}{rgb}{1.0, 0.0, 0.5}
\definecolor{amber}{rgb}{1.0, 0.49, 0.0}
\definecolor{bananayellow}{rgb}{1.0, 0.88, 0.21}
\definecolor{capri}{rgb}{0.0, 0.75, 1.0}
\definecolor{coral}{rgb}{1.0, 0.498, 0.314}

\newcommand{\rvir}{R_{200}}
\newcommand{\rfive}{R_{500}}
\newcommand{\cGpc}{h^{-1}\,{\rm cGpc}}
\newcommand{\cMpc}{h^{-1}\,{\rm cMpc}}
\newcommand{\ckpc}{h^{-1}\,{\rm ckpc}}
\newcommand{\Gpc}{{\rm Gpc}}
\newcommand{\Mpc}{{\rm Mpc}}
\newcommand{\kpc}{{\rm kpc}}
\newcommand{\msun}{{\rm M}_{\odot}}
\newcommand{\msunh}{h^{-1}\,{\rm M}_{\odot}}

\newcommand{\erosita}{\textit{eROSITA}\xspace}

\begin{document}

\title{The \erosita view of the Abell 3391/95 field: a case study from the Magneticum cosmological simulation}

\author{Veronica~Biffi\inst{1,2,3}\thanks{e-mail:         
    \href{mailto:veronica.biffi@inaf.it}{\tt veronica.biffi@inaf.it},
    \href{mailto:biffi@usm.lmu.de}{\tt biffi@usm.lmu.de}}
    \and
    Klaus~Dolag\inst{1}
    \and
    Thomas~H.~Reiprich\inst{4}
    \and
    Angie~Veronica\inst{4}
    \and
    Miriam~E.~Ramos-Ceja\inst{5}
    \and 
    Esra~Bulbul\inst{5}
    \and
    Naomi~Ota\inst{4,6}
    \and
    Vittorio~Ghirardini\inst{5}
    }
    
    \titlerunning{A case study for A3391/95 from simulations}
    \authorrunning{V. Biffi et al.}

\institute{Universitaets-Sternwarte Muenchen,         Ludwig-Maximilians-Universit\"at M\"unchen,
Scheinerstr. 1, D-81679 Munich, Germany
    \and
    INAF - Osservatorio Astronomico di Trieste, via Tiepolo 11, I-34143 Trieste, Italy
    \and
    IFPU - Institute for Fundamental Physics of the Universe, Via Beirut 2, I-34014 Trieste, Italy
    \and
    Argelander-Institut f\"ur Astronomie (AIfA), Universit\"at Bonn, Auf dem H\"ugel 71, 53121 Bonn, Germany
    \and      
    Max-Planck-Institut f\"ur extraterrestrische Physik, Gießenbachstraße 1, D-85748 Garching, Germany
    \and
    Department of Physics, Nara Women's University, Kitauoyanishi-machi, Nara, 630-8506, Japan
    }

   \date{Received ; accepted}

% \abstract{}{}{}{}{} 
% 5 {} token are mandatory

  \abstract
   { Clusters of galaxies reside at the nodes of the Cosmic Web, interconnected by filamentary structures that contain tenuous diffuse gas, especially in the warm-hot phase. Galaxy clusters grow by merging of smaller objects and gas, mainly accreted through these large-scale filaments. For the first time, the large-scale cosmic structure and a long gas emission filament have been captured by SRG/\erosita in a direct X-ray observation of the A3391/95 field.
   }
   { We aim at investigating the assembly history of an
    A3391/95-like system of clusters and the thermo-chemical properties
    of the diffuse gas in it, connecting simulation predictions to
    the \erosita
    observations, to constrain the origin
    and nature of the gas in the pair interconnecting bridge.
      }
   { We analysed the properties of a system resembling A3391/95,
    extracted from the $(352\,\cMpc)^3$ volume of the Magneticum
    Pathfinder cosmological simulations at $z=0.07$.  
        We tracked back in time the main progenitors of the pair clusters
    and of surrounding groups to study the assembly history of the
    system and its evolution. 
   }
   { Similarly to the observed A3391/95 system, the simulated cluster pair is
    embedded in a complex network of gas filaments, with structures
    aligned over more than $20$ projected $\Mpc$ and the whole region
    collapsing towards the central overdense node.  The spheres of
    influence ($3\times\rvir$) of the two main clusters already
    overlap at $z=0.07$, but their virial boundaries are still physically
    separated.  The diffuse gas located in the interconnecting bridge
    closely reflects the WHIM, with typical temperature of $\sim
    1\,$keV and overdensity $\delta\sim 100$, with respect to the mean
    baryon density of the Universe, and lower enrichment level
    compared to the ICM in clusters.  We find that most of the bridge gas
    collapsed from directions roughly orthogonal to the intra-cluster gas
    accretion directions, and its origin is
    mostly unrelated to the two cluster progenitors.  We find clear
    signatures in the surrounding groups of infall motion towards the pair, such as significant radial velocities 
    and slowdown of gas compared to dark matter.  These
    findings further support the picture of the Northern Clump
    (MCXC~J0621.7-5242) cluster infalling along a cosmic gas filament
    towards Abell 3391, possibly merging with it.
   }
   { We conclude that, in such a configuration, the pair clusters of
    the A3391/95-like system are in a pre-merger phase, and did not
    interact yet. The diffuse gas in the interconnecting bridge is
    mostly warm filament gas, rather than tidally-stripped cluster gas.
   }


   \maketitle
%
%-------------------------------------------------------------------

\section{Introduction}

The formation and evolution of the large scale structure (LSS) has
{long} been a key target of astrophysical and cosmological
investigations.  Simulations predict the existence of a thin
filamentary structure, the so-called Cosmic Web~\cite[][]{bond1996},
connecting the knots where galaxy clusters reside.  Observationally, a
detailed study of the Cosmic Web structure and evolution can be
pursued by investigating the properties of the visible matter tracing
the underlying dark matter (DM) distribution. This promoted several
investigations aiming for a census of cosmic
baryons~\cite[][]{persic1992,fukugita1998}.  Compared to observational
estimates, theoretical studies predict that about half of the baryons
in the Universe must be indeed undetected~\cite[][]{cen1999,dave2001}.
In particular, {independently of the use of identification tools
  used to identify the Cosmic Web constituents
}~\cite[e.g.][]{liebeskind2018} --- namely filaments, sheets and voids
--- cosmological simulations consistently predict that a significant
fraction ($\sim40\%$) of the cosmic baryon budget is in the form of
cool intergalactic medium and warm-hot intergalactic medium
\cite[WHIM,][]{cui2019,martizzi2019}, with most of it located in
filaments~\cite[][]{tuominen2020}.  Properties of cosmic filament
populations have also been studied from a statistical point of view,
in order to characterize their evolution and global properties in
correlation with their tracers, namely gas and
galaxies~\cite[][]{gheller2015,gheller2016,cui2018,martizzi2019}.
Recent investigations of various state-of-the-art cosmological
simulations, including the Magneticum Pathfinder suite analysed in the
present study, show that statistically different populations exist:
longer thinner filaments typically connecting smaller structures, and
shorter bridge-like filaments usually denser and connected to massive
objects~\cite[][]{galarraga2020a,galarraga2020b}.

Observational reconstructions of the Cosmic Web are therefore tightly
connected to the extensive search for the diffuse cold gas and WHIM in
the filaments, which is still largely eluding our
detections~\cite[][]{bregman2007}.  {Reconstructions of the LSS}
through galaxy mapping have been successfully performed over the last
decades~\cite[see, e.g.,][]{tempel2014,malavasi2020}.  {For what
  concerns the gaseous content of the LSS, observations have been
  carried out mainly in the far-UV~\cite[see][for a
    review]{bregman2007,peroux2020}, whereas, in the X rays,} direct
observations of the gaseous Cosmic Web has proven to be very
challenging~\cite[][]{nicastro2008}.  The main reason for this lies in
the difficulty of observing the tenuous warm plasma residing in the
filaments, whose low densities ($< 10^{-4}\,{\rm cm}^{-3}$) and
temperatures ($T\sim 10^5$--$10^7$\,K) make its X-ray emission very
faint.  Only recently significant progresses have been done in this
direction, starting with the first detection of the WHIM reported
by~\cite{nicastro2018}, obtained from the absorption lines of highly
ionized oxygen (OVII) in high signal-to-noise spectra of a quasar at
$z>0.4$, using the XMM-Newton Reflection Grating Spectrometer (RGS).
Overall, direct observations of warm gas in filaments have been mainly
limited to special cases, such as specific lines of sight in the
direction of high-redshift quasars for the detection of WHIM in
intervening
absorbers~\cite[][]{nevalainen2015,nicastro2018,nevalainen2019,kovacs2019}
or regions in between close pairs of clusters or close to cluster
outskirts where the WHIM emission could be
enhanced~\cite[e.g.][]{briel1995,finoguenov2003,durret2004}. Longer
gaseous filamentary structures, up to tens of $\cMpc$, have been
instead investigated mostly through statistical approaches, such as
stacking of thermal-SZ and X-ray
observations~\cite[][]{degraaff2019,tanimura2020a,tanimura2020b,lim2020}.
{Recently,~\cite{khabibullin2019} investigated the potential of
  \erosita stacking approaches in characterising the
  WHIM. Specifically, accounting for the contribution of the
  resonantly scattered cosmic X-ray background and comparing the WHIM
  imprints in X-ray absorption and emission, they show that truly
  diffuse gas, filling cosmic filaments, can be significantly
  characterised and better differentiated from denser matter clumps.}

A promising strategy for direct detections of the WHIM X-ray emission
is focusing on specific regions encompassing multiple cluster systems
(like merging clusters and superclusters) or on the prolongation of
cluster
outskirts~\cite[][]{fujita2008,planck2013,eckert2015,bulbul2016,hattori2017,akamatsu2017,parekh2017,connor2018,connor2019,ghirardini2020}.
In fact, a number of cluster binary systems in the local Universe have
been throughly investigated to find evidences for the presence of
cosmic diffuse gas in the bridges connecting the member clusters. Well
known case studies in the low-redshift Universe are represented by
A3556 and A3558 in the Shapley supercluster ($z=0.048$,
\citealt{mitsuishi2012,ursino2015}); RXC J1825.3+3026 ($z\sim 0.065$)
and CIZA J1824.1+3029~\cite[$z\sim 0.071$;][]{botteon2019}; the
A399--A401 system (at redshifts $z\sim0.0724$ and $z\sim0.0737$,
respectively~\citealt{sakelliou2004,fujita2008}; see also thermal SZ
observations by~\citealt{bonjean2018}); the merging pair
A222--A223~\cite[$z\sim0.21$;][]{werner2008}, or the multiple merging
system Abell~1758~\cite[$z\sim0.28$;][]{durret2011}.

Among these, also the cluster system A3391--A3395 has been widely
studied in the literature, with special attention to the gaseous
bridge between the pair members~\cite[][]{tittley2001}.  This system
comprises two main clusters, {the northern one, Abell~3391, and the
  southern one, Abell~3395~\cite[a double system
    itself;][]{reiprich2002}} at $z\sim 0.05$, separated by about
$50'$ {(approximately $3\,\Mpc$)} on the sky.  {Both systems
  have X-ray temperatures around $T_X \sim
  5$\,keV~\cite[][]{reiprich2002,vikhlinin2009}.  Early results
  by~\cite{reiprich2002}, combining \textit{ROSAT} and \textit{ASCA}
  data, indicated masses of $M_{500}\sim 5\times10^{14}\msun$ for
  A3391 and $M_{500}\sim8.8\times 10^{14}\msun$ for both A3395s and
  A3395n (assuming an Einstein--de~Sitter cosmology and a Hubble
  constant $H_0 = 50\,{\rm km\,s}^{-1} \Mpc^{-1}$). Especially for
  A3395, and despite the uncertainties, these values were likely
  overestimated, probably due to the merging phase of the system.
  More recent estimates by~\cite{piffaretti2011}, based on
  \textit{ROSAT} X-ray luminosities, report lower masses of about
  $M_{500}\sim 2\times 10^{14}\msun$ for both A3391 and A3395.  }

{From previous studies, debated conclusions have been drawn on the
  dynamical status of the pair, with different interpretations of the
  nature of the gas observed between the two main
  clusters~\cite[][]{tittley2001,planck2013,sugawara2017,alvarez2018}.
  \cite{sugawara2017} and~\cite{alvarez2018}, for instance, report the
  presence of hot gas in the bridge region, concluding that this could
  have been tidally stripped or heated by the interaction between
  A3391 and A3395. This is in contrast with earlier {\it ASCA} and
  {\it ROSAT} results from~\cite{tittley2001}, rather favouring a
  scenario in which the filament connecting the two systems is in fact
  longer and aligned lengthwise with the line of sight (l.o.s.). The
  geometry of the system with respect to the l.o.s.\ and the
  temperature of the interconnecting gas are key aspects to interpret
  the nature of the bridge, where the detection of cold-warm gas with
  $T\lesssim 1\,$keV can indicate the presence of true filament gas.
}

More recently, the A3391/95 system has also been targeted by the
German-built X-ray telescope on board the Spectrum-Roentgen-Gamma
(SRG) mission, the extended ROentgen Survey with an Imaging Telescope
Array~\cite[\erosita;][]{predehl2020}, during the Performance
Verification (PV) phase observational
campaign~\cite[][]{reiprich2020}.  Thanks to the high sensitivity in
the soft X-ray band, large field of view and good spatial
resolution~\cite[][]{merloni2012}, these \erosita observations allowed
to find hints of emission from warm gas in the pair bridge, in
addition to the known hotter gas emission.  {In the \erosita
  observation, the bridge emission qualitatively spans $\sim 3$\,Mpc
  (projected)~\cite[][]{reiprich2020}, which} cannot be entirely
attributed to the known galaxy group (ESO~161-IG~006) located in
between the two main clusters. Furthermore, the emission from hot and
warm diffuse gas is detected well beyond the projected virial
boundaries of the main clusters, allowing for the first time to map a
continuous warm-hot emission filament in the X rays, spanning {from
  north to south} $4$~degrees across the observed field. This WHIM
filament extends north of A3391 and south of A3395 for a total
projected length of $\sim15\,$Mpc, at the median redshift of the
system, and its presence is confirmed through the Planck SZ map and
DECam (optical) galaxy density distribution as well.  With this
unprecedented X-ray detection of a WHIM filament in the observation of
a single system reported by~\cite{reiprich2020}, the new \erosita PV
observations are also capturing the formation of the LSS {in the
  region} where the system is located. In fact, several groups and
clusters are discovered in its local environment (i.e. at the same
redshift) and there are indications of the substructures {moving}
towards the A3391/95 system~(see Veronica et al.,~submitted;
Ramos-Ceja et al.,~in prep.).

With the primary goal of accompanying the observational results
obtained by the \erosita PV observations of the Abell 3391/95 system
and its surroundings, and help their interpretation, we resort here to
cosmological hydrodynamical simulations.  {Specifically, we focus
  on a simulated pair of galaxy clusters extracted from the Magneticum
  cosmological hydrodynamical simulation, sharing several similarities
  with the observed A3391/95 system.  The simulated analog allows us
  to investigate the intrinsic properties of an A3391/95-like system
  and its formation and evolution. In particular, we aim at
  reconstructing the system assembly history considering the main
  clusters and the additional structures in the surrounding
  environment. Furthermore, we can make use of the simulations to
  study the thermal and chemical properties of the diffuse gas
  component in the system, both at the selection redshift ($z=0.07$)
  and throughout the cosmic evolution. This will allow us to explore
  the expected properties of the WHIM and of the bridge gas, and to
  compare them against the ICM characteristics within the
  clusters. Especially, we can directly investigate the origins of the
  gas in the region between the pair clusters, to determine whether
  this can be filament gas or rather tidally-stripped intracluster
  gas.  }

The paper is organized as follows.  In Sec.~\ref{sec:sims} we present
the simulation suite used for the present analysis, namely the
Magneticum Pathfinder cosmological hydrodynamical
simulations. Interested in multiple cluster systems in the local
Universe, we focus on one simulation snapshot at low redshift
($z=0.07$) and explore in Sec.~\ref{sec:pairs} the cluster pair
candidates available, depending on three-dimensional and projected
separation, as well as on the mass and mass ratio of the
members. Among those, we select and introduce the best cluster pair
selected to represent the theoretical analog of the observed A3391/95
system (Sec.~\ref{sec:analog}).  In Sec.~\ref{sec:results} we
illustrate the main results of our study.  The three-dimensional
assembly of the pair system in the Cosmic Web, with its local
evironment, is presented in Sec.~\ref{sec:orig} together with the
results on the origin of the diffuse gas residing in the bridge
between the pair member clusters.  In Secs.~\ref{sec:chem}
and~\ref{sec:chem_evol} we discuss instead the thermal and chemical
properties, and their time evolution, of the diffuse gas in various
thermal phases and spatial regions in and around the pair.  In
Sec.~\ref{sec:clumps} we then investigate the groups aligned along
large-scale filaments connected to the pair, in order to find the
signatures of their infall towards the overdense Cosmic Web knot where
the pair is located.  Finally, we summarize and conclude in
Sec.~\ref{sec:conclusion}.

\section{Magneticum Pathfinder Simulations}\label{sec:sims}

We employ a large cosmological volume, simulated including a variety
of physical processes, to study the properties of multiple galaxy
cluster systems and the evolution of the local Cosmic Web in which
they are embedded.

To this scope, we consider the Magneticum Pathfinder
simulations\footnote{\texttt{www.magneticum.org}}, a set of
state-of-the-art cosmological smoothed-particle hydrodynamics (SPH)
simulations, comprising boxes of different volumes and resolution.
The simulations have been performed with the TreePM/SPH code
P-Gadget3, an extended version of P-Gadget2~\cite[][]{springel2005}.
This includes several improvements of the SPH formulation, such as the
treatment of viscosity and artificial conduction, and the use of
higher-order kernels~\cite[][]{dolag2005,beck2016}.  The code also
accounts for a large variety of physical processes describing the
evolution of the baryonic components.  These comprise radiative
cooling and heating from a uniform time-dependent ultraviolet (UV)
background~\cite[][]{haardt2001}, as well as a sub-resolution model
for star formation~\cite[][]{springel2003}.  Following detailed
stellar evolution models, a description of chemical enrichment is also
included as in~\cite{tornatore2004,tornatore2007}. Specifically,
metals are produced from stellar sources depending on their mass and
typical lifetimes, assuming an initial mass function (IMF) according
to~\cite{chabrier2003} and a mass-dependent lifetime function
from~\cite{padovani1993}.  The three main enrichment channels are
supernovae Type Ia (SNIa) and Type II (SNII) and low- and
intermediate-mass stars undergoing the Asymptotic Giant Branch (AGB)
phase.  Eleven different chemical elements are explicitely traced
(namely, H, He, C, Ca, O, N, Ne, Mg, S, Si, and Fe) assuming stellar
yields from~\cite{vandehoek1997}, for AGB stars, \cite{thielemann2003}
for SNIa and \cite{WW1995} for SNII.  Gas radiative cooling depends on
the local gas metallicity in a self-consistent way, as described
in~\cite{wiersma2009}.  SNII also contribute to thermal and kinetic
energy feedback by driving galactic winds, with a mass loading rate
proportional to the star formation rate (SFR) and a resulting wind
velocity of $v_{w} = 350\,{\rm km\,s}^{-1}$~\cite[][]{springel2003}.
The simulations further account for black hole (BH) growth and gas
accretion, powering energy feedback from Active Galactic Nuclei (AGN),
based on the implementation by~\cite{springeldimatteo2005}
and~\cite{dimatteo2005} and with modifications as
in~\cite{fabjan2010}.  As shown in previous studies, the Magneticum
simulations successfully reproduce many observed properties of cosmic
structures, such as kinematical and morphological properties of
galaxies~\cite[][]{teklu2017,remus2017,schulze2018,remus2021},
chemical properties of both galaxies and
clusters~\cite[][]{dolag2017}, statistical properties of the AGN
population at various
redshifts~\cite[][]{hirschmann2014,steinborn2016,biffi2018}.  At
cluster scales, the Magneticum set has been employed to constrain the
scatter in the observable X-ray luminosity-temperature relation via
intra-cluster medium (ICM) velocity diagnostics~\cite[][]{biffi2013}
as well as the pressure profile of the ICM~\cite[][]{gupta2017}, and
to explore the level of contamination of ICM emission due to the
central AGN, expected for synthetic \erosita
observations~\cite[][]{biffi2018}.

Given the number of well known observed cluster pairs at low redshift,
and motivated in particular by the \erosita PV observation of the
A3391--A3395 system, we focus on the simulation snapshot corresponding
to redshift $z=0.07$.  The specific simulation considered is one of
the larger ones, namely the Magneticum ``Box2'' cosmological box at
high resolution (``hr'').  This box comprises a comoving volume of
$(352\,\cMpc)^3$ and is resolved with $2\times 1584^3$ particles,
corresponding to a mass resolution of $m_{\rm DM}=6.9\times
10^8\,\msunh$ and $m_{\rm gas}=1.4\times 10^8\,\msunh$, for dark
matter and gas particles respectively~\cite[see also][]{biffi2018}.
Since the gas can spawn up to four stellar particles decreasing its
mass, or accrete metals from chemical pollution, the mass of the gas
particles can vary in time.  At this resolution, the softening length
for DM and gas particles is $\epsilon_{\rm DM,gas}=3.75\,h^{-1}\,$kpc
and for stellar particles $\epsilon_{\rm stars}=2\,h^{-1}\,$kpc.  The
simulations assume a standard $\Lambda$CDM cosmological model with the
Hubble parameter set to $h=0.704$, the density parameters for matter,
dark energy and baryons equal to $\Omega_M=0.272$,
$\Omega_\Lambda=0.728$ and $\Omega_b=0.0451$, and $\sigma_8 = 0.809$,
for the normalization of the fluctuation amplitude at
$8\,$Mpc~\cite[according to the seven-year results of the Wilkinson
  Microwave Anisotropy Probe, WMAP,][]{wmap7}.  The data of the
simulation volume used for the present analysis, with several
pre-computed halo properties and mock X-ray data~\cite[generated with
  the PHOX X-ray photon simulator,][]{biffi2012,biffi2013}, are
publicly available on the Cosmological Web
Portal\footnote{\texttt{https://c2papcosmosim.uc.lrz.de}}~\cite[][]{ragagnin2017}.

In order to identify subhalos, we employ the SUBFIND substructure
finding algorithm~\cite[][]{springel2001,dolag2009}.  The main halo
detection is based on a standard friends-of-friends
algorithm~\cite[FoF;][]{davis1985}.  Taking into account the presence
of baryons as well~\cite[][]{dolag2009}, SUBFIND also identifies all
self-bound substructures within main halos, around local density
peaks.  In particular, the subhalos of a cluster are defined as all
the substructures identified within its virial
radius~\cite[{$R_{\rm vir}$}, computed with the
  spherical-overdensity approach using the top-hat model
  by][]{eke1996}.  SUBFIND also allows to identify for each
substructure the corresponding content of gravitationally bound DM,
gas, stars etc.  The center of each halo is identified with the
position of the minimum of the potential well, considering the member
particles.

For each main structure, characteristic quantities are computed, such
as {characteristic radii and enclosed masses.  For data comparison
  purposes, we compute the $\rfive$ radius, i.e.\ the radius of the
  sphere encompassing an average density that is $500$ times the
  critical density of the Universe ($\rho_{\rm cr}$), and the mass
  within it, $M_{500}$.  The estimate of $\rvir$ used in the following
  corresponds for simplicity to $1.5\times\rfive$, and we verified
  that this matches within a few percents the value computed at the
  actual overdensity of $\Delta=200$.  }

For all metallicity values reported in this work, we refer for
simplicity to the solar abundance pattern by~\cite{angr89} as a
reference. Given that the solar unit is purely used to normalize our
estimates, this can be easily re-scaled for other solar reference
values, such as~\citealt{asplund2009}\footnote{For instance, in order
  to adopt the \citealt{asplund2009} solar pattern instead of
  \cite{angr89}, the Fe abundances reported in the following Sections
  should be multiplied by a factor $Z_{\rm Fe,\odot}^{\rm AG89}/Z_{\rm
    Fe,\odot}^{\rm Aspl09} = 4.68\times10^{-5}/3.16\times10^{-5} \sim
  1.5$}.

\section{Galaxy cluster pair candidates}\label{sec:pairs}

In the Magneticum ``Box2/hr'' box at $z=0.07$, there are a total of
$10429$ identified haloes with masses in the range $10^{13} < M_{500}
[\msun] < 1.44\times 10^{15}$.  In particular, we identify $448$
objects with $M_{500} > 10^{14}\,\msun$, from which we can therefore
expect $\sim 10^5$ candidate pairs of individually-identified
cluster-size haloes.

Motivated by the properties of observed pairs of galaxy clusters, for
which a variety of observations have been dedicated to investigating
their properties, the stage of interaction and the possible detection
of filamentary structures (bridges) physically connecting them, we
further restrict the search to the ``close'' pairs, in projection.  We
find that only $535(135)$ pairs appear closer than $10(5)$\,Mpc in at
least one projection (considering the three Cartesian axes $x$, $y$
and $z$ as possible l.o.s.\ directions).  By inspecting the
three-dimensional spatial distribution of all the cluster-size haloes
in the simulated volume, we find nevertheless that only for $146$,
$88$ and $40$ unique pairs the 3D distance between the two haloes is
actually smaller than $20$, $15$ and $10$ megaparsecs, respectively
{(see Table~\ref{tab:pairs_stat}, in Appendix~\ref{app:pairs})}.
The number of close pairs decreases further if we impose additional
constraints on the mass ratio between the two haloes, or on the
initial mass-range selection of the clusters.  We see in particular
that for haloes with masses comprised in the range $1.5$--$3.5 \times
10^{14}\,\msun$, only 25 candidate pairs are in fact closer than
$d_{\rm 3D} < 20$\,Mpc, with $\sim 30\%$ of them actually closer than
$d_{3\rm D} < 10$\,Mpc.

These results for Box2/hr at $z=0.07$ indicate that we expect roughly
$\sim 400$ pairs per $\Gpc^3$ with cluster size members
($M_{500}>10^{14}\msun$) separated by a physical distance smaller than
$10\,\Mpc$. In most of the cases ($\sim 70\%$) the mass ratio of the
member clusters does not exceed 2, and in $20\%$ of them the members
have in fact comparable masses (${\rm M1/M2}\sim 1$).  In order to
confirm this, we investigated also larger cosmological volumes (namely
boxes with $0.9\,\cGpc$ and $2.7\,\cGpc$ per side) at similar
resolution, that are part of the Magneticum simulation set.  In the
local Universe ($z\sim0$), we consistently find $\sim 300$ cluster
pairs per $\Gpc^3$ volume with three-dimensional separation lower than
$10\,\Mpc$ and member masses $M_{500}>10^{14}\msun$.  By restricting
to member clusters with $1.5 < M_{500}^{1,2}\,[10^{14}\,\msun] < 3.5$,
we expect to find $\sim 50$ close ($d_{\rm 3D}\lesssim 10\,\Mpc$)
pairs per $\Gpc^3$ --- with $50\%$ of them having a mass-ratio ${\rm
  M1/M2}\lesssim 1.2$--$1.3$.

In order to compare against the specific case of A3391/95, we choose
one of the 7 close pairs at $z=0.07$ (member separation of $d_{3\rm D}
< 10\,\Mpc$ and mass ratio $1 < {\rm M1/M2}\lesssim 1.2$) for deeper
investigation, as a study case in the simulations that fairly
resembles the observed A3391--A3395 system.  The specific selection
has been performed based on a visual inspection of each candidate,
taking into consideration also the projected distances between the
member clusters in the three main projections of the simulation box
(namely, we restrict for simplicity to those along the $x$, $y$ and
$z$ Cartesian axes).  We remark nonetheless that each candidate among
those with similar member masses, mass ratio and physical separation
could have been in principle chosen, by inspecting among all the
possible l.o.s.\ the projection which resembles as close as possible
the observed system.

\subsection{A3391--A3395: a simulation analog}\label{sec:analog}

The close galaxy cluster pair, selected at $z=0.07$ as a case study,
comprises two haloes with masses $M_{500}^1=1.96\times 10^{14}\,\msun$
(GC1) and $M_{500}^2=1.67\times 10^{14}\,\msun$ (GC2), respectively,
namely with a mass ratio of $\sim1.17$.  The two pair members are
physically separated by $d_{\rm 3D} = 4.54$\,Mpc and the projected
distance is $2.6$, $4.5$ and $3.8$ megaparsecs in the $xy$, $xz$ and
$yz$ projection planes, respectively.  The physical distance between
their centers corresponds roughly to three times the sum of their
$\rfive$ radii, {respectively $\rfive^1=862\,\kpc$ and
  $\rfive^2=819\,\kpc$.  This is equivalent to $d_{\rm 3D}\sim 2
  \times (\rvir^1 + \rvir^2)$, considering $\rvir \approx
  1.5\times\rfive$.  The top-hat virial radii, $R_{\rm vir}$, are
  similar and both approximately $50$\% larger than $\rvir$ (namely,
  $R_{\rm vir}^1=1.88\,\Mpc$ and $R_{\rm vir}^2=1.87\,\Mpc$).  } The
spheres approximating $3\times \rvir$ are therefore already
overlapping, physically and, therefore, also in projection.

\begin{figure*}
\centering
    \includegraphics[width=.99\textwidth]{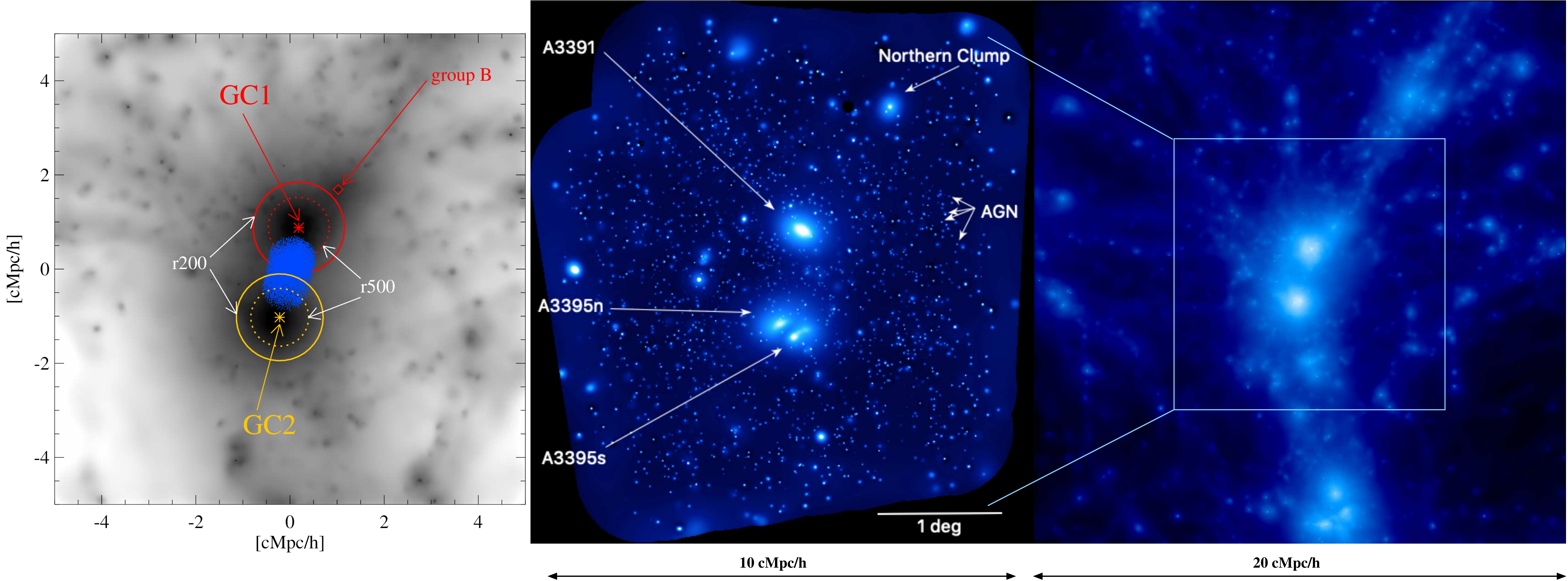}
    \caption{Comparison between the simulated galaxy pair candidate
      {at $z=0.07$} and the observed A3391/95 system.  {\it Left:}
      zoom onto the projected gas density map of the simulated cluster
      pair. The image is $10\,\cMpc$ per side and projected for
      $10\,\cMpc$ along the l.o.s.  {We label the two clusters GC1
        and GC2 and an infalling group (``group B''), and mark their
        extent. In blue, is shown the gas in the interconnecting
        bridge.}  {\it Middle:} \erosita PV observation of a $\sim
      15\,{\rm deg}^2$ region around the multiple A3391/95 system,
      with the main clusters and Northern Clump
      marked~\cite[][]{reiprich2020}. Here, $1\,{\rm deg} \approx
      3.9\,\Mpc$, at the redshift of A3391 and for the cosmology
      adopted in~\cite{reiprich2020} {(for comparison to
        simulations, we also report the comoving scale).}  {\it
        Right:} {projected gas density map of the simulated cluster
        pair in its local environment. The map encloses a
        $(20\,\cMpc)^3$ volume, spanning $\sim 26\,\Mpc$ (physical) in
        projection, for the chosen redshift and cosmology.}  }
    \label{fig:sims_obs}
\end{figure*}

We select this pair, in particular, in the reference $xy$ projection
(i.e.\ with the l.o.s. along the $z$-axis of the cosmological box),
motivated by the similarities with the A3391--A3395 system --- as
shown in Fig.~\ref{fig:sims_obs}.
In the central panel of the Figure, we report the \erosita PV
observation of the $\sim 15\,{\rm deg}^2$ region around
A3391/95~\cite[][]{reiprich2020}, with all the main systems labelled.
In the right panel of Fig.~\ref{fig:sims_obs} we report the gas
density map of the simulated analog at $z=0.07$, comprising a cubic
volume of $20\,\cMpc$ per side.  Compared to the observed system, the
GC1 and GC2 clusters have a broadly similar orientation, and are
surrounded by other systems aligned in a filamentary structure from
top to bottom over $20\,\cMpc$ in projection (approximately $26$
physical Mpc at this redshift).  Fig.~\ref{fig:sims_obs}-left shows a
simulation zoom onto the central $10\,\cMpc$-size region, roughly
covering the same size of the eROSITA field. Here we label the two
main clusters and their sizes, an infalling group-size halo (see
Sec.~\ref{sec:clumps}), and mark in blue the interconnecting gas
bridge.  On the plane of the sky, they appear to be at a relatively
similar distance ($\sim 3\,\Mpc$) compared to the observed A3391/95,
in which the projected $3\times \rvir$ extent of the two main systems
also overlap.

Compared to the simulations, the mass of the observed A3391/95 system
reported by~\cite{piffaretti2011} is similar, although $\sim 25\%$
more massive overall, with $M_{500}=2.16\times 10^{14}\,\msun$ and
$M_{500}=2.4\times 10^{14}\,\msun$ for A3391 and A3395,
respectively~\cite[][]{piffaretti2011,alvarez2018}.  Similarly to
those works, for the purpose of comparing to simulations, we will
mainly consider the A3391/95 system as a cluster pair, treating A3395
as a whole although it is a double-peaked merging
cluster~\cite{reiprich2002,reiprich2020}. We note nonetheless that the
southern GC2 cluster in the simulations shows a broad similarity with
A3395, hosting two massive substructures that recently merged (see
further discussion in Sec.~\ref{sec:discussion}).

\begin{figure*}
\centering
    {\large $z=1$ \hspace{.26\textwidth} $z=0.5$ \hspace{.26\textwidth} $z=0.07$}\\[3pt]
	\includegraphics[width=.32\textwidth]{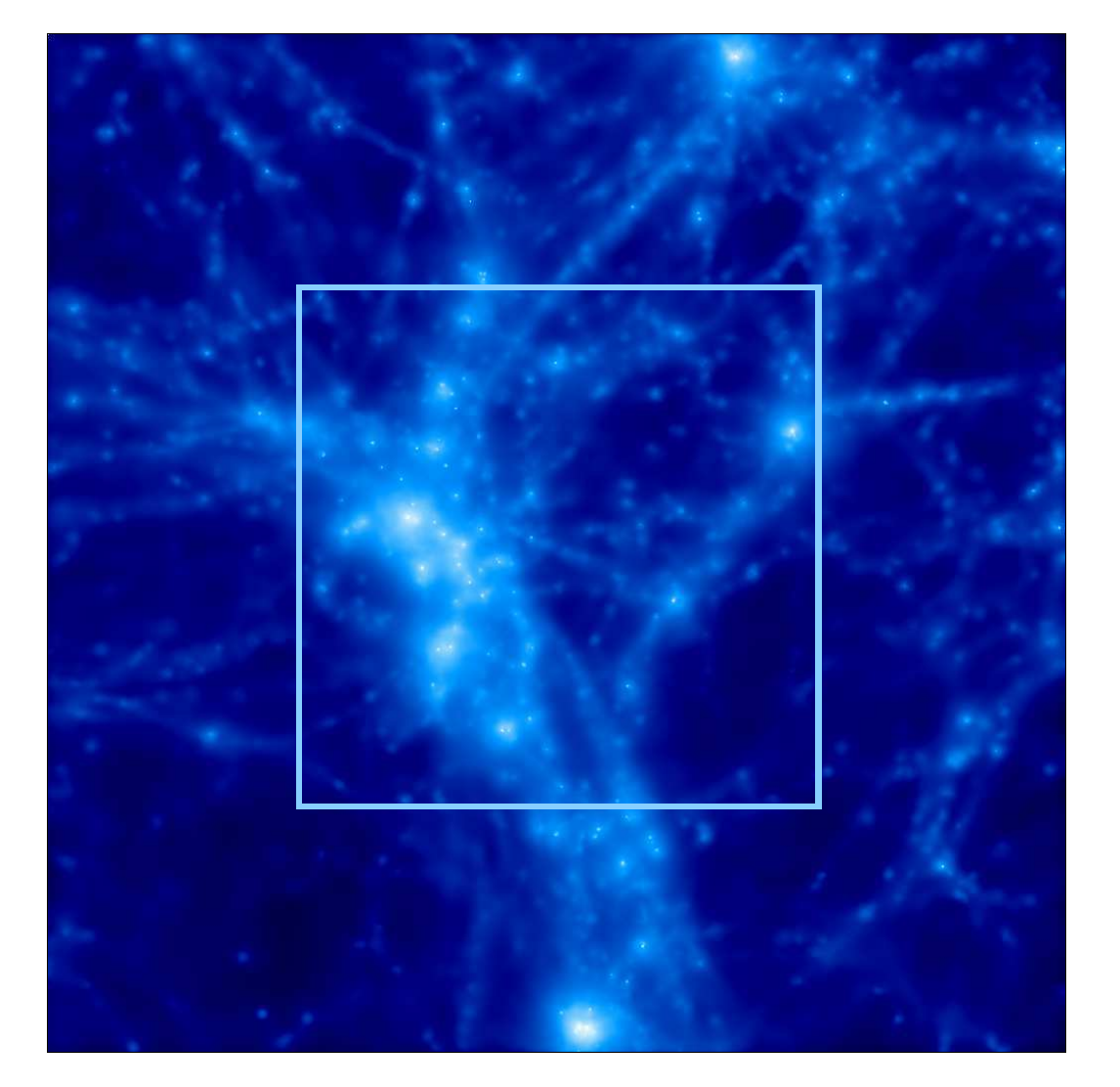}
	\includegraphics[width=.32\textwidth]{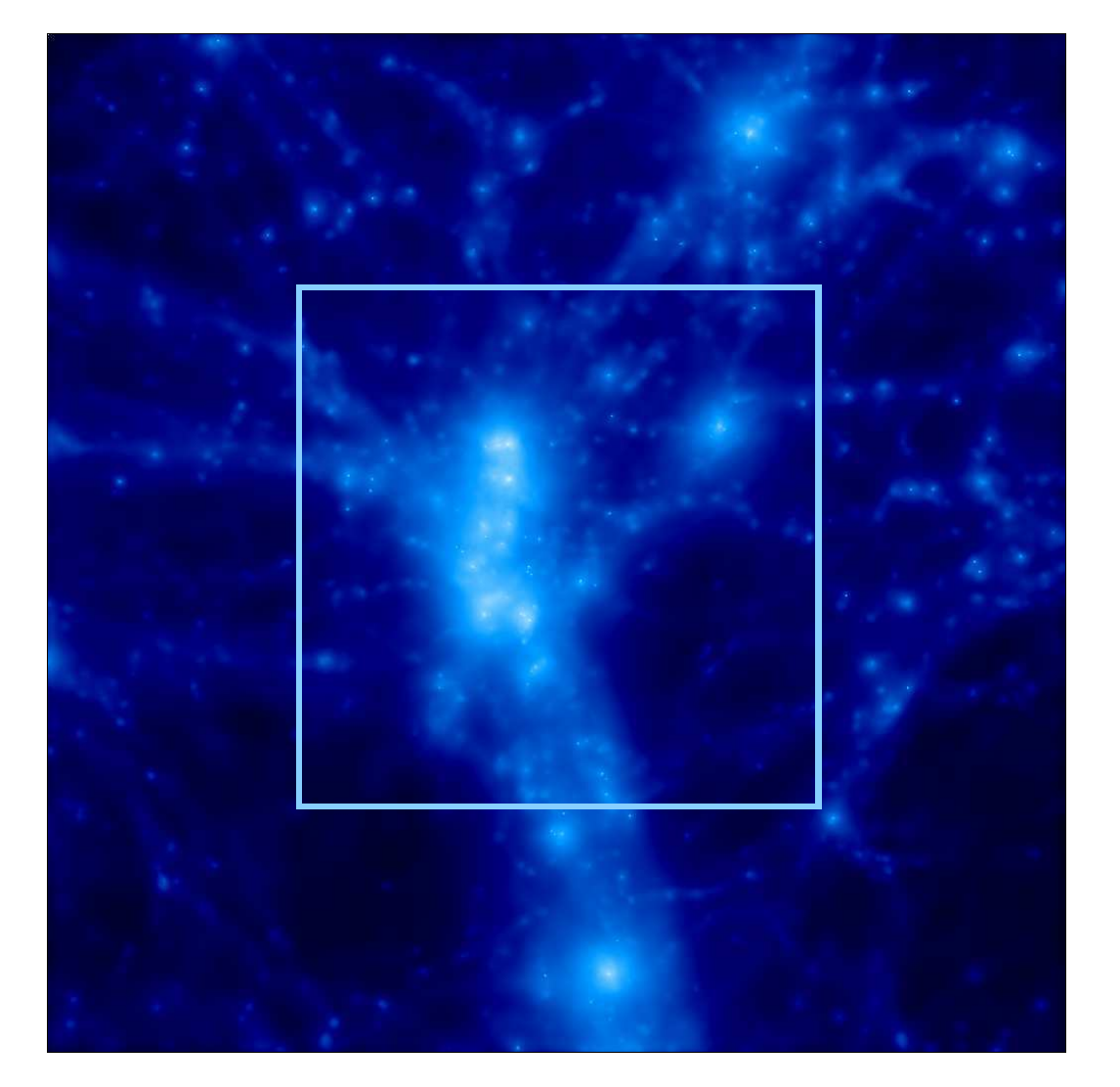}
	\includegraphics[width=.32\textwidth]{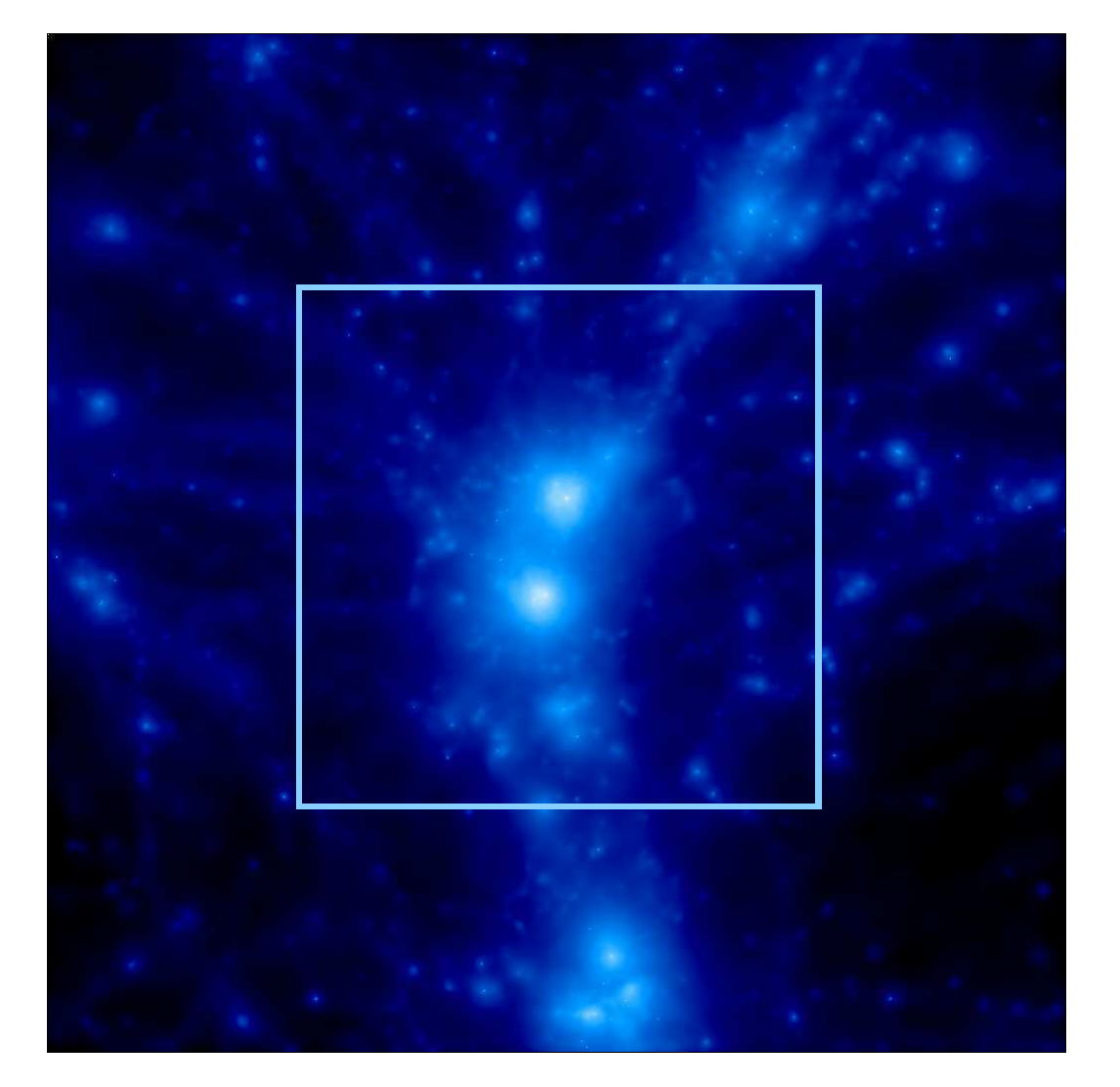}\\[3pt]
	(a) gas density\\[7pt]
    \includegraphics[width=.32\textwidth]{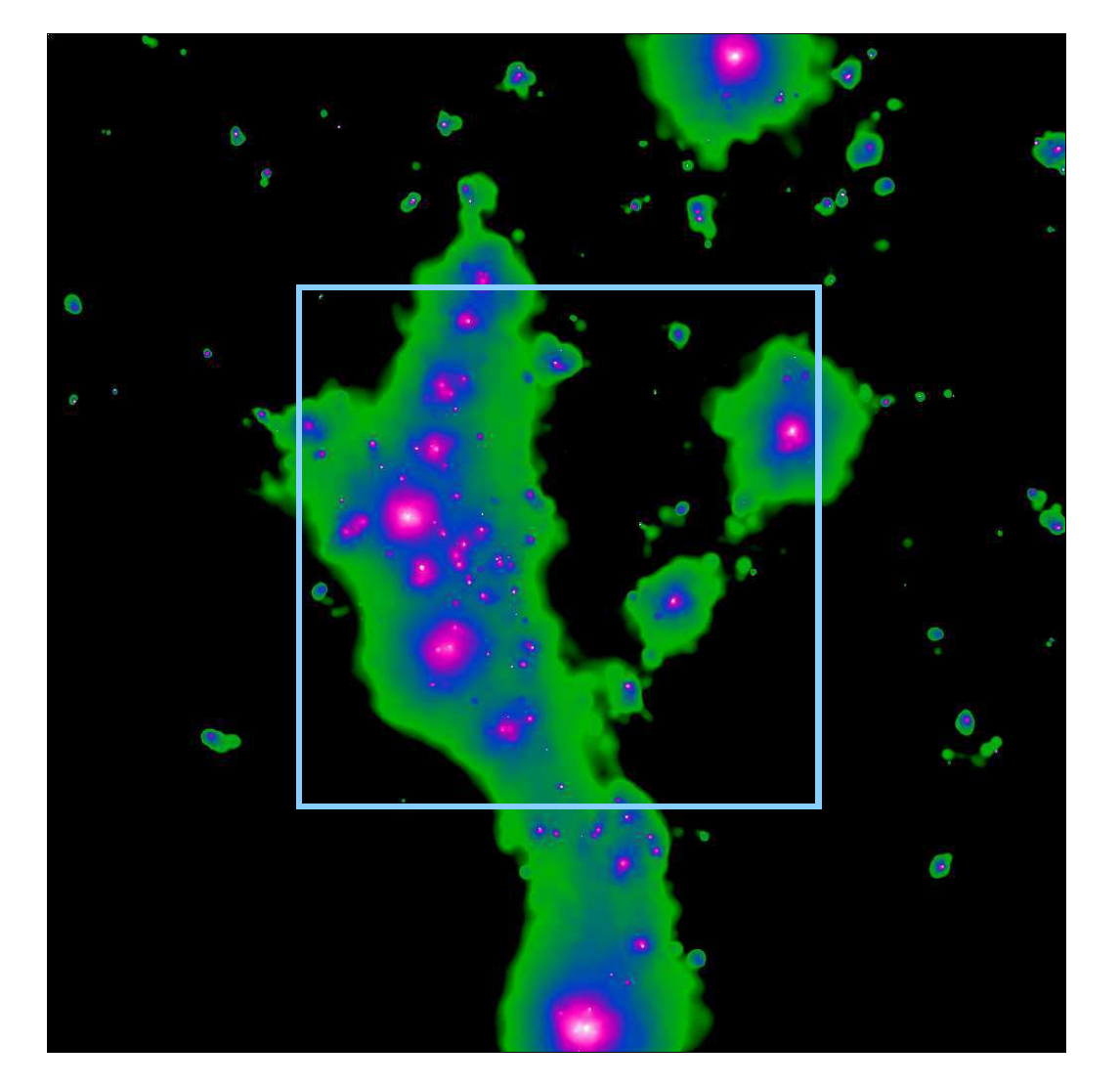}
	\includegraphics[width=.32\textwidth]{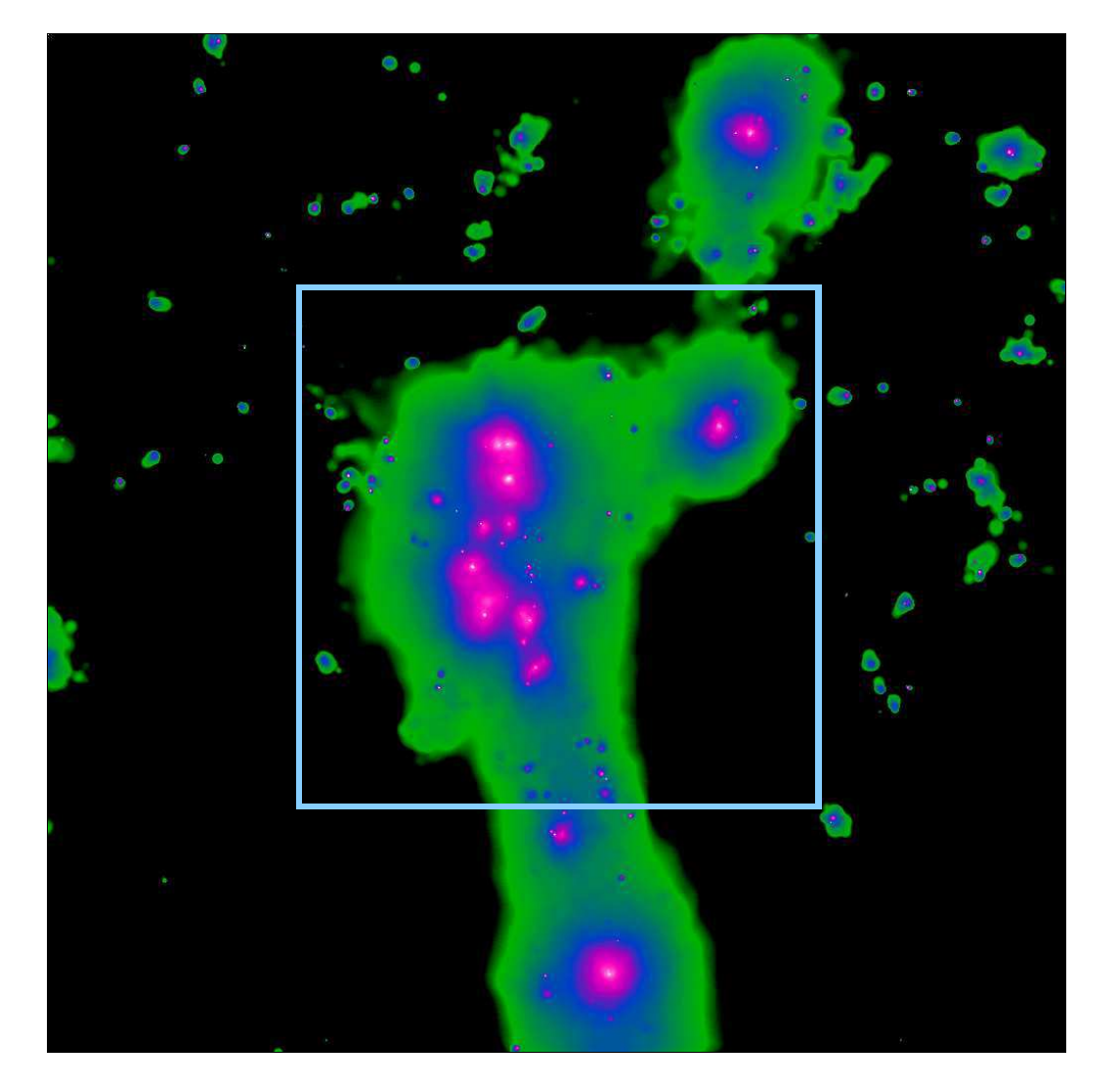}
	\includegraphics[width=.32\textwidth]{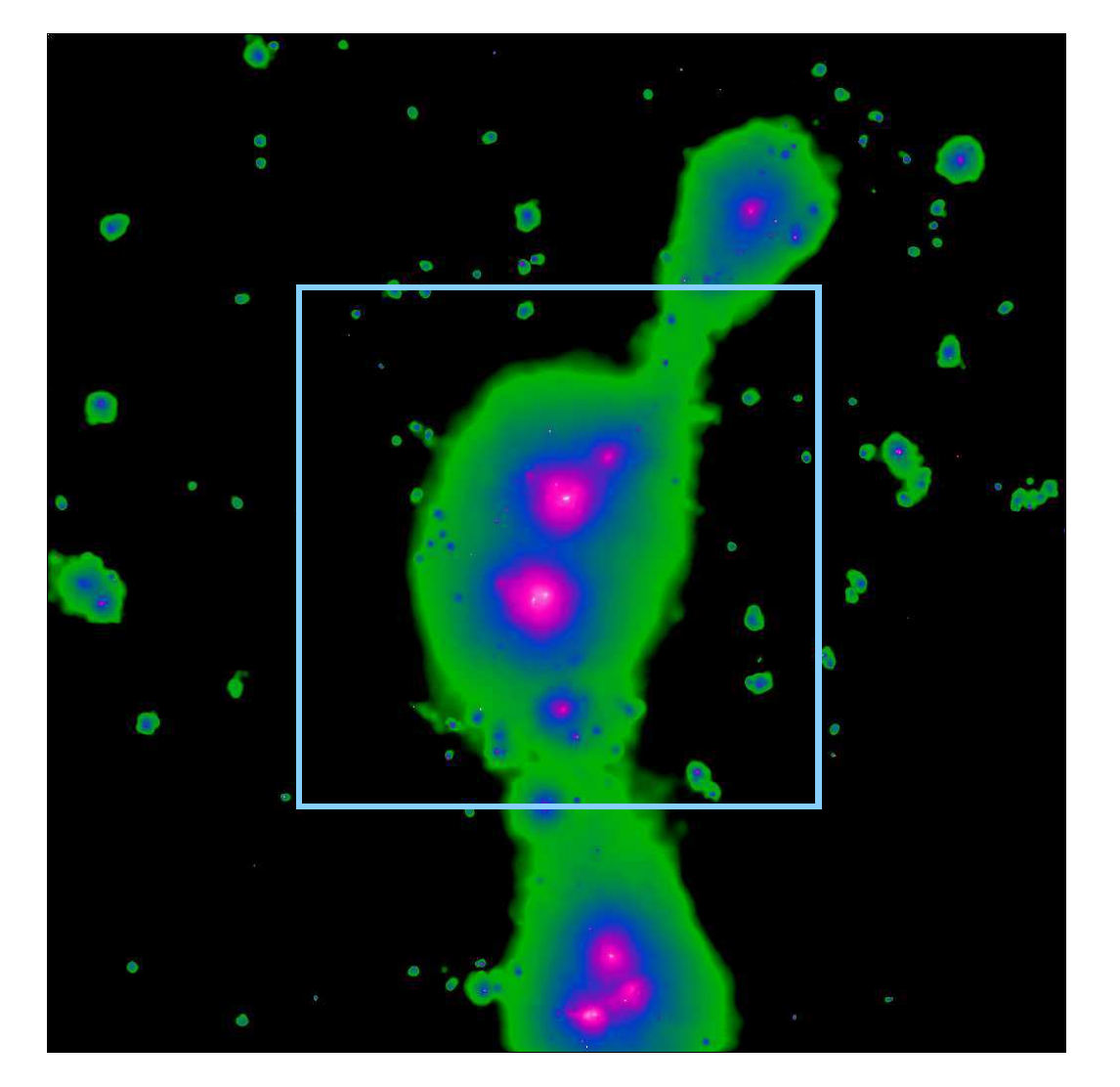}\\[3pt]
	(b) soft X-ray emission\\[3pt]
   \caption{Projected maps of gas density and ideal X-ray emission in the $[0.5$--$2]$\,keV band.
   {Each map is $20\,\cMpc$ per side, integrated for $20\,\cMpc$ along the l.o.s. ($z$-axis) and centered on the pair center of mass at $z=0.07$. The squares ($10\,\cMpc$ per side) approximate the size of the eROSITA observation.}
   \label{fig:evol_maps}}
\end{figure*}

In Fig.~\ref{fig:evol_maps} we show the evolution of the system in its
local environment from $z=1$ down to $z=0.07$ (from left to right).
The maps show the gas surface density (upper row) and ideal X-ray
emission in the $[0.5$--$2]$\,keV band (lower row) at three
representative redshifts $z=1.04$, $z=0.47$ and $z=0.07$, projected
along the $z$-axis and centered on the pair center of mass.  In order
to visualize the local environment around the system, these maps
comprise a comoving volume of $(20\,\cMpc)^3$.  Overdense and
higher-emission regions are marked with lighter colors in the top
panels and blue-magenta colors in the bottom ones.

The evolution of this region shows how the assembly of structures
proceeds along filaments, by the accretion and merging of smaller
structures and diffuse matter.  {From the X-ray emission
  (Fig.~\ref{fig:evol_maps}, lower panels), one} can notice that the
main systems are aligned along a top-down direction, and the two pair
members are enclosed by a common higher-emissivity envelope, tracing
the X-ray emitting diffuse gas.  {From the left-most panel, we see
  that at $z\sim1$ the cores of the main cluster progenitors are
  already distinguishable, as well as the other main haloes in the
  surroundings.}  Compared to the gas density maps, this allows us to
better single out the principal filamentary structure defined by the
main massive structures aligning in the field and embedded in the
diffuse gaseous component.  In fact, minor gas filaments that are
visible in the density maps are too faint to significantly emit in the
X-ray, even when the soft band is considered.  The X-ray maps in
Fig.~\ref{fig:evol_maps} indicate a remarkable resemblance with the
\erosita PV observations of the A3391/95 field, where a $\sim15\,\Mpc$
gas emission filament has been detected thanks to \erosita large FoV
($\sim 1\,$deg) and superior soft X-ray effective area. The observed
filamentary gas emission extends north and south from the A3391/95
system spanning $\sim 4$ degrees in the sky and connecting other
extended structures, namely groups and clusters, all at the same
redshift~\cite[][]{reiprich2020}.

\subsubsection{Gas bridge between the pair clusters.}
\label{sec:bridge}

A debated aspect of the observed A3391/95 system regards the nature of
the gas between the two main clusters.  Following the common
terminology, the gas filament between clusters in close pairs is
typically referred to as ``bridge''.  Depending on the interaction
stage of the two systems, this could be either stripped gas from the
outer cluster atmospheres or actual filament gas.

Distinguishing pure filament gas from outer cluster atmospheres is
{\it per se} very challenging, in observations but also in
simulations. Therefore, it is difficult to define the exact boundaries
of the bridge.  In our simulations, we define the interconnecting
bridge as a three-dimensional cylinder-like volume in between the two
main clusters. For simplicity, we define the bridge main axis as the
three-dimensional line connecting the centers of GC1 and GC2.  For the
radius of the cylinder, we assume $\sim 660\,\kpc$.  Along the main
axis, the bridge is limited to the region outside of $\rvir$ of each
cluster. Given the physical separation of the two clusters, this
corresponds to $\sim 2\,\Mpc$ along the spine direction.  We verified
that either taking a right cylinder (with flat ends) or modifying the
main axis direction to account for asymmetries in the mass
distribution of the clusters does not affect our conclusions.

We primarily consider the volume within a radius of $500\,\ckpc$ from
the spine, namely $\sim 660\,\kpc$ for the chosen redshift and
cosmology, to focus on the core of the bridge and investigate the gas
origin and properties. This radius approximately corresponds to $\sim
0.8\times\rfive$ for both clusters.  We adopt here a similar
definition as in \cite{brueggen2020}, who investigate the bridge of
thermal gas between A3395 and A3391 that was recently observed with
eROSITA, to search for signatures of radio synchrotron
emission. Indeed, they also define the bridge as a cylindrical volume
with a radius of $600\,\kpc$, and a length of $1.3\,\Mpc$ (see
Sec.~\ref{sec:discussion} for further discussion).

\begin{figure}
    \centering
    \includegraphics[width=0.8\columnwidth,trim=0 0 0 20,clip]{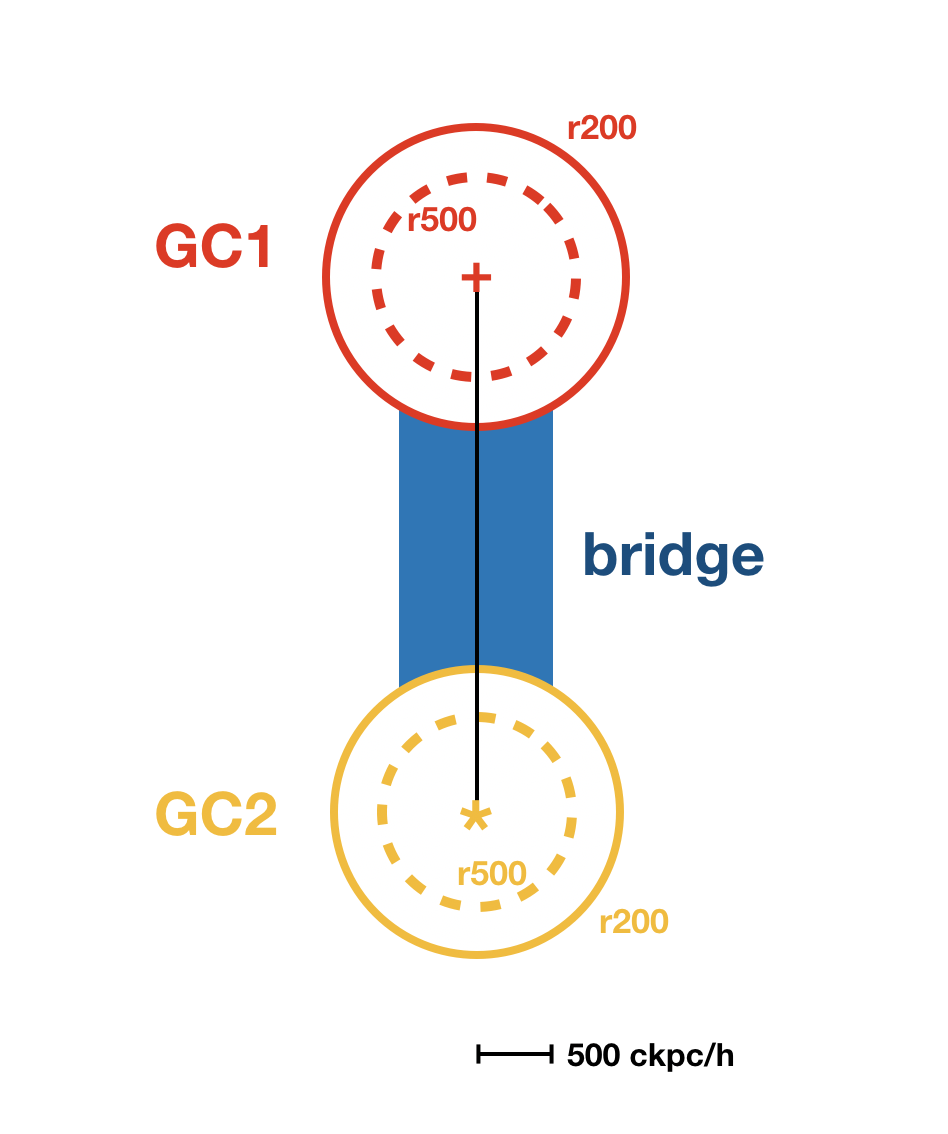}
    \caption{{Schematic view of the simulated system and bridge, rotated to maximize the physical separation between GC1 and GC2.}}
    \label{fig:bridge}
\end{figure}

A schematic view of the system and the bridge, rotated to visualize
the maximum physical distance between GC1 and GC2, is reported in
Fig.~\ref{fig:bridge}. The bridge gas thus defined is also marked in
blue in Fig.~\ref{fig:sims_obs}. The gas mass in the bridge thus
defined is roughly $\sim10\%$ of the gas mass comprised within $\rvir$
of either cluster in the pair.

\section{Results}\label{sec:results}

\subsection{Origin of the pair system}\label{sec:orig}

\begin{figure*}
    \centering
    \begin{tabular}{ccc}
    {$z$ projection}  & {$y$ projection} & {$x$ projection}\\
    \includegraphics[width=.3\textwidth,trim=70 5 70 18,clip]{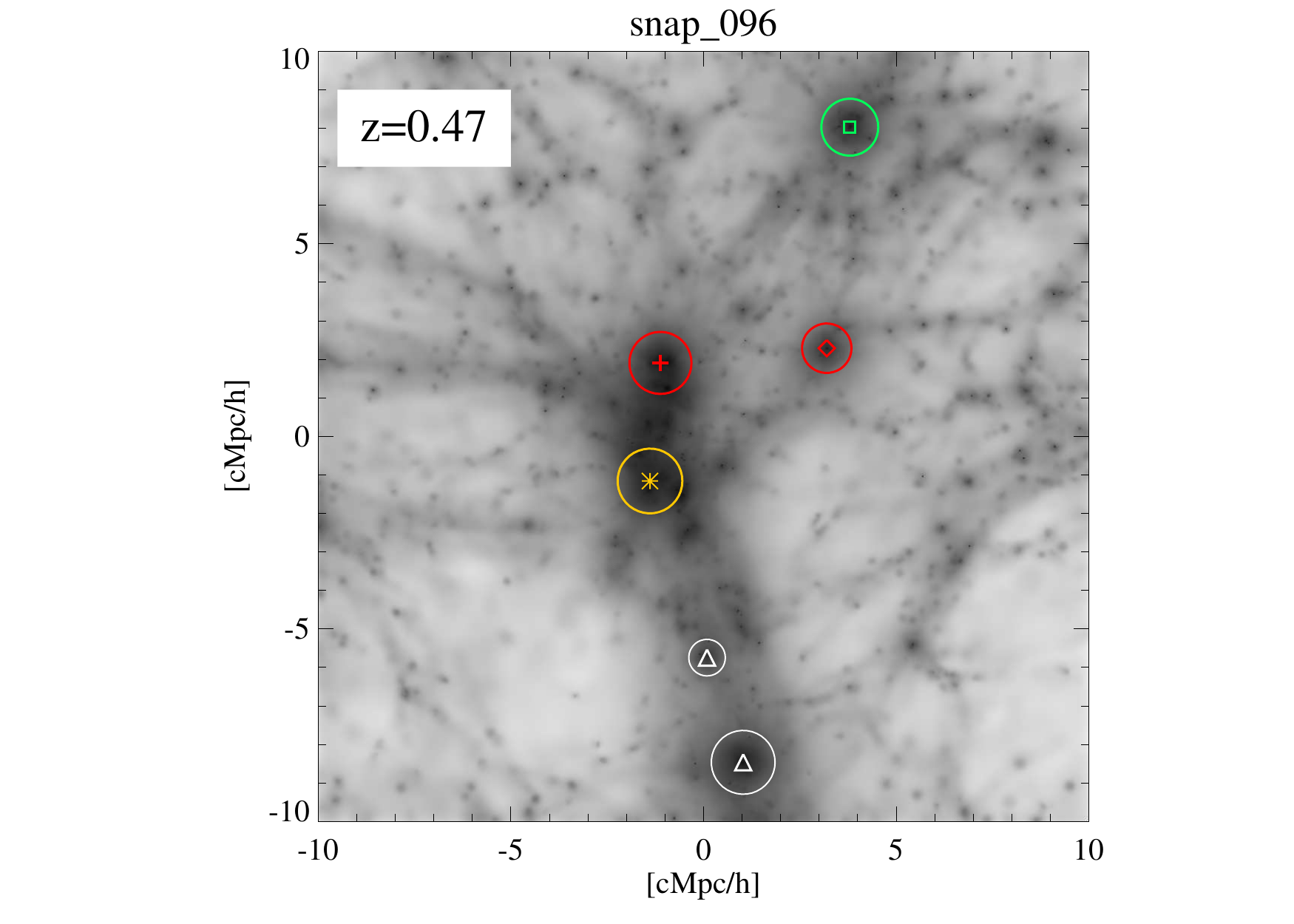}&
    \includegraphics[width=.3\textwidth,trim=70 5 70 18,clip]{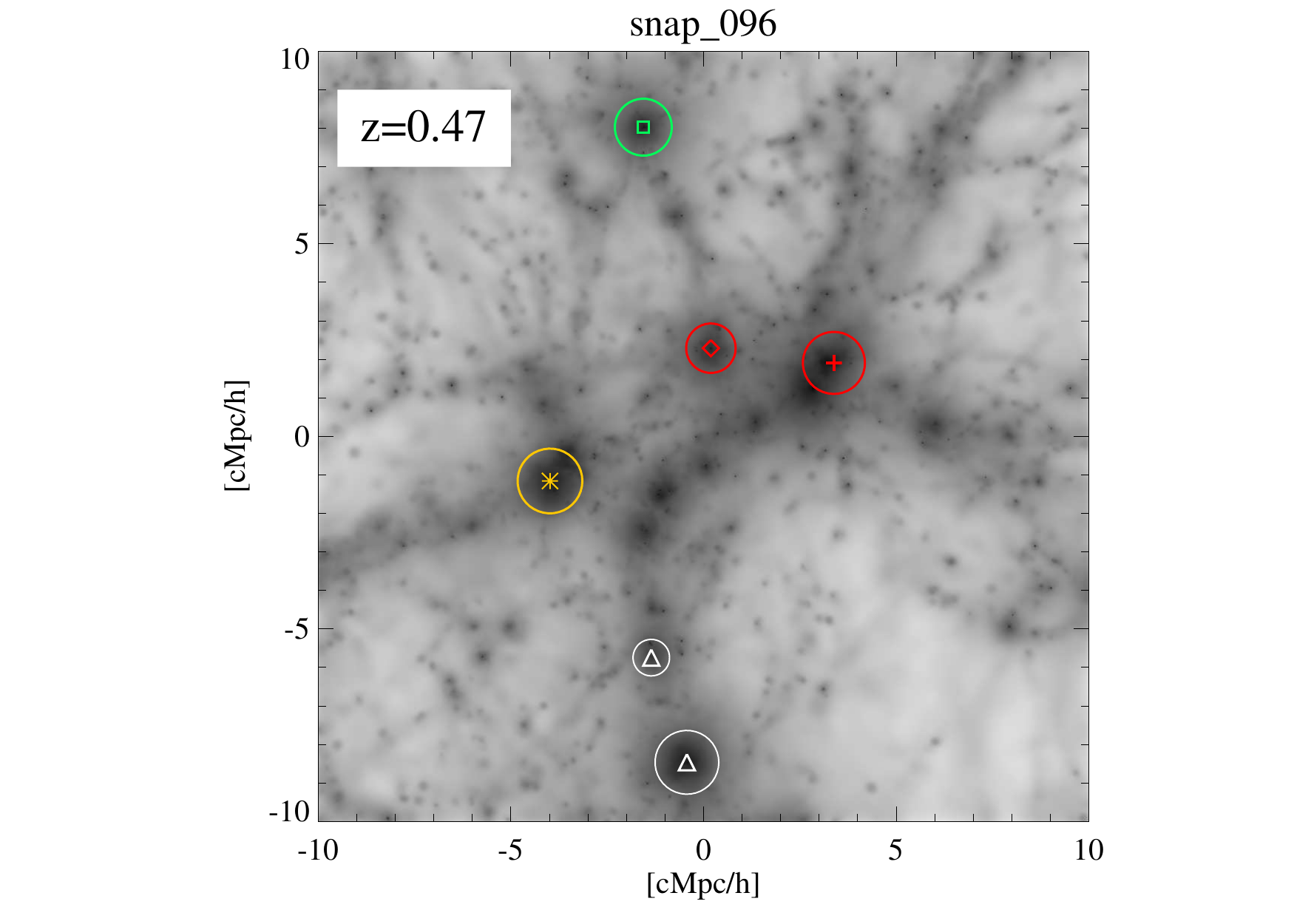}&
    \includegraphics[width=.3\textwidth,trim=70 5 70 18,clip]{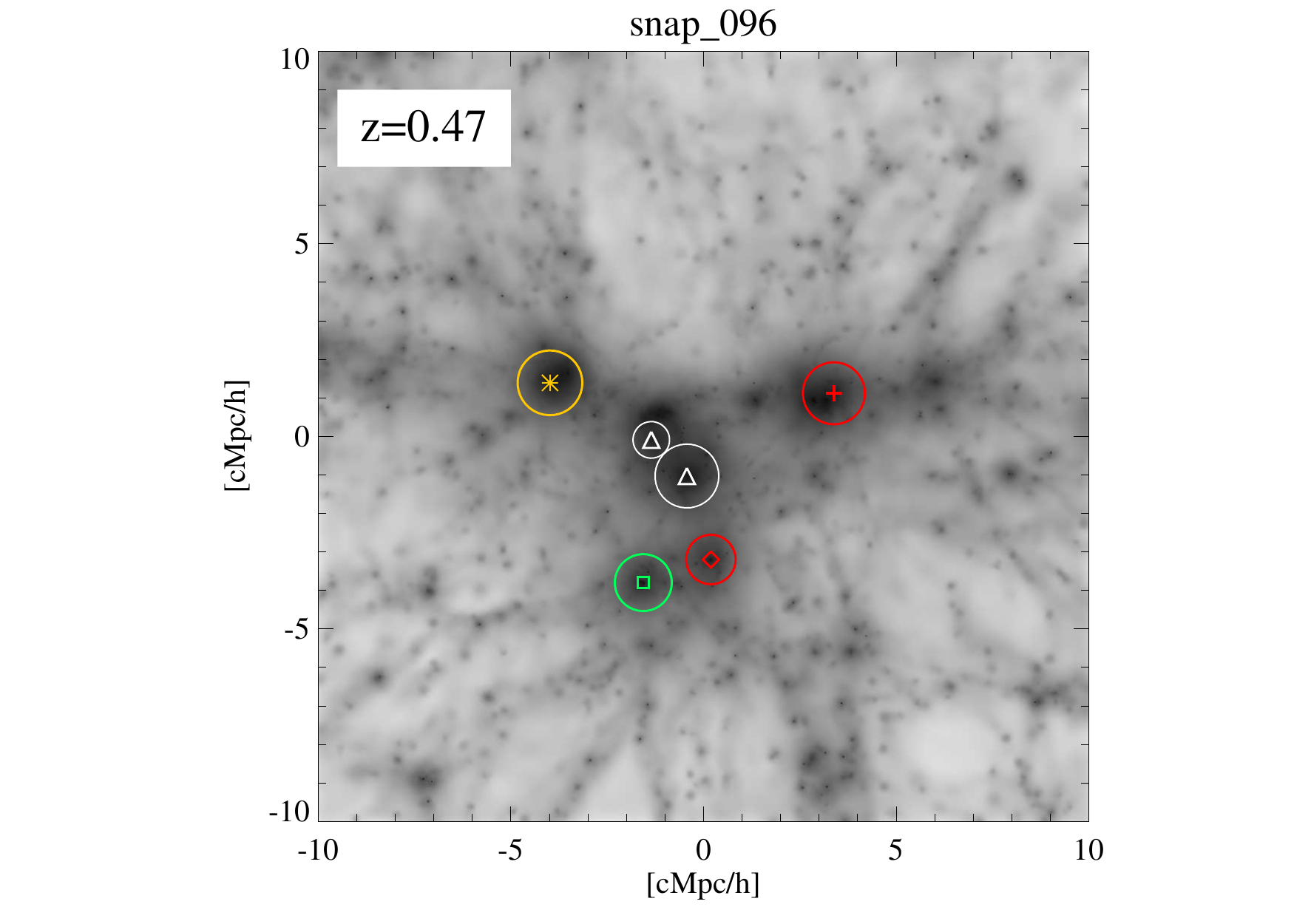}\\
    \includegraphics[width=.3\textwidth,trim=70 5 70 18,clip]{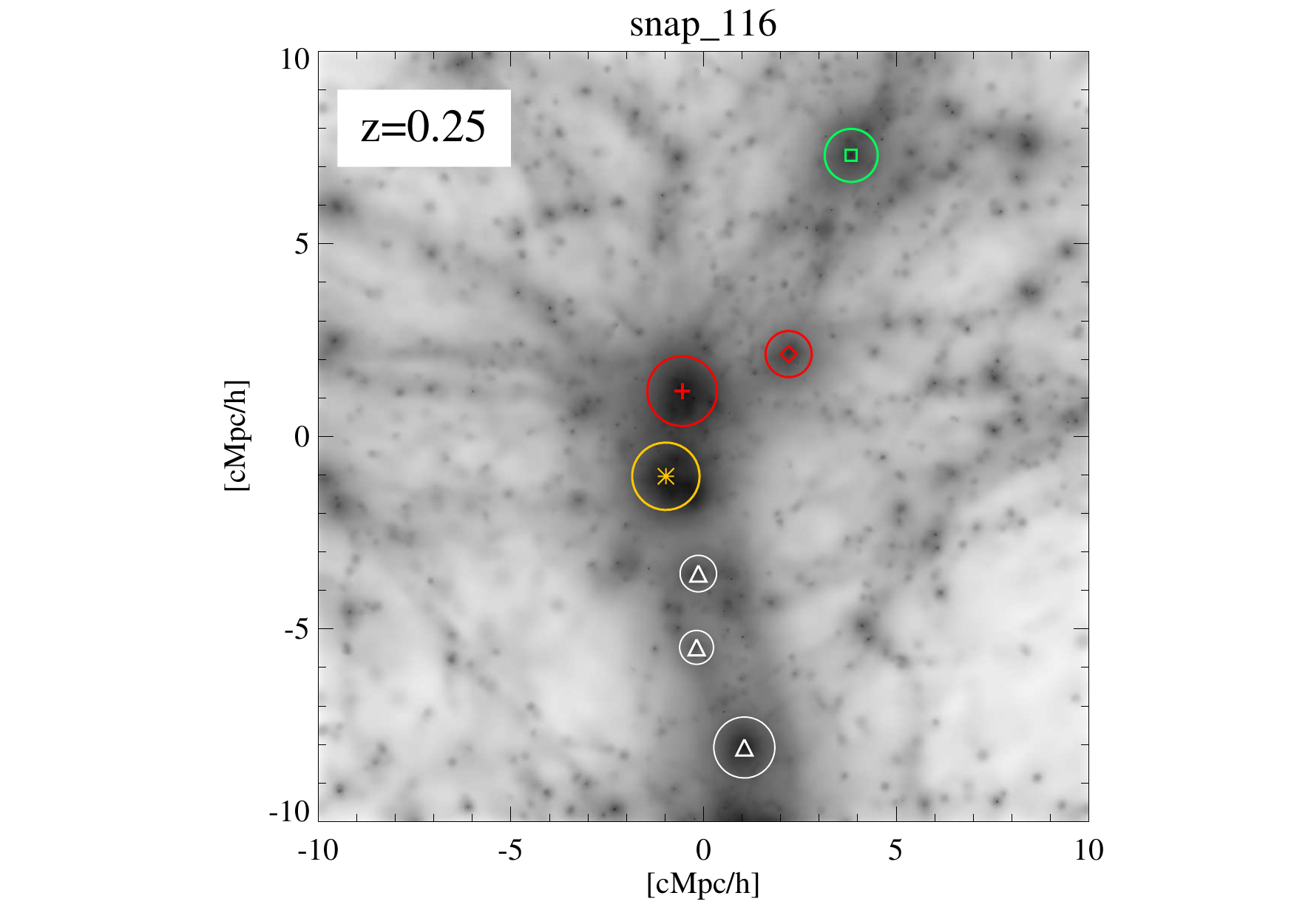}&
    \includegraphics[width=.3\textwidth,trim=70 5 70 18,clip]{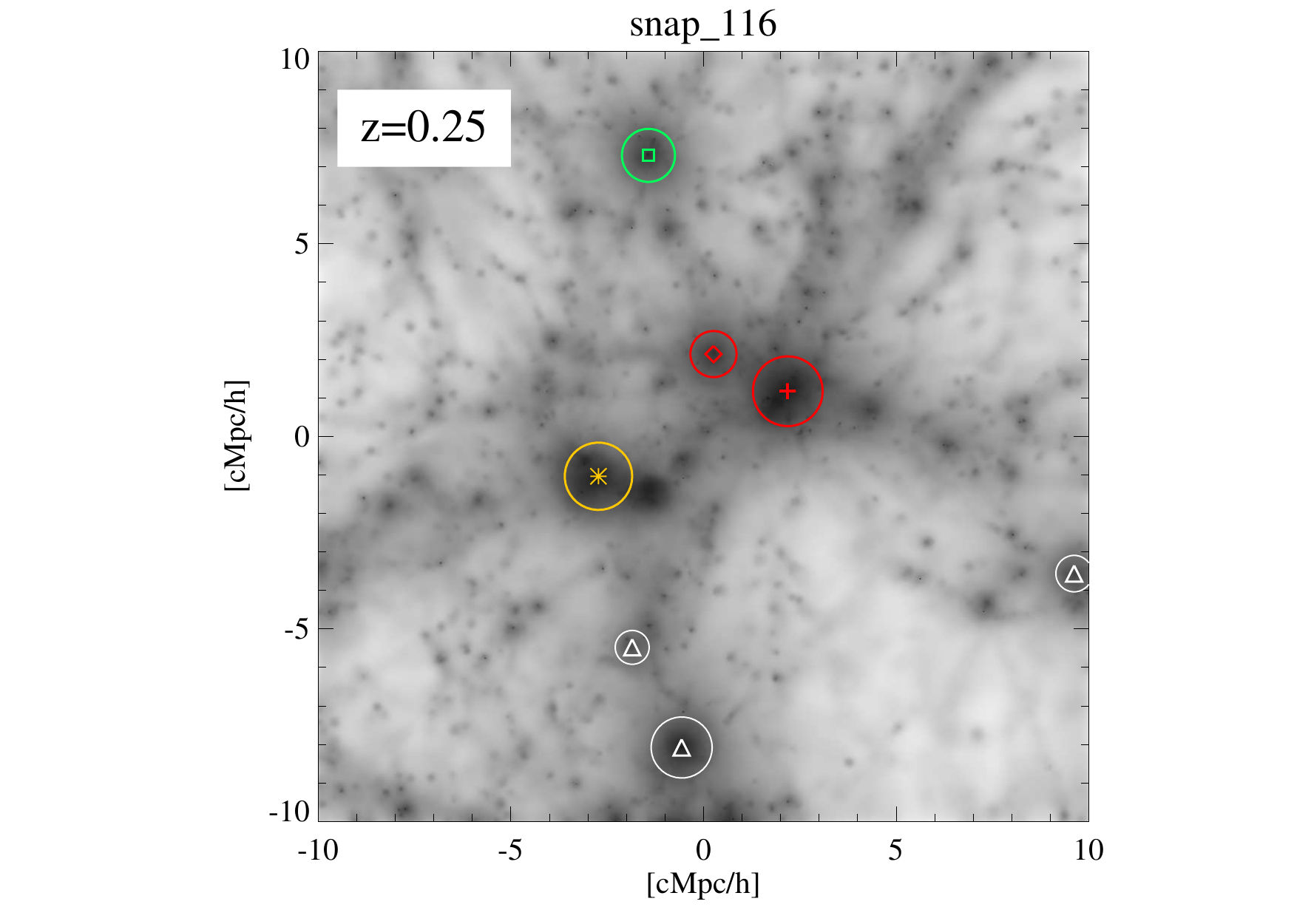}&
    \includegraphics[width=.3\textwidth,trim=70 5 70 18,clip]{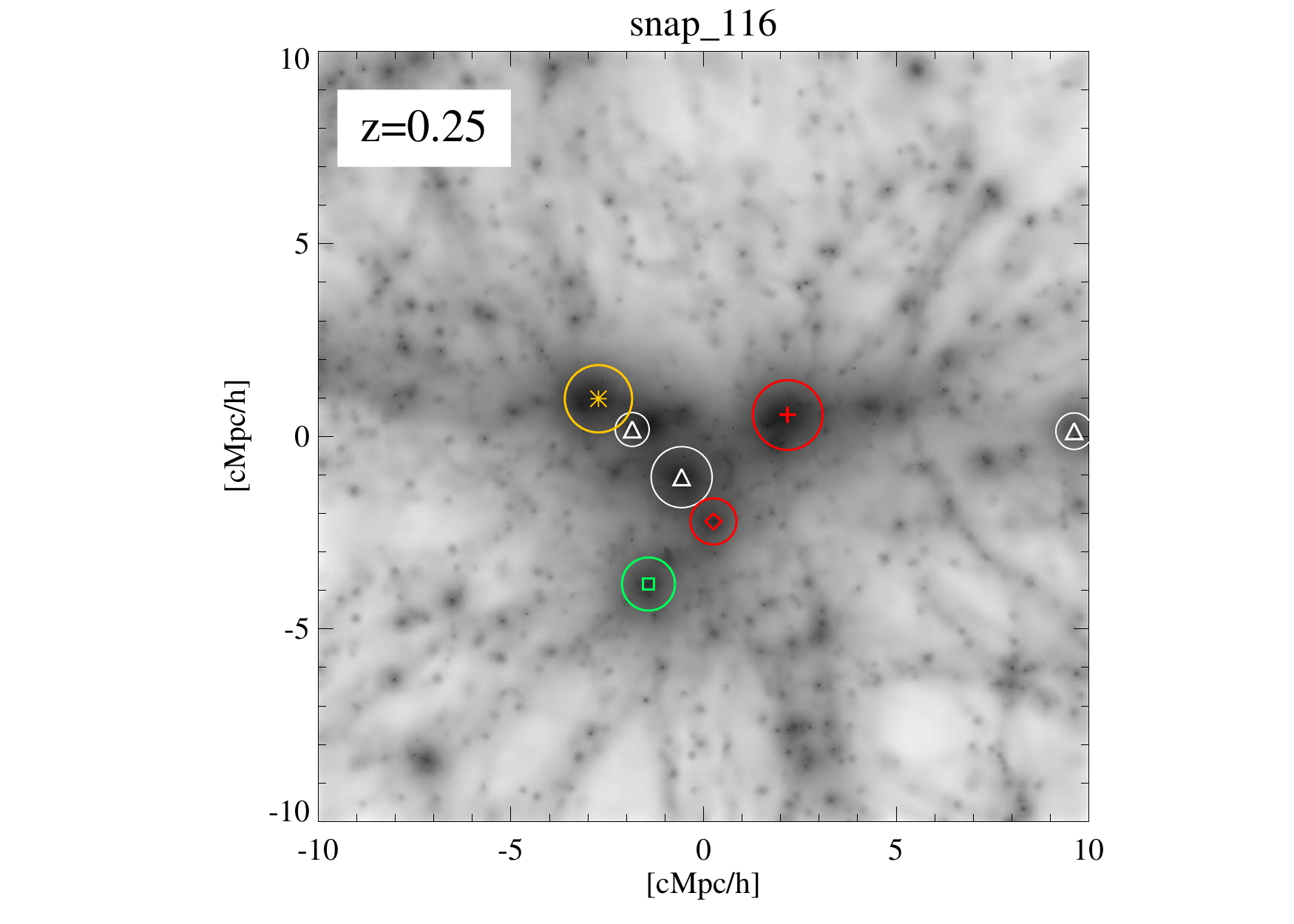}\\
    \includegraphics[width=.3\textwidth,trim=70 5 70 18,clip]{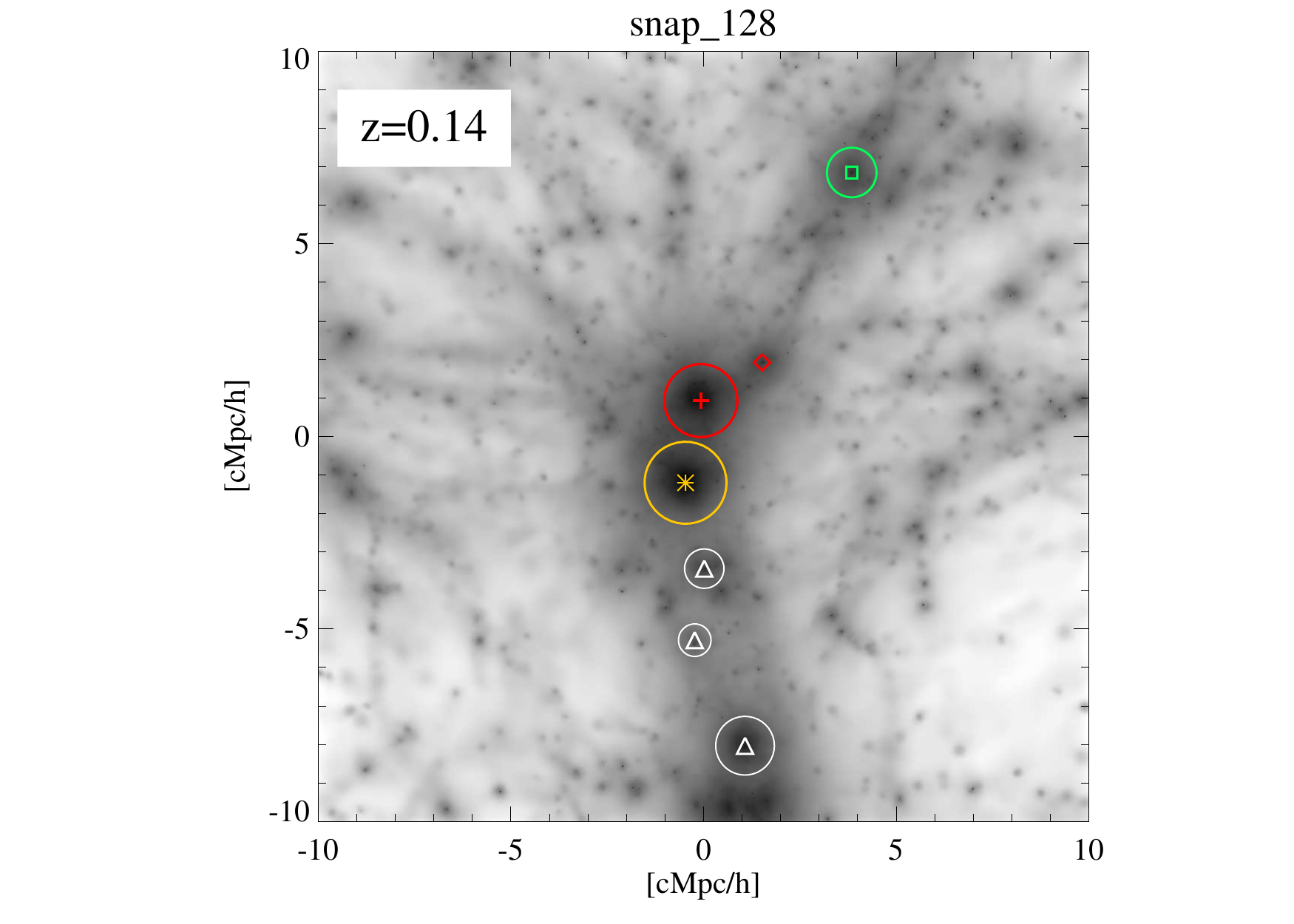}&
    \includegraphics[width=.3\textwidth,trim=70 5 70 18,clip]{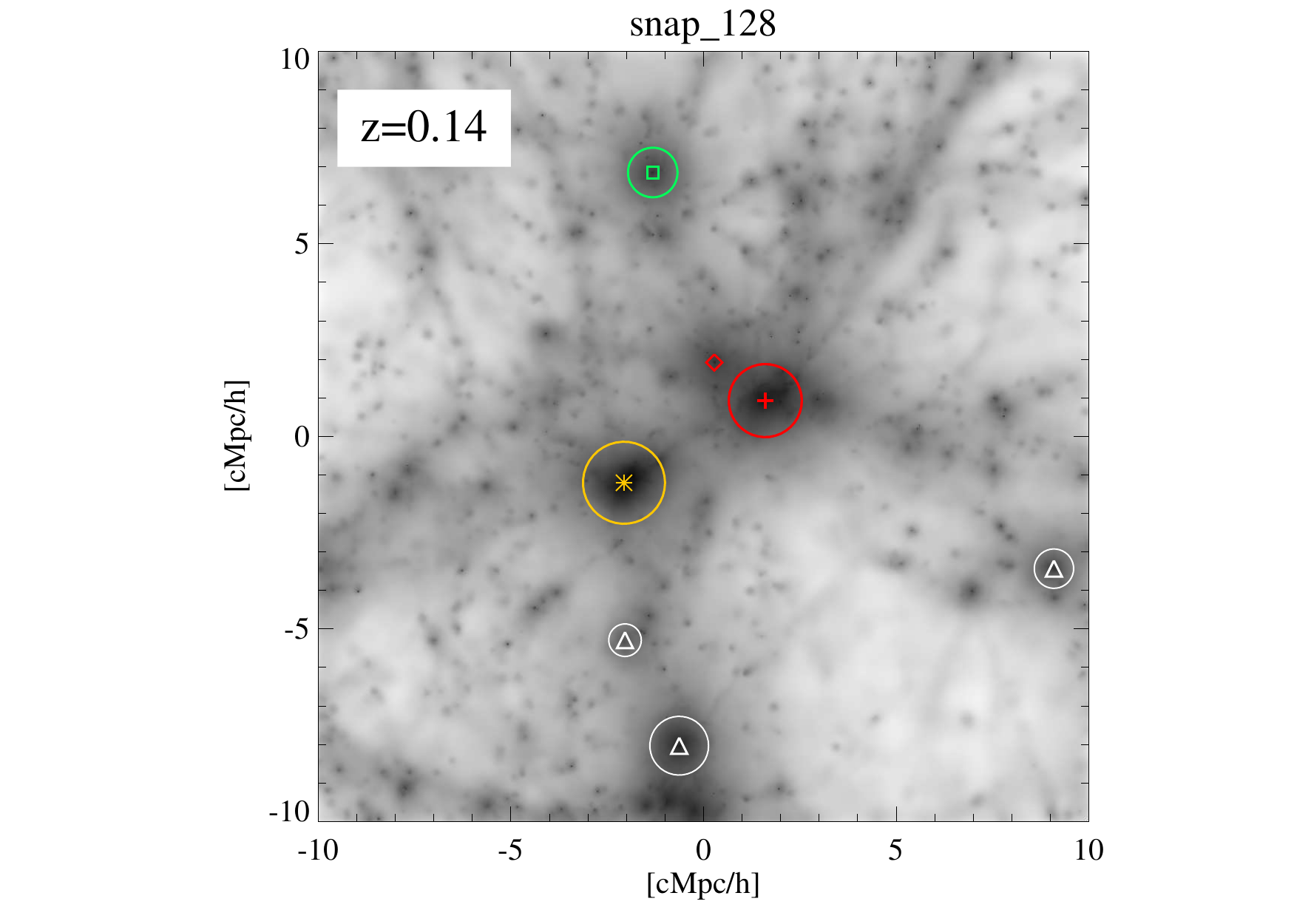}&
    \includegraphics[width=.3\textwidth,trim=70 5 70 18,clip]{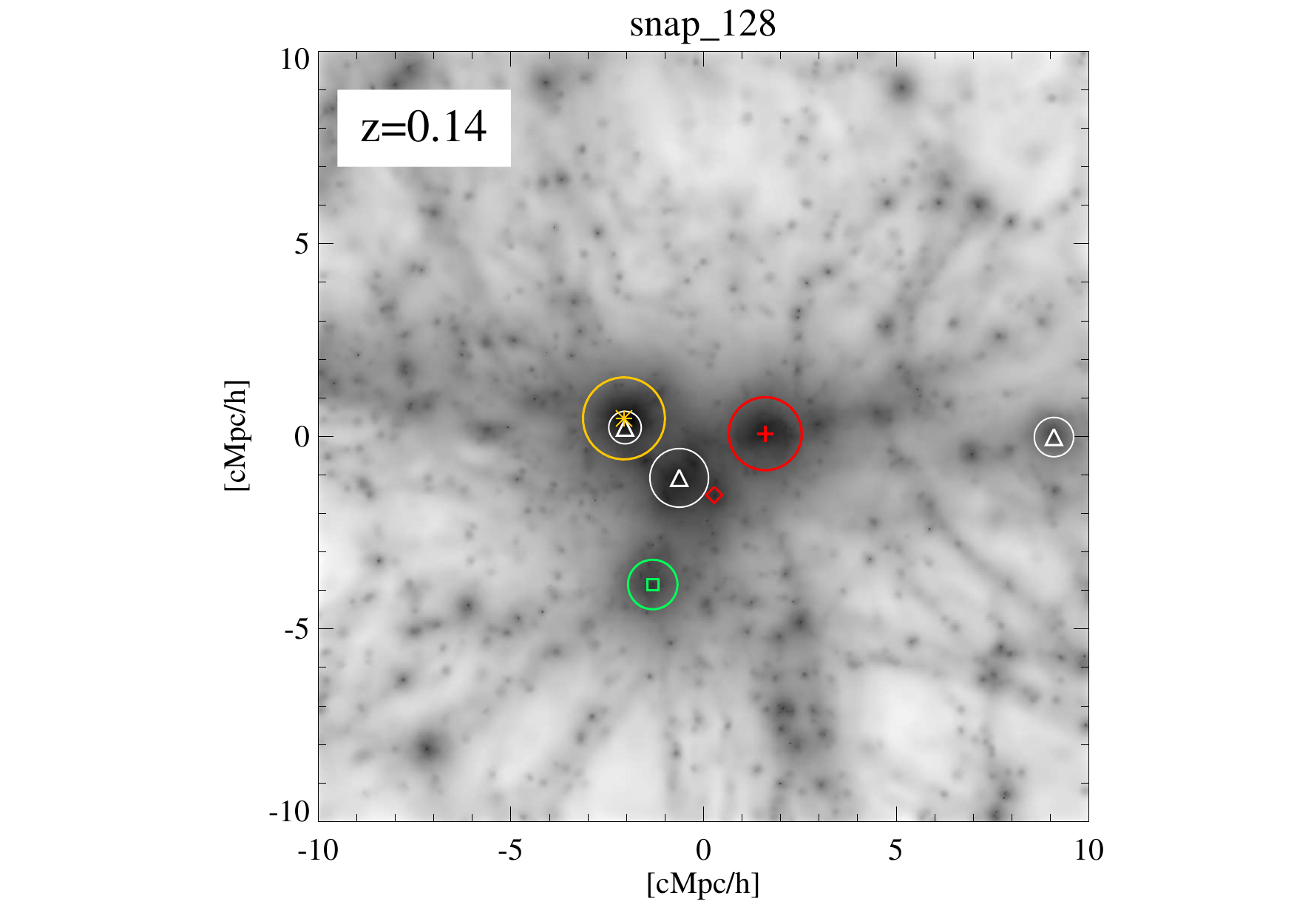}\\
    \includegraphics[width=.3\textwidth,trim=70 5 70 18,clip]{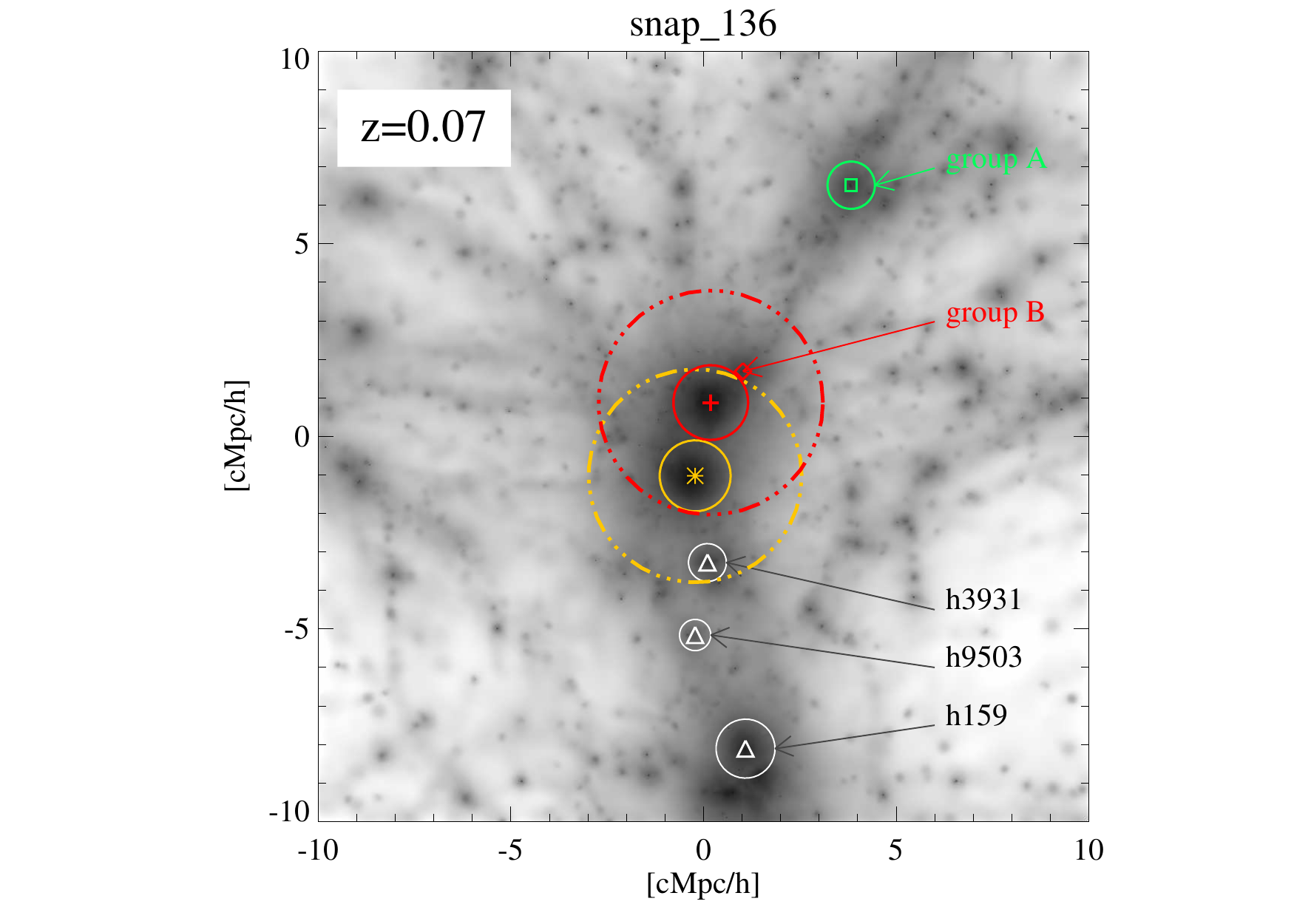}&
    \includegraphics[width=.3\textwidth,trim=70 5 70 18,clip]{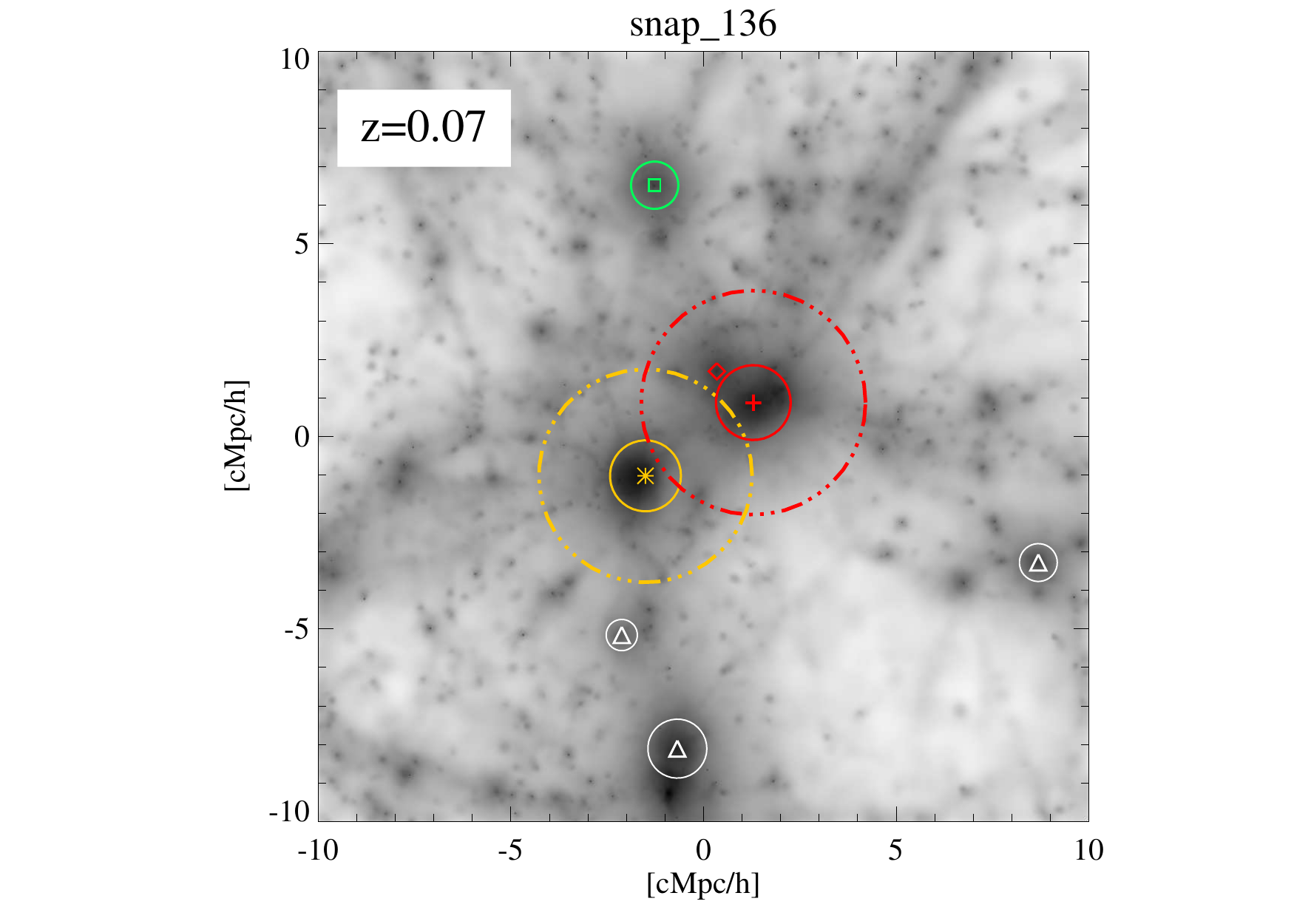}&
    \includegraphics[width=.3\textwidth,trim=70 5 70 18,clip]{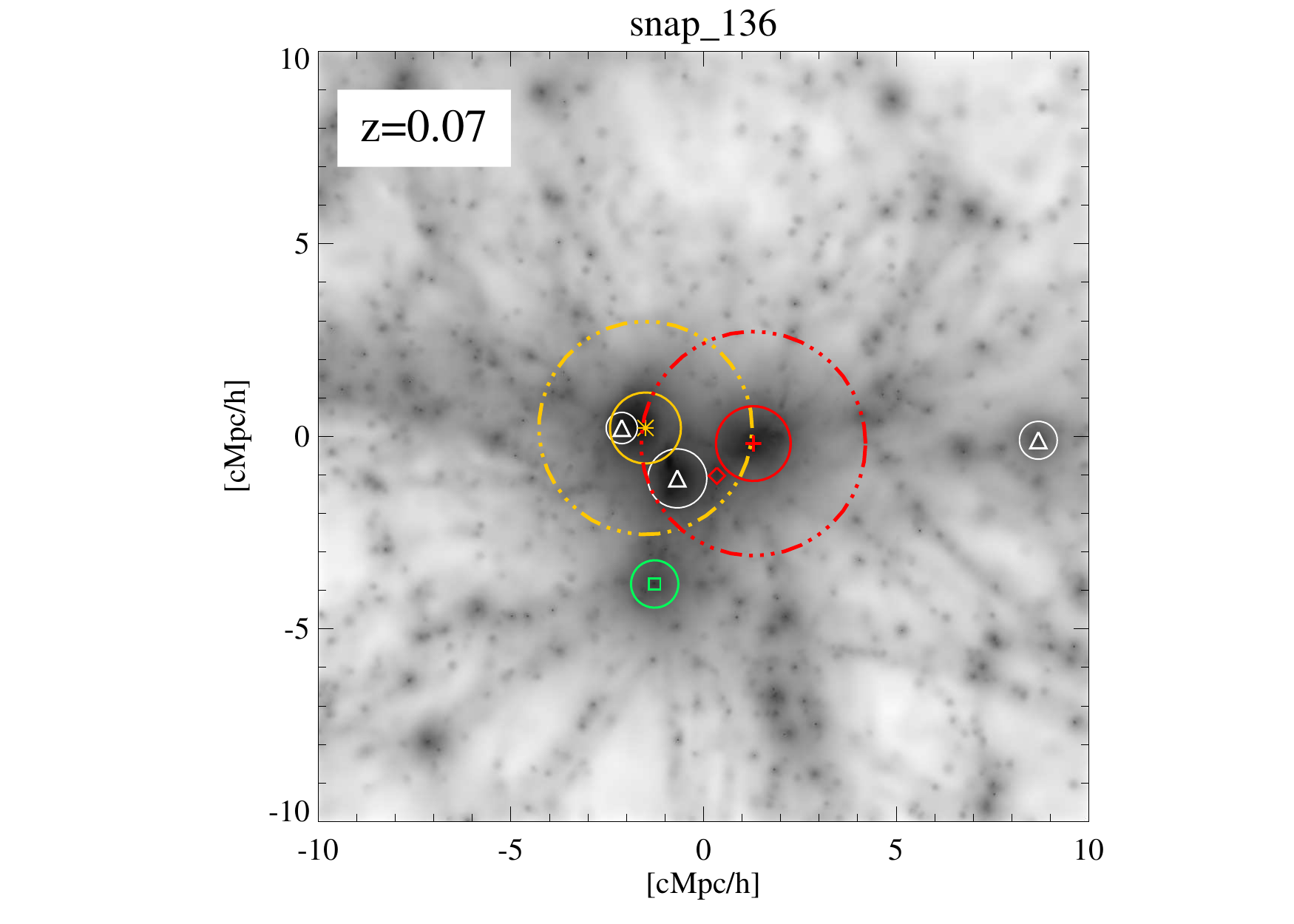}
    \end{tabular}
    \caption{{Evolution of the main haloes in the pair region, in
        the three main projection axes ($z$, $y$ and $x$, from left to
        right). The maps show projected gas density within a
        $(20\,\cMpc)^3$ comoving volume centered on the final (at
        $z=0.07$) pair center of mass, } for different redshifts
      between $z=0.47$ and $z=0.07$ (from top to bottom).  We mark GC1
      and GC2 (red {cross and yellow asterisk}, respectively), {the
        two northern groups A and B ({green square and red
          diamond}), and the additional main group-size haloes in the
        region (white triangles), and their progenitors.}  For each
      halo, circles approximate their $\rvir$ extent, except for group
      B at $z\lesssim0.16$, when it is identified as a substructure
      gravitationally bound to GC1.  {For GC1 and GC2, the dot-dashed
        circles at $z=0.07$ correspond to $3\times \rvir$.}
    } \label{fig:trck_gas_prog}
\end{figure*}

In Fig.~\ref{fig:trck_gas_prog} we show the evolution of the system at
four redshifts between $z\sim0.47$ and $z=0.07$.  The gas density
maps, projected along the three major axes of the simulations ($z$,
$y$ and $x$ from left to right), comprise the local environment within
a comoving {volume of $(20\,\cMpc)^3$ centered} on the pair center
of mass.  In addition to the pair members GC1 and GC2, {we also
  mark} the position and $\rvir$ extent of the most massive groups
{identified at $z=0.07$} in the region (see Sec.~\ref{sec:clumps},
in the following).  The maps allow to follow the assembly of the knot
region in the last $\sim 4$ gigayears, and the complicated
three-dimensional geometry connecting the multiple haloes and
filaments around it.

Filamentary structures in the gas distribution are observed in all
maps, with a complex three-dimensional geometry.  In fact, several
gaseous filaments can be noticed, even without significant group-size
haloes located in them, such as those in the regions left and right of
the main halo chain in the $z$-projection (left panels). These are
also recognizable in the other two projections.  {Such minor
  filaments, however, are expected to be faint in the X rays, as
  noticed from the lower panels in Fig.~\ref{fig:evol_maps}.}  From
the redshift evolution in Fig.~\ref{fig:trck_gas_prog} we {follow}
the whole region, filaments and haloes, collapsing to form the central
Cosmic Web knot, where the cluster pair is finally located.

Focusing in particular on the pair system, simulations allow us to
explicitly investigate the origin of the gas between the two member
clusters.  {It is evident, from the evolution of the halo
  progenitors shown in Fig.~\ref{fig:trck_gas_prog}, that the pair
  members GC1 and GC2 have not merged yet by $z=0.07$.}  Despite the
$\rvir$ of GC1 and GC2 almost overlap in the reference projection (see
Fig.~\ref{fig:sims_obs}, left), we remind here that the two clusters
are indeed physically separated by a larger distance (namely, by two
times the sum of their {$\rvir$} radii, as described in
Sec.~\ref{sec:analog}) --- which might also be the case for the
observed A3391/95 system~\cite[][]{tittley2001,reiprich2020}, although
it is difficult to assess the true physical separation from redshift
arguments only.  In the simulations, we can explicitely visualize this
from the additional projections along the $y$ and $x$ axes in
Fig.~\ref{fig:trck_gas_prog} (last row, central and right-hand-side
panels).

Here, we aim at assessing whether the gas in the bridge can be
stripped gas from the outer atmosphere of the two clusters or rather
true filament-like gas.  To this {end}, we select the gas that
resides within GC1, GC2 and in the interconnecting bridge at $z=0.07$,
and track it back in time.  For the two GC1 and GC2 clusters, we
consider all the gas particles enclosed within their $\rfive$.
{The gas bridge is selected in the region between GC1 and GC2, as
  defined in Sec.~\ref{sec:bridge}}

\begin{figure*}
    \centering
    \includegraphics[width=.42\textwidth,trim=20 5 10 0,clip]{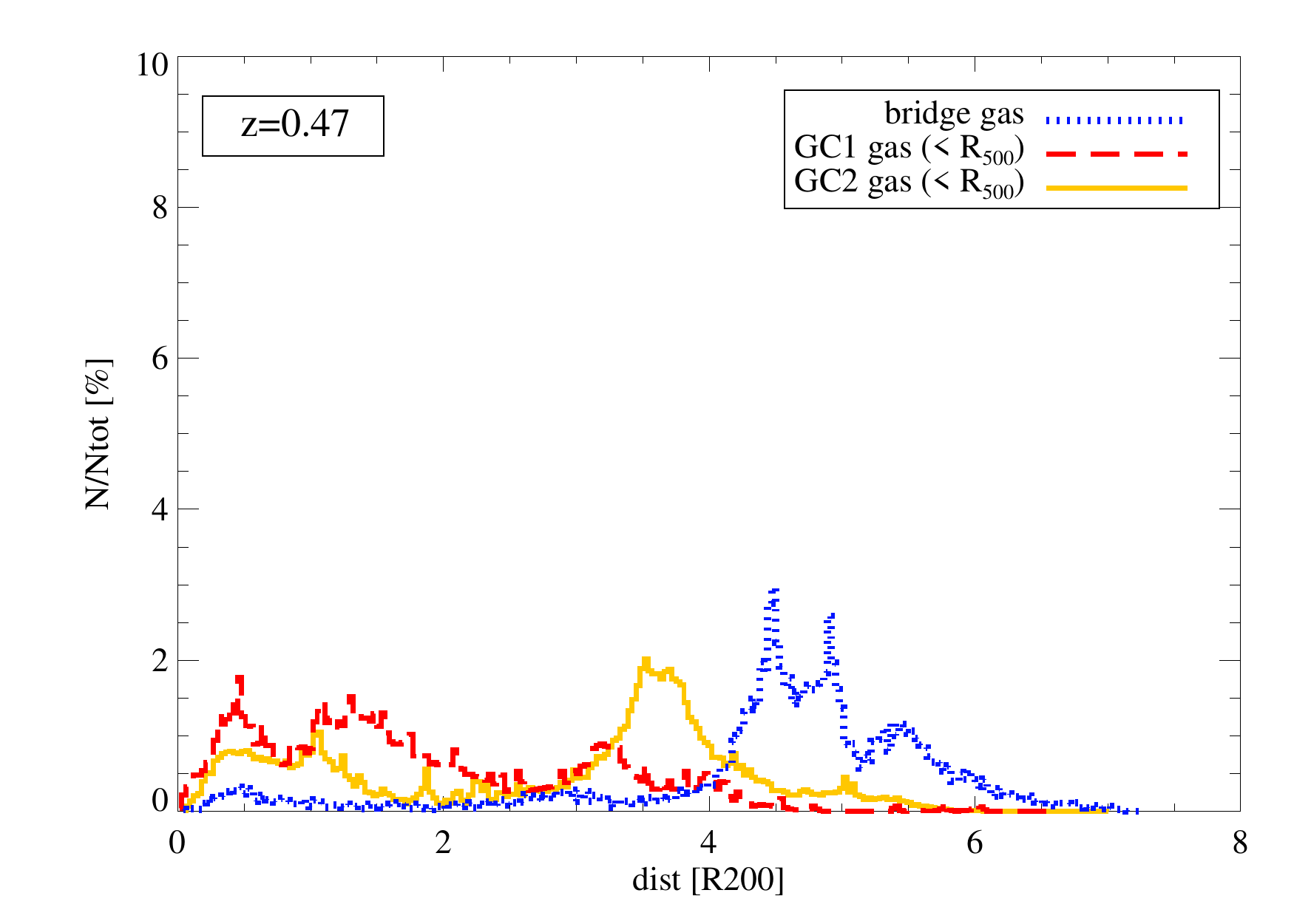}\quad
    \includegraphics[width=.42\textwidth,trim=20 5 10 0,clip]{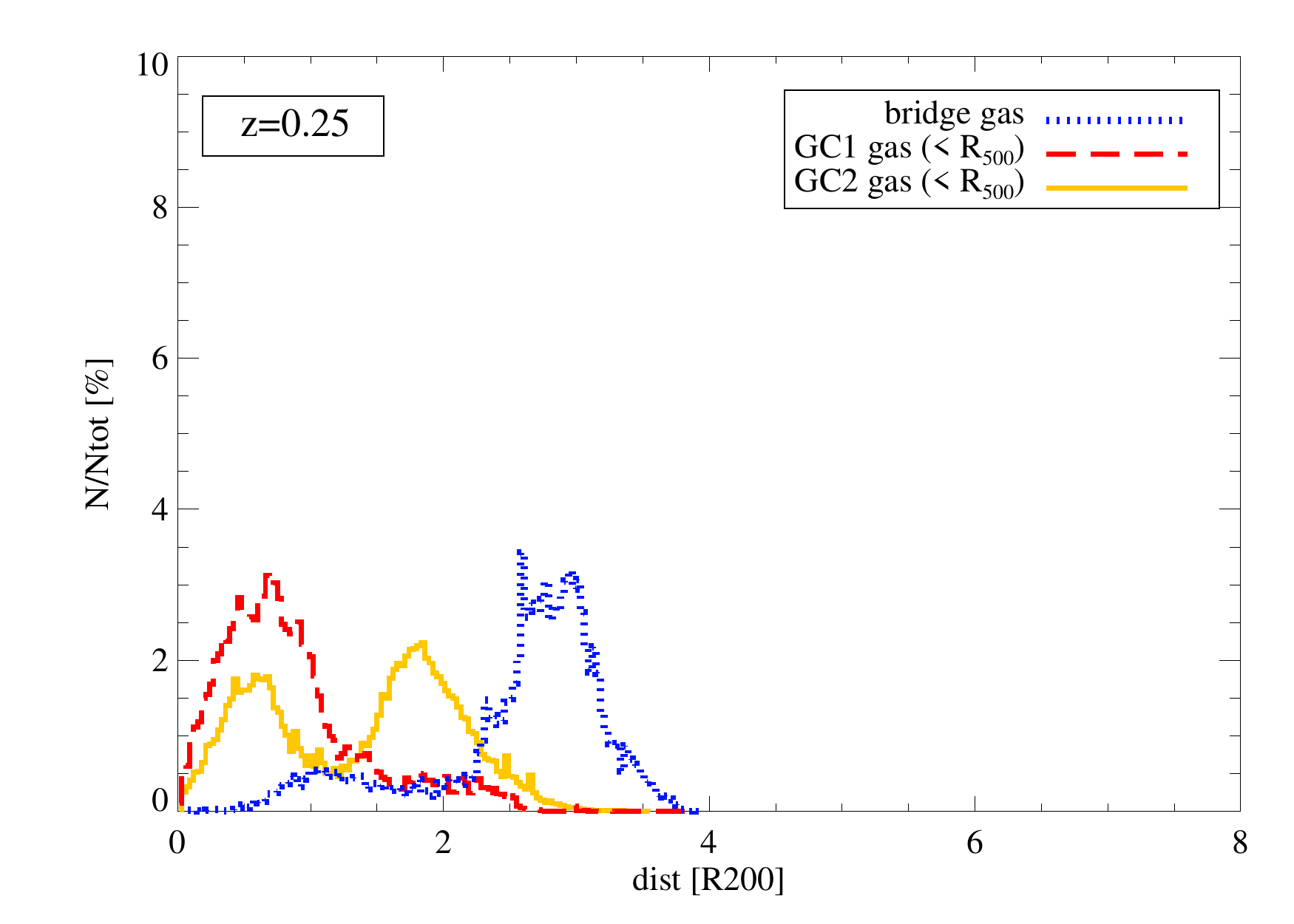}\\[5pt]
    \includegraphics[width=.42\textwidth,trim=20 5 10 0,clip]{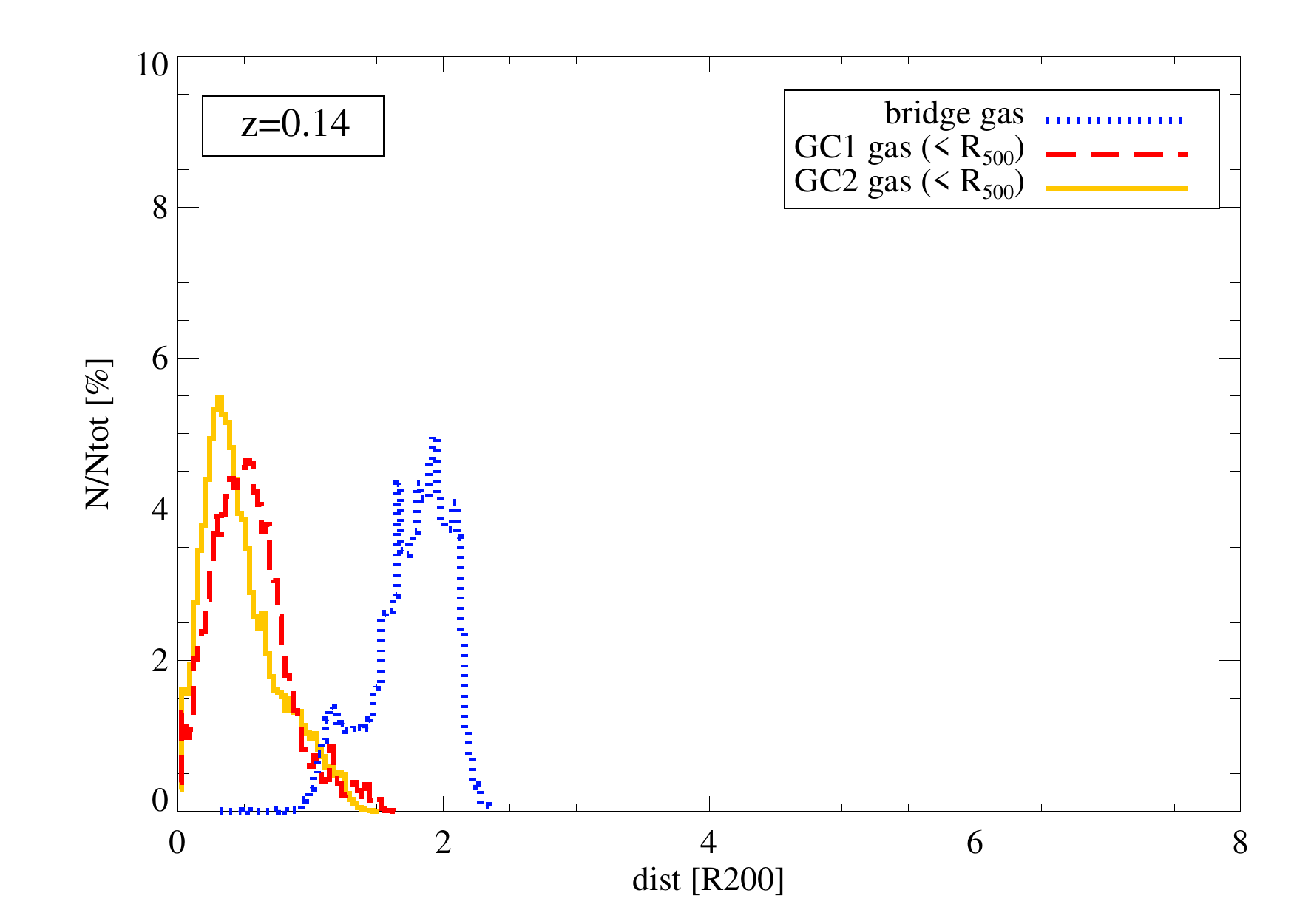}\quad
    \includegraphics[width=.42\textwidth,trim=20 5 10 0,clip]{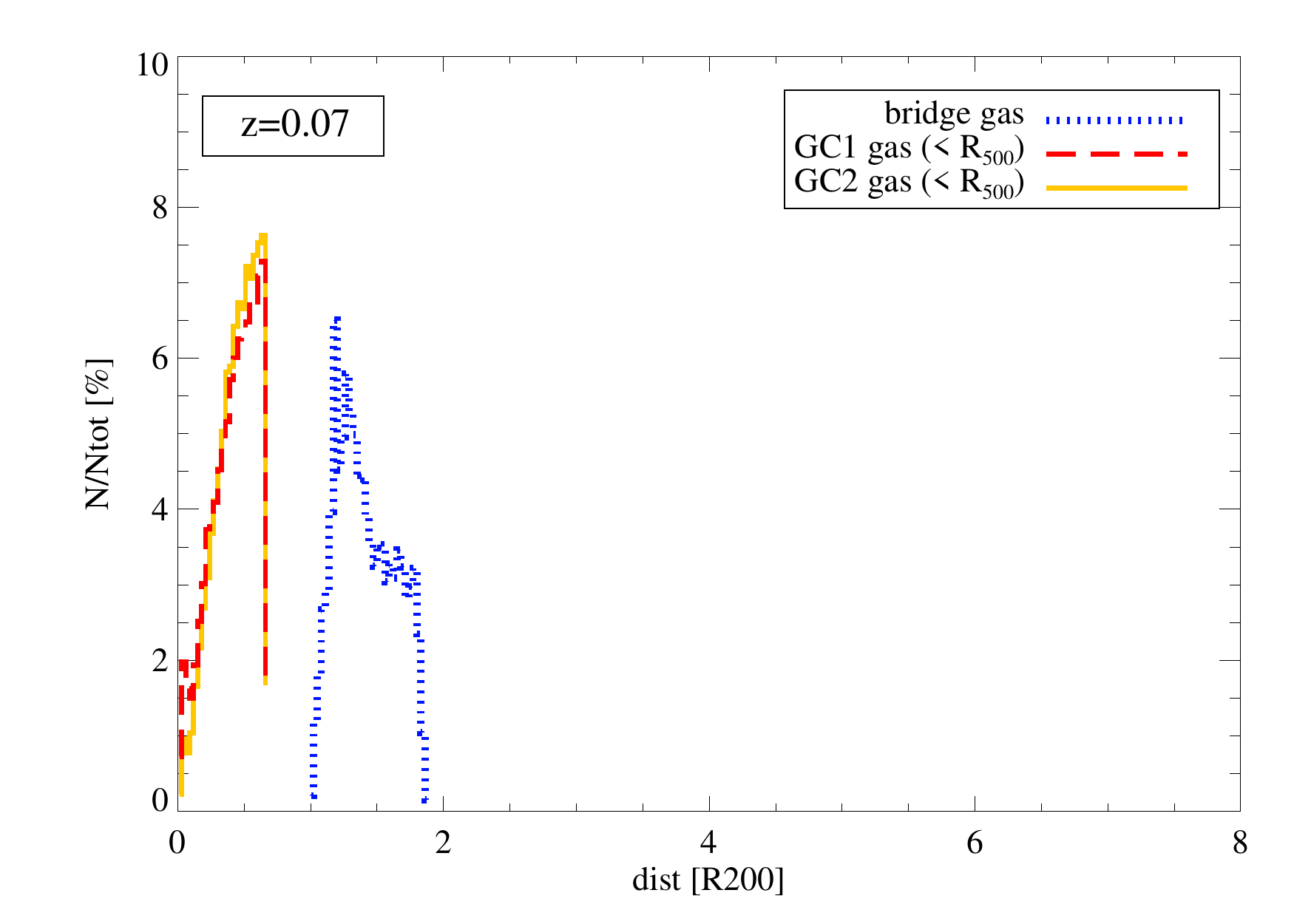}
    \caption{Evolution between $z\sim 0.47$ and $0.07$ of the gas distance with respect to the closest progenitor of the two pair members, in units of its $\rvir$ at the given redshift. Gas is selected at $z=0.07$ within the bridge (blue dotted line) or within $\rfive$ of each cluster (GC1: red dashed line; GC2: yellow solid line).
    } \label{fig:trck_gas_dist}
\end{figure*}

For each selection, {Fig.~\ref{fig:trck_gas_dist} shows the
  evolution of the gas distances from the closest cluster progenitor,
  between redshift $z=0.47$ and $z=0.07$}.  In particular, the {red
  dashed and yellow solid lines} refer to the gas residing within
$\rfive$ from GC1 and GC2, respectively. {The blue dotted line
  marks} the gas selected within the pair bridge.

We find that the bulk (90--95\%) of the gas in the bridge at $z=0.07$
was never inside $\rvir$ of either cluster progenitor, and a large
majority ($\gtrsim 80\%$) was actually always beyond $2\times \rvir$
down to $z\sim 0.25$.  {Since $R_{\rm vir}\sim 1.5\times\rvir$,
  $\gtrsim 80\%$ of the bridge gas was essentially never within the
  virial radius ($R_{\rm vir}$) of the clusters main progenitors.}

At lower redshifts, when the pair system approaches the final
configuration, the gas bridge is more and more confined between
$\rvir$ and $2\times \rvir$, where it is finally identified.  {From
  our analysis, we conclude that} the diffuse gas in the bridge in
this case is mostly filament-like and comprises only a minor fraction
of gas which was mixed in from the outer atmosphere of the two main
clusters.

Compared to the bridge gas, Fig.~\ref{fig:trck_gas_dist} shows that
the gas selected within $\rfive$ in both clusters has been more
smoothly accreted during their formation history. Below $z\sim 0.25$
it almost entirely resides already within $2\times \rvir$.  Both GC1
and GC2 undergo significant mergers, which occur at $z\lesssim 0.5$
for GC1 and more recently, at $z\sim 0.2$, for GC2.  The latter, in
particular, is clearly shown in the top-right panel of
Fig.~\ref{fig:trck_gas_dist}, where the two clear components in the
gas distance distribution of GC2 (yellow {solid line}) approach and
merge between $z\sim 0.25$ and $z\sim 0.14$ (also visible from the gas
density maps in Fig.~\ref{fig:trck_gas_prog}). In the case of GC2, a
significant fraction of the gas selected within $\rfive$ at $z=0.07$
comes from the merger, as $\sim 55(34)\%$ of it is still beyond
$\rvir$ at $z\sim 0.25(0.18)$.  Differently, most (70\%) of the gas
finally residing within $\rfive$ of GC1 is already within its virial
boundary ($<\rvir$) by $z=0.25$, whereas only $\sim 5\%$ is farther
than $2\times\rvir$.  From this, we therefore also conclude that the
gas within the $\rfive$ of GC1 does not come from the northern group
B, which is finally entering the atmosphere {($R_{\rm vir}$)} of
GC1 at $z\sim 0.16$ for the first time.  {Compared to the observed
  A3391/95 system, this is consistent with a picture where the ICM of
  the Northern Clump, located at $3\times\rvir$ relative to A3391, has
  not yet been stripped and mixed in with the ICM of A3391.}

\begin{figure*}
    \centering
    \includegraphics[width=.33\textwidth,trim=85 10 65 15,clip]{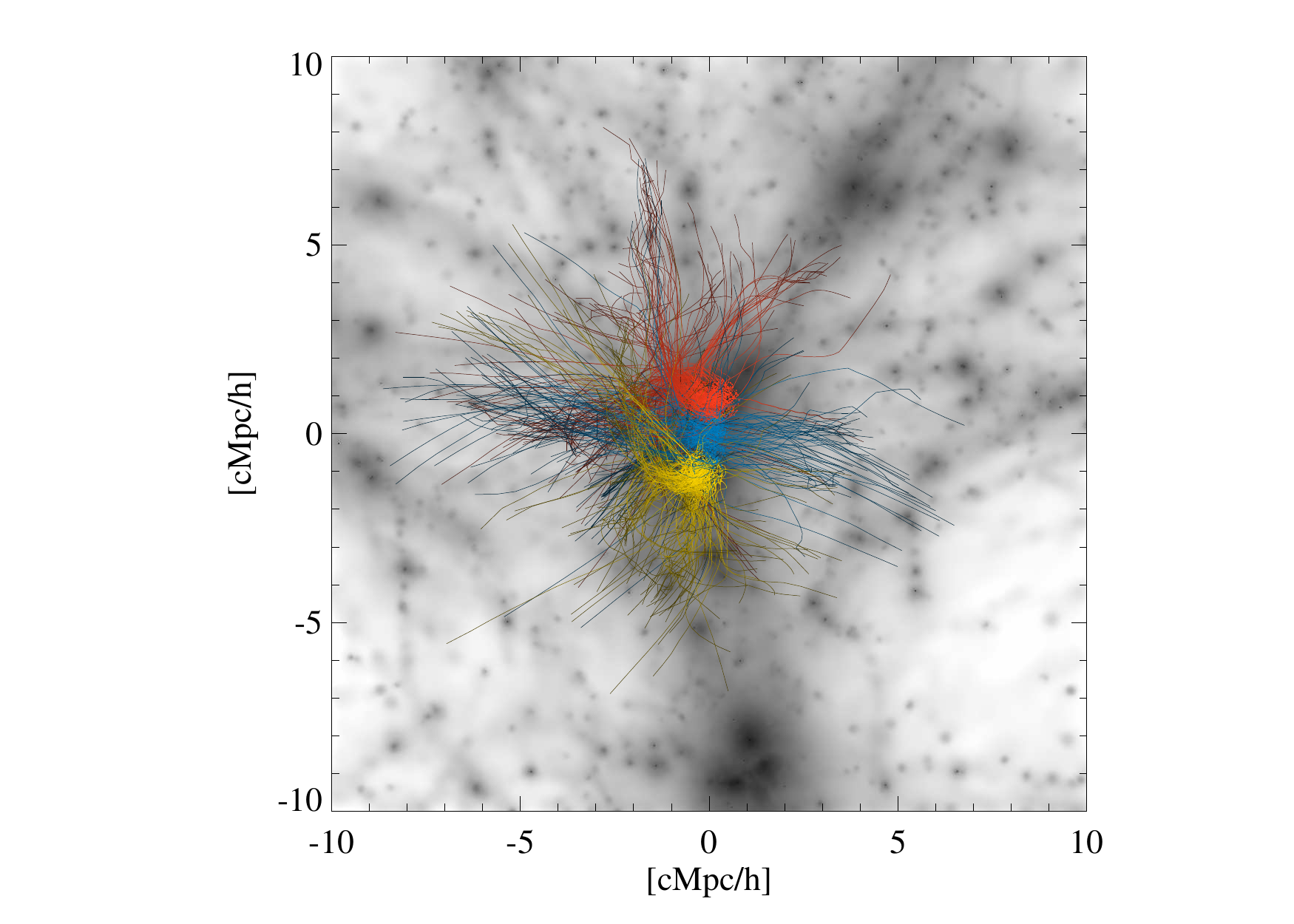}
    \includegraphics[width=.33\textwidth,trim=85 10 65 15,clip]{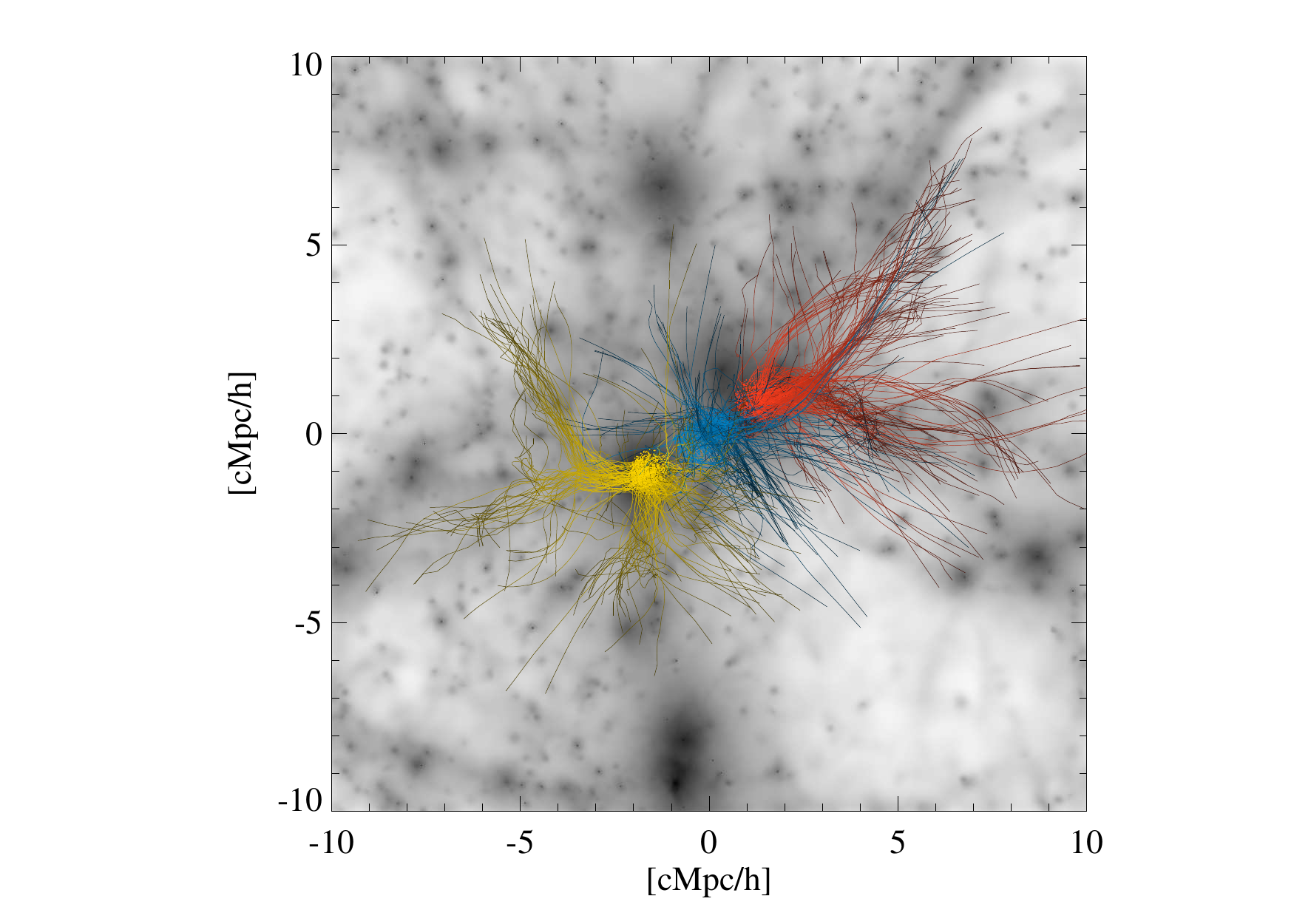}
    \includegraphics[width=.33\textwidth,trim=85 10 65 15,clip]{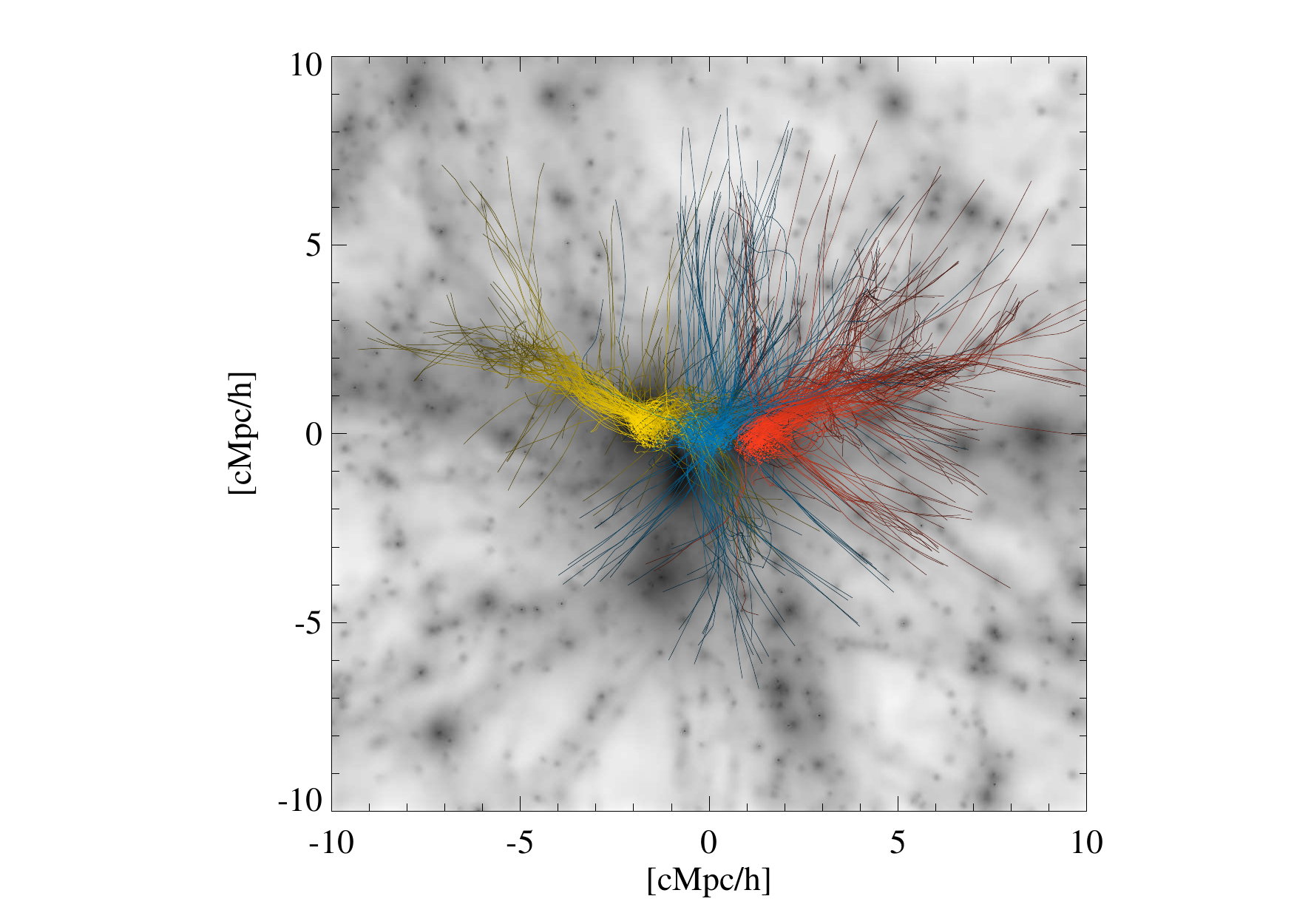}
    \caption{Gas density maps in the three main projection axes ($z$, $y$ and $x$ simulation axis, from left to right) at $z=0.07$. Each map, centered on the cluster pair, encloses a cubic volume of $20\,\cMpc$ per side.
    Overplotted in red, yellow and blue, the spatial trajectories of the gas particles finally selected within $\rfive$ of GC1 and GC2 and in the bridge at $z=0.07$. Each trajectory shows a color gradient with time, from dark at $z\sim 2$ to light at $z=0.07$.} \label{fig:trck_gas_traj}
\end{figure*}

Fig.~\ref{fig:trck_gas_traj} visually shows the spatial origin of the
gas particles of the three aforementioned components through their
trajectories towards their final positions at $z=0.07$. In the
background, we report the projected gas density maps in the
$(20\,\cMpc)^3$ comoving volume centered on the pair (same as
Fig.~\ref{fig:trck_gas_prog}-bottom row).  The lines mark the
trajectories of the gas in the different components, from their
position at $z\sim 2$ till $z=0.07$.  The colors distinguish the gas
in GC1, GC2 and in the bridge between them (red, yellow and blue
respectively; as in Fig.~\ref{fig:trck_gas_dist}), with a color
gradient from early times (dark) till $z=0.07$ (light).  In order to
ease the visualization, we only show the trajectories for 1000
randomly-selected gas particles, for each gas component.  {From the
  visual inspection of the trajectories, we note that the gas in the
  bridge (blue lines) is actually collapsing from directions almost
  orthogonal to the main accretion directions of the gas selected from
  the clusters (red and yellow lines).}

\subsection{Thermal and chemical properties of the diffuse gas}\label{sec:chem}

{Given the geometrical configuration presented by this system, with a
  large physical separation relative to the $\rvir$ of the member
  clusters and a mostly independent origin of the majority of gas in
  the bridge, this is an optimal target to study the thermo-chemical
  properties of the warm-hot gas populating cosmic filaments in
  comparison to ICM in clusters.  Observationally, hints for the
  presence of colder gas in the interconnecting bridge of the A3391/95
  system have been in fact found by \erosita~\cite[][]{reiprich2020}
  as well, encouraging a deeper dedicated spectroscopic analysis of
  the warm gas in the bridge (Ota et al.~2021,~in prep.).  }

In the simulated system, GC1 and GC2 are small-size clusters with
mass-weighted average temperatures of $T^1_{\rm mw, 500} = 2.60\,$keV
and $T^2_{\rm mw, 500}=2.93\,$keV, respectively.  {As often done in
  simulation analysis for a better comparison to X-ray observations,
  we can estimate a spectroscopic-like ICM temperature ($T_{\rm sl}$),
  i.e.\ an average temperature weighted by $\rho^2
  T^{-0.75}$~\cite[][]{mazzotta2004} instead of gas mass. In this
  case, we obtain $T^1_{\rm sl, 500} = 2.03, \,$keV and $T^2_{\rm sl,
    500}=2.64\,$keV, for GC1 and GC2 respectively.}

\begin{figure}
    \centering
    \includegraphics[width=.95\columnwidth,trim=55 0 40 0,clip]{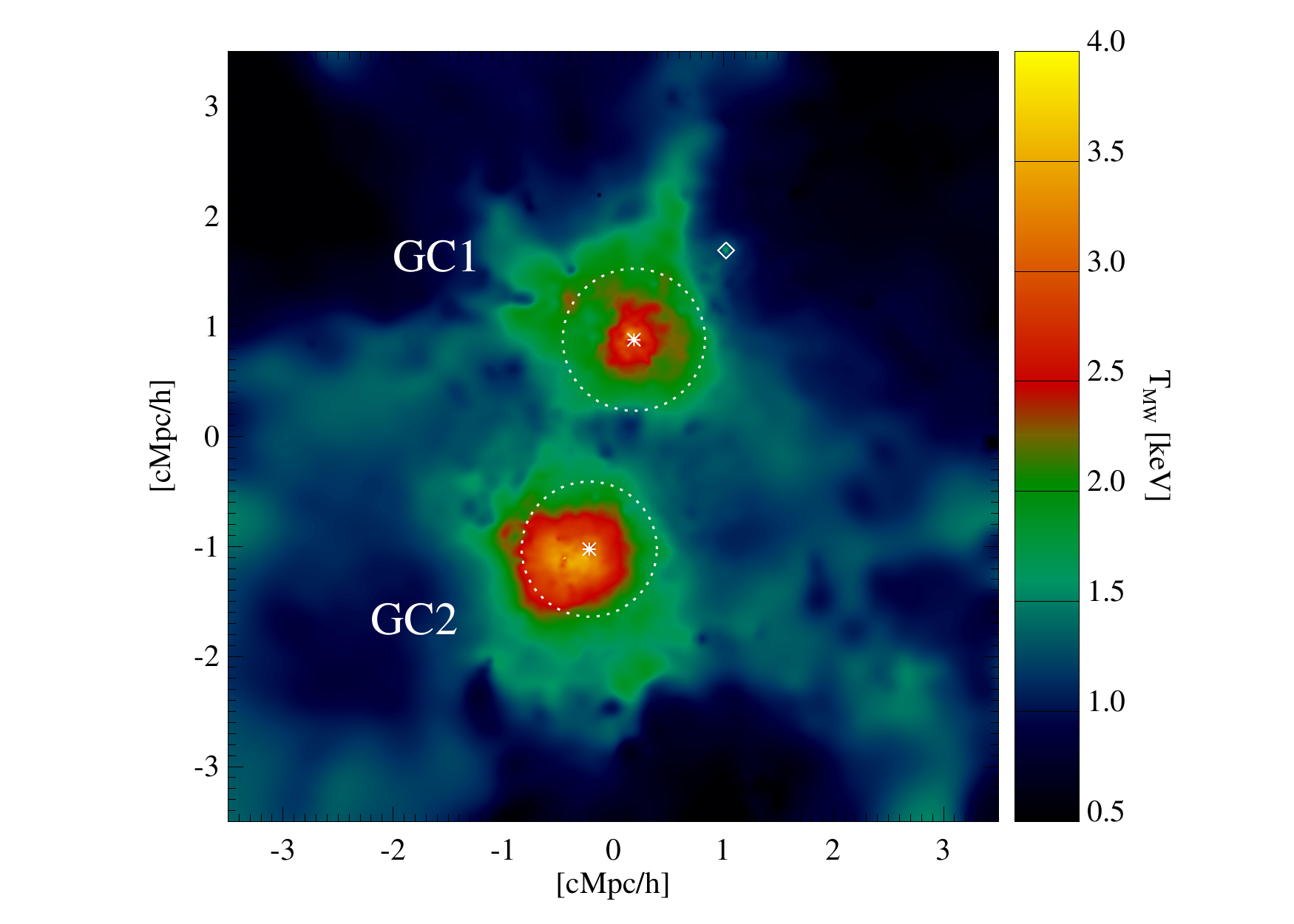}
    \caption{Projected mass-weighted temperature map of the galaxy cluster pair at $z=0.07$. The map is centered on the system center of mass and is $7\,\cMpc$ per side, and the integration along the l.o.s. is also performed over $7\,\cMpc$. 
    {The white dotted circles idicate the $\rfive$ radii of GC1 and GC2.}
    The position of group B (diamond) is also marked.}
    \label{fig:t-map}
\end{figure}

In Fig.~\ref{fig:t-map}, we show a zoom-in mass-weighted temperature
map of the system at $z=0.07$, projected along the $z$-axis, with the
two main clusters GC1 and GC2 and their $\rfive$ radii marked. The map
comprises a region of $7\,\cMpc$ per side, centered on the cluster
pair center of mass. From the map we note that a radial gradient in
temperature is present in both systems with higher temperatures
($\gtrsim3.5$\,keV) in the cores. As for GC2, we also note an
asymmetry in the innermost region, with a more extended
high-temperature core. This is related to the presence of two main
substructures, i.e. galaxies, that recently merged ($\sim 3$\,Gyr ago;
see also Fig.~\ref{fig:trck_gas_dist} and according discussion).
{Within $\rfive$, we note in fact that GC2 presents an overall
  hotter ICM with respect to GC1, despite having a smaller mass.}  In
the region between the two clusters, we notice from the color code the
presence of colder (bluer) gas, which can be linked to the bridge
connecting the two clusters~\cite[see also][]{reiprich2020}.  In
Fig.~\ref{fig:t-map}, one can further see the colder region on the top-right corner,
which corresponds to the smaller group-size halo that is approaching
$\rvir$ of the GC1 cluster (diamond; group B). At $z=0.07$, this
sub-halo is found to be already gravitationally bound to GC1, with the
merging starting at $z\sim0.16$

\begin{figure}
\centering
	\includegraphics[width=.95\columnwidth,trim= 50 10 40 0, clip]{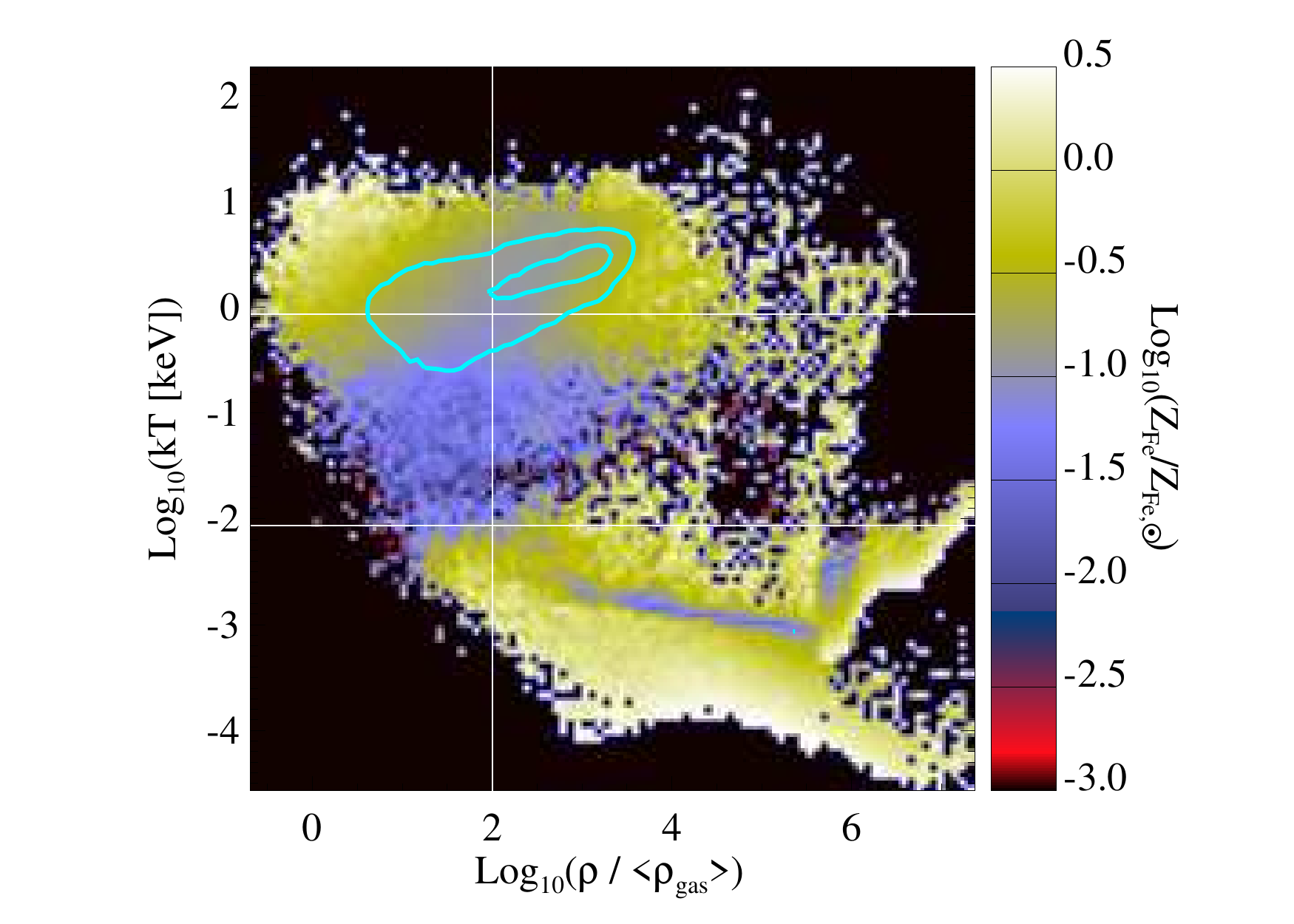}
	\caption{{Density-temperature phase-space diagram of the gas in the $(7\,\cMpc)^3$ region centered on the cluster pair at $z=0.07$, color-coded by gas Fe abundance (in solar unit w.r.t.~\protect\cite{angr89}). Overplotted, }
	the contours (cyan lines) enclosing $\sim 25\%$ (inner) and $\sim 75\%$ (outer) of the {selected} gas mass.
	\label{fig:rhot}}
\end{figure}

In Fig.~\ref{fig:rhot} we report the {density-temperature
  phase-space} diagram of the gas {particles} enclosed in the
$(7\,\cMpc)^3$ region already shown in Fig.~\ref{fig:t-map}.  The gas
density is here reported in terms of overdensity w.r.t.\ the mean
background baryon density of the simulation, $\delta \equiv \rho /
<\!\rho_{\rm gas}\!>$.  The vertical line in the diagram corresponds
to an overdensity of $\delta=100$, whereas the horizontal lines mark
the temperature thresholds $T=10^5$\,K and $T=10^7$\,K.  In this way,
we can easily separate interesting gas phases depending on its
temperature and density~\cite[see
  also][]{cen2006,ursino2010,martizzi2019}.  We define ``hot'' gas the
phase with $T > 10^7$\,K (roughly corresponding to $T> 1$\,keV). This
is typically the gas within deep potential wells, such as those of
galaxy clusters or groups, as well as in shocks.  The warm gas,
instead, is defined as the gas with $10^5 < T [{\rm K}] < 10^7$.  In
terms of density, the diffuse gas that is expected to fill the
filaments and bridges interconnecting clusters in the large-scale
structure is typically characterised by overdensity of $\sim 100$ or
lower. In particular, we refer here to the so-called warm-hot
intergalactic medium (WHIM) as the gas with $10^5 < T [{\rm K}] <
10^7$ and $\delta < 100$.

The color-code in Fig.~\ref{fig:rhot} marks instead the gas Fe
abundance, weighted by the gas mass in each temperature-density bin.
From the inspection of the phase diagram, we note that the hot phase
and the dense and cold gas are typically characterised by the highest
abundance level, with Fe abundances around the solar value or higher.
The low-density gas is characterised by a lower Fe abundance, $Z_{\rm
  Fe,\odot} \sim 0.01$--$0.1$ (see also following Sections),
especially when the warm phase ($10^5 < T [{\rm K}] < 10^7$) is
considered.

{We overplot contour lines (cyan curves) to mark the regions where
  $\sim 25\%$ and $\sim 75\%$ (inner and outer curves, respectively)
  of the total selected gas mass is located. From these, } we notice
that the majority of the gas mass ($\sim 75\%$) in the pair and its
surroundings is in the warm and hot phase, and with overdensities
between a few and $\sim 2500$. The gas in smaller haloes (low
temperatures and high densities), {and} the hot low-density gas
associated to feedback processes, only contribute $\sim 10\%$ of the
gas mass in the whole region.

\begin{figure}
    \centering
    \begin{subfigure}[b]{\columnwidth}
        \includegraphics[width=.95\columnwidth,trim= 50 10 40 0, clip]{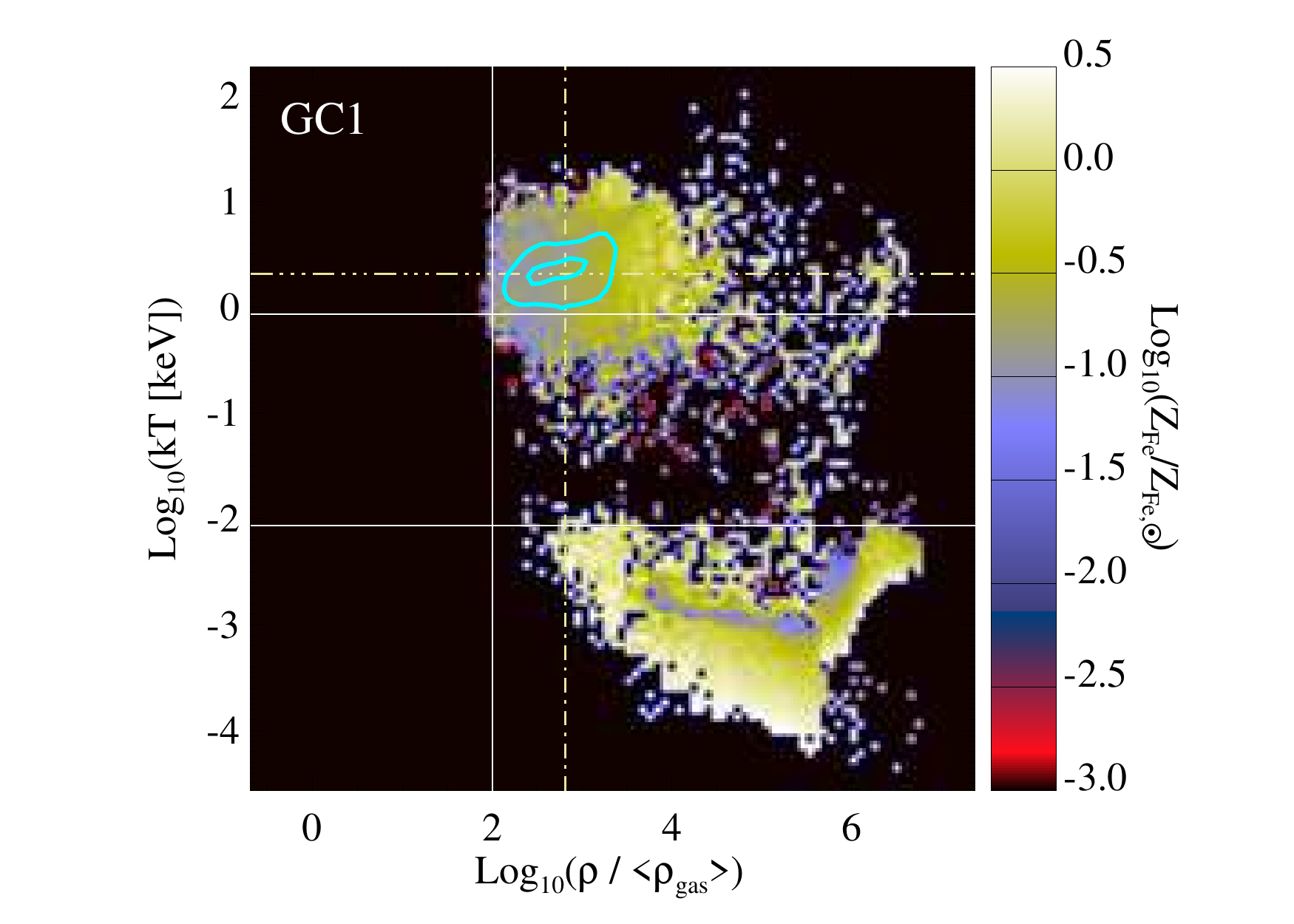}
        \caption{GC1.\label{fig:rhot-GC1}}
    \end{subfigure}
	\begin{subfigure}[b]{\columnwidth}
	    \includegraphics[width=.95\columnwidth,trim= 50 10 40 0, clip]{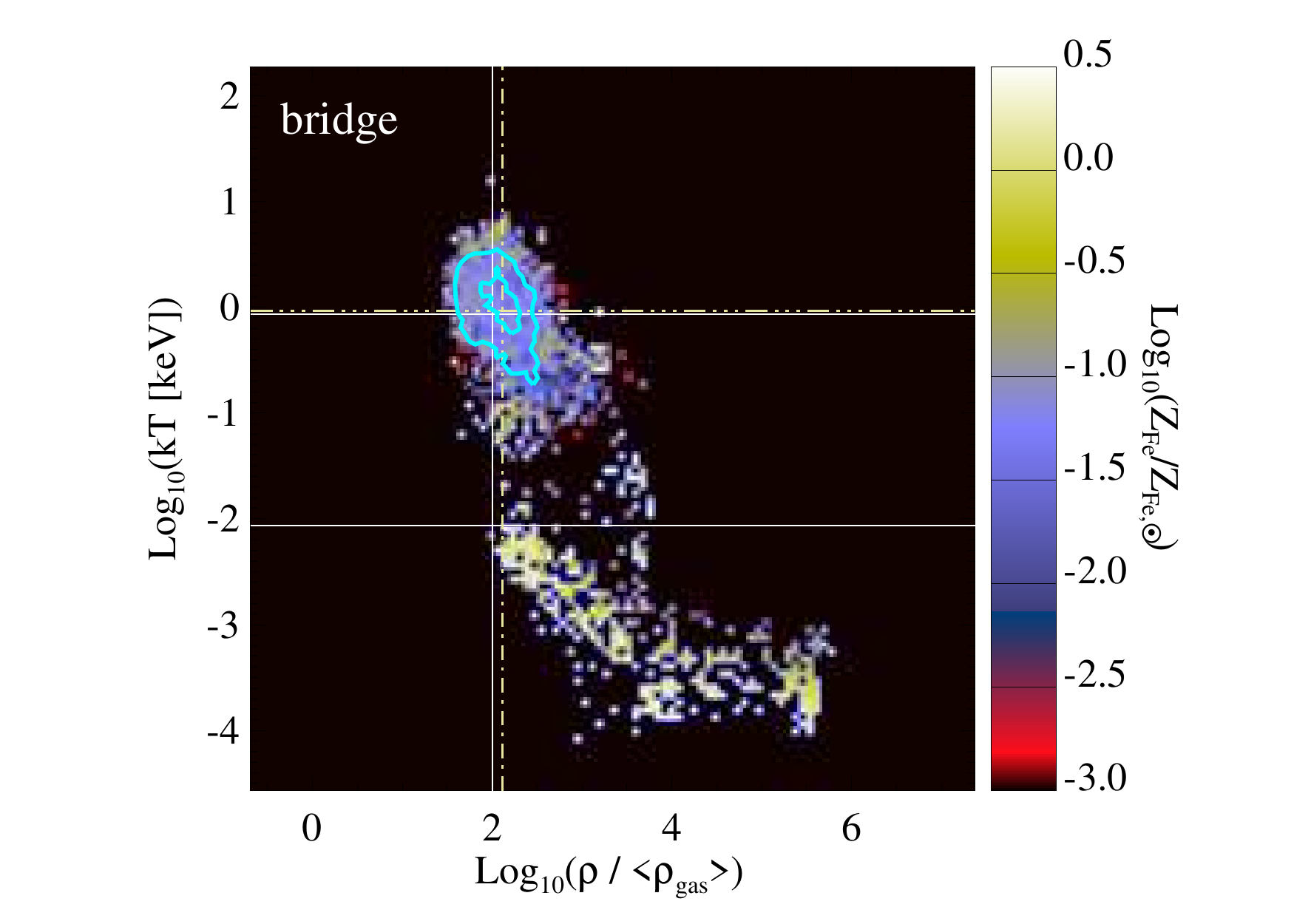}
	\caption{Bridge.\label{fig:rhot-bridge}}
	\end{subfigure}
	\caption{Same as Fig.~\protect\ref{fig:rhot}, but only for the gas within $\rvir$ for 
	{one of the member clusters, namely GC1} (upper panel) and for the gas in the bridge (lower panel) between the pair clusters, at $z=0.07$.
	{Dot-dashed lines mark the average temperature and overdensity values (mass-weighted estimate) of the selected gas.}
	\label{fig:rhot-GC1-bridge}}
\end{figure}

For comparison, in Fig.~\ref{fig:rhot-GC1-bridge} we show instead the
{density-temperature} diagram for the gas included in one of the
two pair members (upper panel), and in the bridge (lower panel).
{As an example, we select the GC1 cluster and show the gas enclosed
  within its $\rvir$ radius, in Fig.~\ref{fig:rhot-GC1}. The
  phase-space} diagram is therefore only populated at overdensities
larger than $\sim 100$. As indicated by the external contour level,
$75$--$80\%$ of the GC1 gas mass within $\rvir$ has $T\gtrsim 10^7$\,K
and median overdensities $\delta \sim 1000$, as marked by the
mass-weighted values reported in the diagram (dot-dashed lines).  From
the temperature-density distribution we also note that the gas hot
phase ($T\gtrsim 10^7$\,K) shows a gradient in the Fe abundance: at
lower densities the typical metallicity is also lower. This
essentially corresponds to the X-ray emitting ICM, for which the
radial abundance profile is indeed expected to be a decreasing
function of the cluster-centric distance~\cite[see][for recent
  observational and numerical reviews]{mernier2018Rv,biffi2018Rv},
with larger values in the central regions (corresponding to higher
overdensities) and lower values in the outskirts ({at} lower
overdensities).  The cluster potential well also comprises
high-density cold gas, which is typically found in (or close to)
star-forming regions, or within substructures,
e.g. galaxies. Typically, this is consequently characterised by large
values of the metallicity, as it has been more easily polluted by
stellar sources.  {Although we only show the case of GC1, as an
  example for the pair clusters, we verified that the
  density-temperature diagram of GC2 presents very similar features.}

Fig.~\ref{fig:rhot-bridge} shows that the majority of the gas in the
core of the pair bridge (i.e.\ closer than $< 500\,\ckpc$ from its
spine and outside the $\rvir$ of either cluster;
Sec.~\ref{sec:bridge}) occupies the upper envelope of the
distribution, as marked by the mass contours.  Specifically, we find
that in the simulated system $\sim75\%$ of the gas mass in the bridge
is in the warm-hot phase, with a typical temperature of $\sim 1\,$keV,
a median overdensity of $\sim 100$ (marked by the dot-dashed lines)
and an homogeneous iron abundance of $Z_{\rm Fe} \sim 0.1\,Z_{\rm
  Fe,\odot}$.  {The colder phase of the WHIM,} with
$T\sim10^5$--$10^6$\,K and $\delta \lesssim 100$, is not significant.
{This is consistent with recent simulation results, e.g.\ by
  \cite{galarraga2020b}, who find the presence of hot gas within
  $1\,\Mpc$ from the spine of short filaments. At larger distances
  from the filament spine they predict the colder WHIM phases to
  dominate.  By enlarging the bridge radius up to $1\,\cMpc$
  (approximately $1.3\,\Mpc$ in physical units), we also confirm a
  shift of the gas mean overdensity towards lower values $\delta
  \lesssim 100$.  Finally, from Fig.~\ref{fig:rhot-bridge}, we find
  that a minor fraction of the bridge gas mass is characterised by
  large values of overdensity and iron abundance, and low temperatures
  ($T\lesssim 10^5$\,K). This is halo gas associated to galaxies,
  either comprised within small substructures or stripped from
  galaxies at earlier times.  The presence of a small, albeit non
  negligible, fraction of halo gas in short filaments is similarly
  reported by~\cite{galarraga2020b}.}

{Overall, in such a system, the bridge gas closely reflects the
  chemo-energetic properties of the WHIM, with average temperature and
  density clearly lower than the ICM within the two clusters. Combined
  with the analysis of its spatial origin (Sec.~\ref{sec:orig}), this
  further supports the conclusion that this is filament-like gas and
  not stripped ICM due to the interaction between the pair clusters.}

\paragraph{Metallicity distribution.}

\begin{figure}
	\includegraphics[width=\columnwidth]{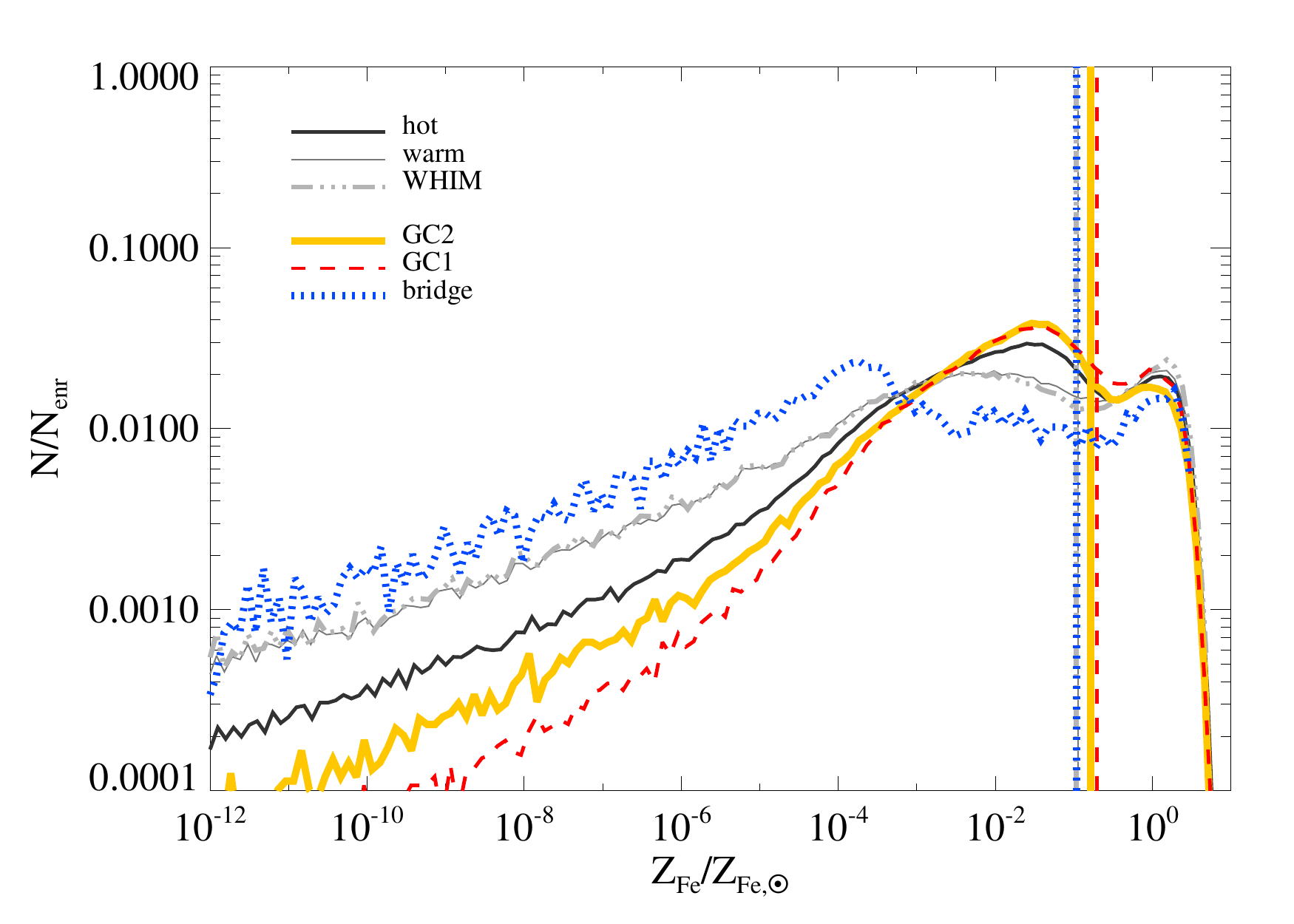}
	\includegraphics[width=\columnwidth]{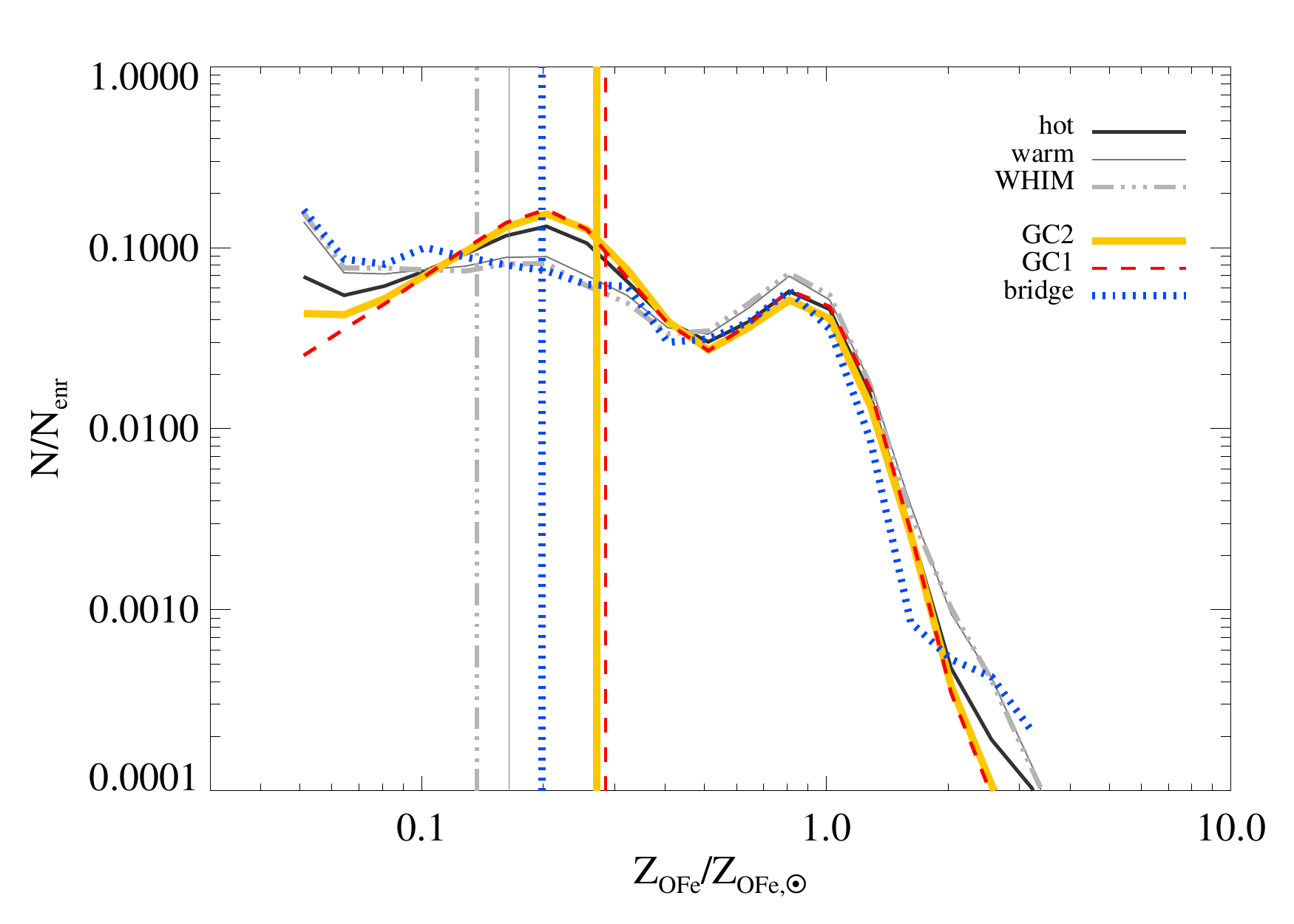}
    \caption{\label{fig:distrib_met} Distributions of Fe abundance and
      O/Fe abundance ratio for the gas selected at $z=0.07$ in
      different spatial regions and thermal phases.  {For each gas
        selection, the corresponding distribution is normalized to the
        total number of enriched (i.e. $Z_{\rm Fe}>0$ or $Z_{\rm
          OFe}>0$) gas particles.}  }
\end{figure}

{Fig.~\ref{fig:distrib_met} shows the iron abundance distribution
  for the gas in the $(7\,\cMpc)^3$ volume around the pair at
  $z=0.07$, for different thermal phases and spatial selections.}  In
particular we consider six cases: all the gas in the hot phase
($T>10^7$\,K; black), the warm gas ($10^5< T[{\rm K}] < 10^7$; solid
grey), the WHIM ($10^5< T[{\rm K}] < 10^7$ and $\delta < 100$;
dot-dashed grey), the X-ray-emitting ICM in GC1 and GC2 (with
$T>0.3$\,keV, and located inside $\rvir$; red and yellow respectively)
and the gas selected in the bridge between them (blue; defined as in
Sec.~\ref{sec:orig}).  In the Figure we report the number distribution
of iron abundance (upper panel) and oxygen-over-iron abundance ratio
(lower panel), normalized to the total number of Fe- or O/Fe-rich
particles in each gas subsample.

Considering the iron abundance (Fig.~\ref{fig:distrib_met}, top), we
note that a broad double-peak feature is common to all curves, with
differences in the gradient of the low-abundance tail and the relative
amplitude of the peaks depending on the gas selection considered. The
two peaks roughly correspond to abundances of $Z_{\rm Fe}\sim
10^{-3}$--$10^{-2}\, Z_{\rm Fe,\odot}$ and to solar values ($\sim
Z_{\rm Fe,\odot}$), respectively.  For the bridge gas, the
low-metallicity peak is further shifted to $\sim 10^{-4}\,Z_{\rm
  Fe,\odot}$, with a more extended valley between the two peaks.
Despite the differences, we note that the mass-weighted abundance of
each sample is always of order $\sim 10^{-1}\,Z_{\rm Fe,\odot}$,
namely between $0.1\,Z_{\rm Fe,\odot}$ for the warmer phases and
$0.2\,Z_{\rm Fe,\odot}$ for the ICM/hot gas\footnote{We note that, for
  the ICM in GC1 and GC2, the global estimate within $\rvir$ is
  dominated by the outer lower-density regions, where the bulk of the
  cluster mass resides --- see also mass contours in
  Fig.~\ref{fig:rhot-GC1}.}.  {Compared to other selections,} the
whole hot-phase gas and the ICM residing within clusters typically
show a similarly steeper low-metallicity tail and a more pronounced
peak at low metallicity relative to the solar-metallicity one.
Overall, a large fraction of the ICM is significantly enriched, with
abundances $\gtrsim 10^{-2}\, Z_{\rm Fe,\odot}$.  Compared to the
hotter gas phases, for the gas in the warm phase and the WHIM, the
low-metallicity tail is shallower and has a higher
normalization~{\cite[consistently with previous studies,
    e.g.][]{cen2006}}, with a shallower peak at abundances $\sim
10^{-3}\, Z_{\rm Fe,\odot}$. This broad low-abundance component
extends down to even lower values, $\sim 10^{-4}\, Z_{\rm Fe,\odot}$,
for the gas in the core of the pair bridge.  Essentially, we can
divide the different gas selections into two broad categories based on
the trends shown in Fig.~\ref{fig:distrib_met}: the hotter gas within
the main clusters and the warmer, lower-density gas phase.

The warm diffuse gas outside the main clusters and in the bridge is
typically poorly enriched. It always comprises, nevertheless, a
highly-enriched component originating from, or still associated to,
{galaxy-size haloes. This is responsible for the distribution peak
  centered on solar abundances.}

A broad similar distribution is found for the oxygen abundance, with
the main difference that both peaks are more pronounced and the peak
at lower abundances is broader and shifted towards lower values.  This
can be investigated via the O/Fe abundance ratio, shown in the lower
panel of Fig.~\ref{fig:distrib_met}.  The oxygen-to-iron ratio
distribution shows in all cases a very similar peak around solar
values, originated by the corresponding solar-abundance peaks present
in both iron and oxygen distributions.  Small differences are still
observed between hot/cluster gas and warmer/bridge plasma in the
left-hand-side tail of the distribution. This is in fact broadly flat
for the warmer {gas phases,} whereas presents a secondary peak for
the hot ICM and a decrease towards $Z_{\rm OFe}\lesssim 0.1\,Z_{\rm
  OFe, \odot}$.  The mass-weighted average O/Fe ratio is three times
lower in the warm phase than within clusters, indicating that a larger
fraction of warmer gas has very small O/Fe abundance ratio.  This is
consistent with the fact that the vast majority of warm diffuse gas is
expected to be located far from dense star-forming sites which are
primarily responsible for oxygen enrichment.

\subsection{Evolution of gas chemical properties}
\label{sec:chem_evol}

\begin{figure*}
\centering
	\includegraphics[height=.23\textheight]{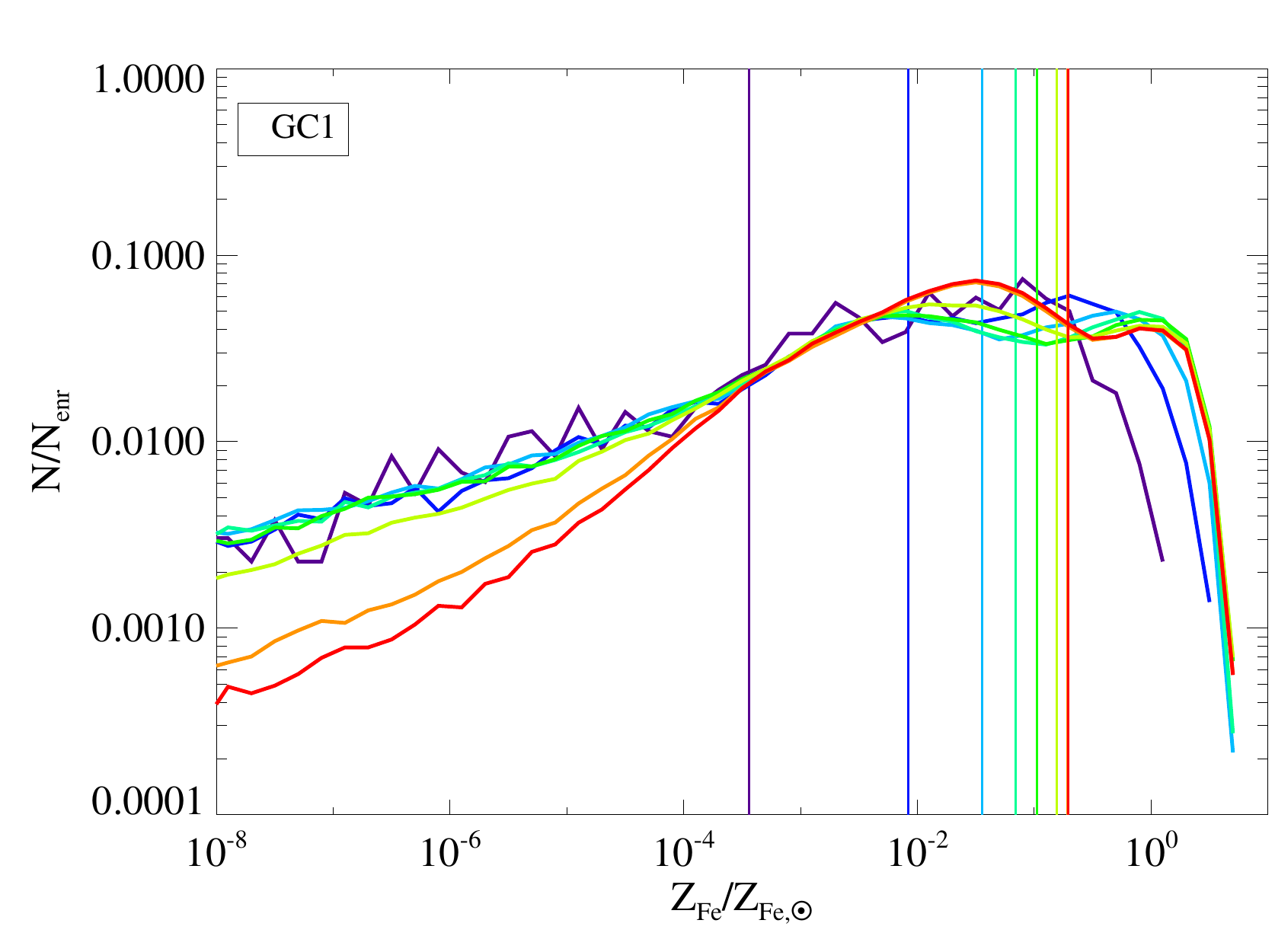}\quad
	\includegraphics[height=.23\textheight]{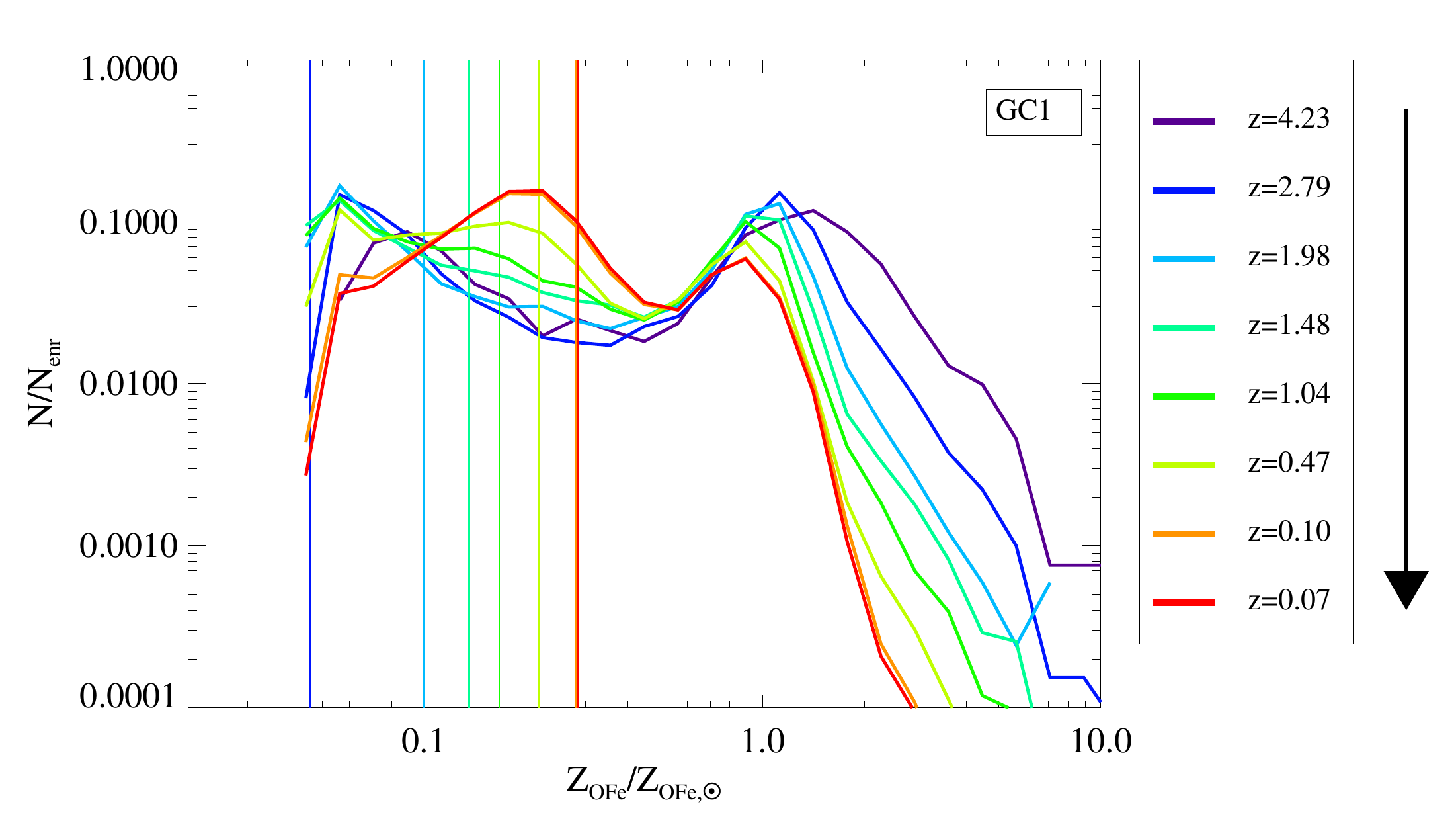}\\
	\includegraphics[height=.23\textheight]{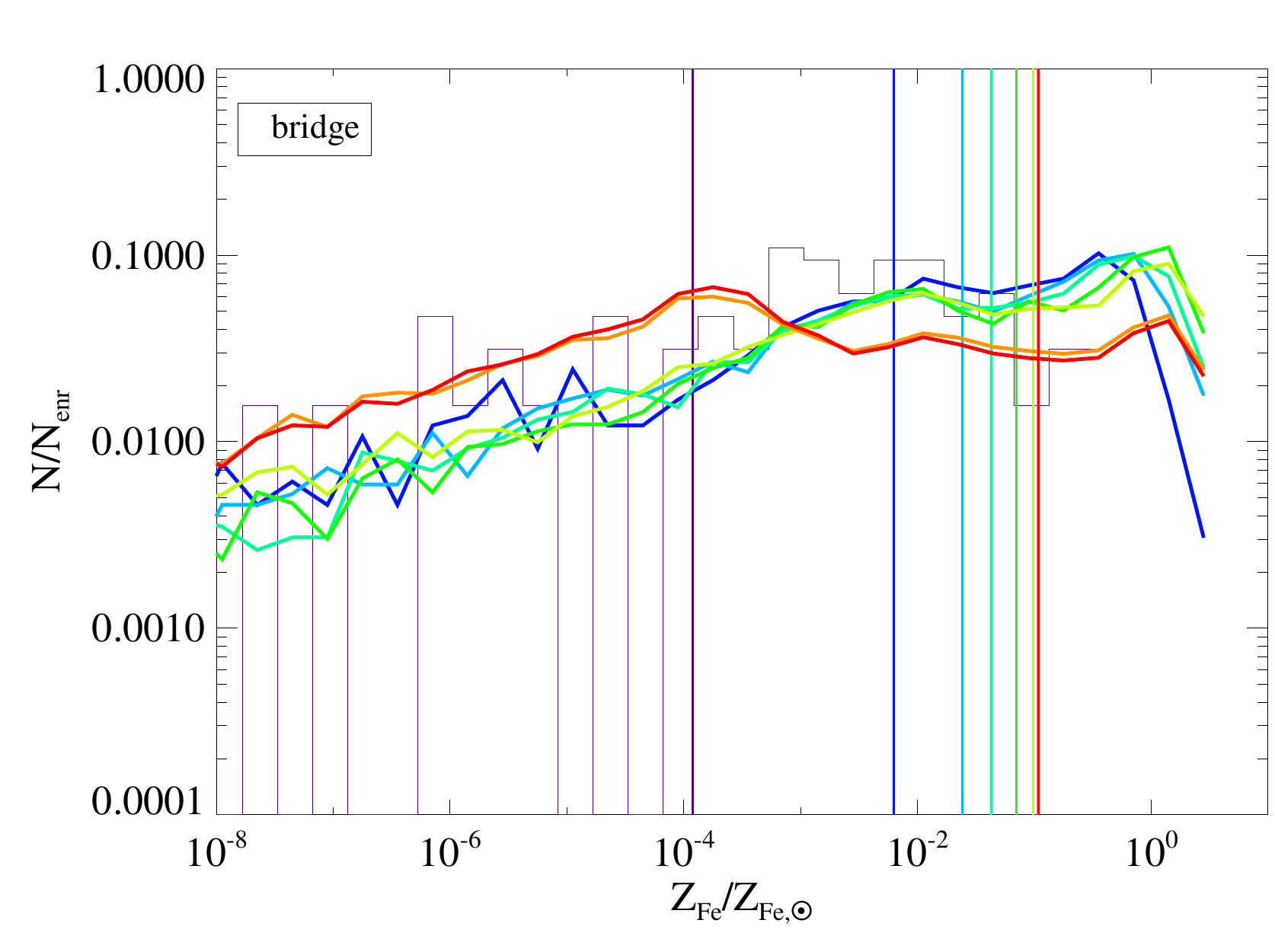}\quad
    \includegraphics[height=.23\textheight]{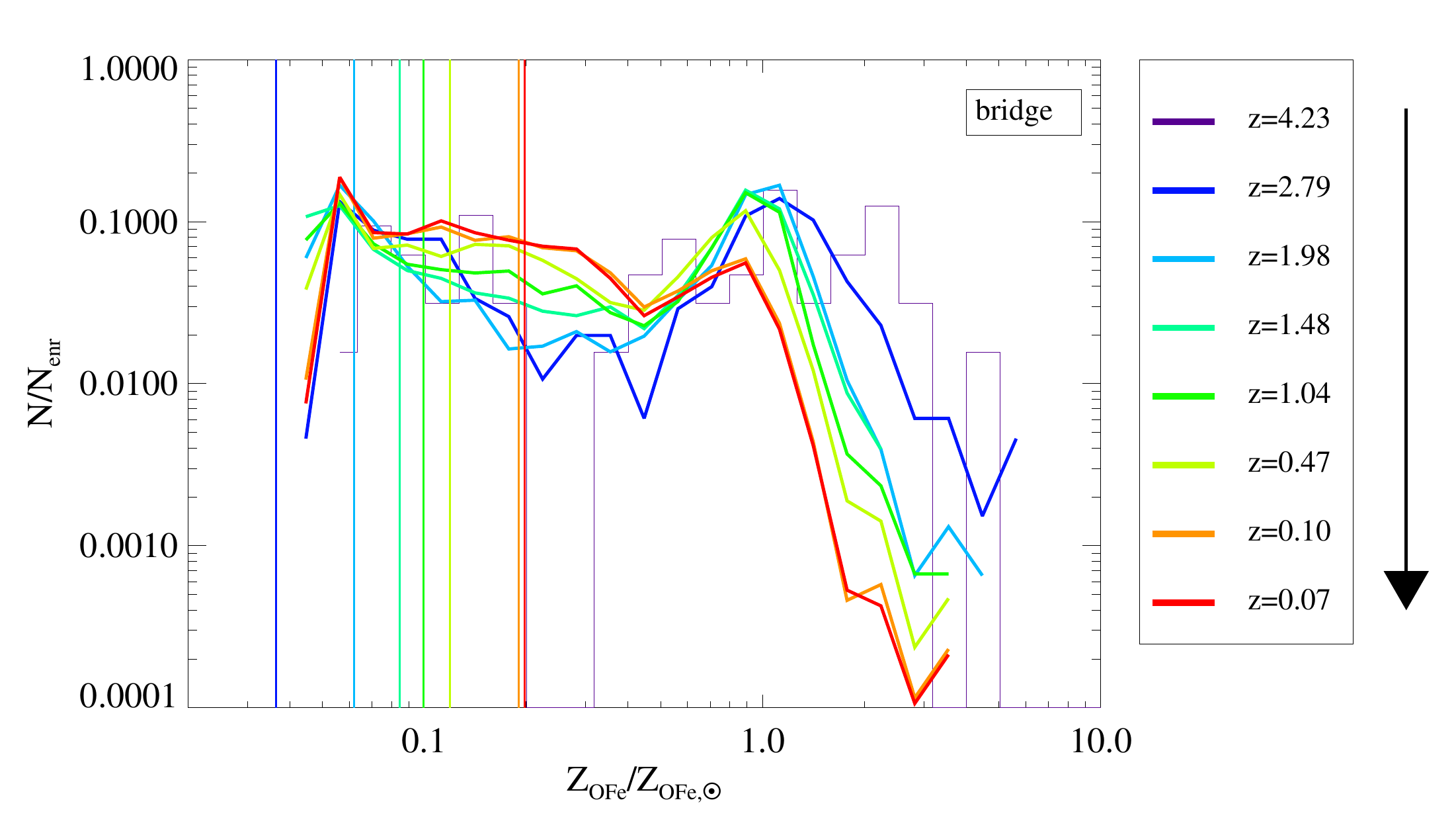}
	\caption{\label{fig:distribN_met_evol}{Redshift evolution of the iron abundance (left) and O/Fe abundance ratio (right) distributions 
    for the ICM in one of the clusters (GC1; top) and the bridge gas (bottom), selected at $z=0.07$ and tracked back in time.} 
    Vertical lines mark the MW-average abundance at any corresponding redshift.
    {The bridge histogram line for $z=4.23$ is thinner to mark the very low statistics of metal-enriched gas particles (only $\lesssim 100$, i.e.\ $\sim 0.6\%$ of the total).}}
\end{figure*}

Given the possibility to track {back in time} the gas particles
selected at $z=0.07$, we investigate the evolution of the gas chemical
properties depending on its {\it final} location or thermal phase.
This is shown in Fig.~\ref{fig:distribN_met_evol} for the gas selected
at $z=0.07$ within GC1 (top) and in the pair bridge (bottom),
{representative of the hotter and warmer phases respectively}.  In
the four panels of the figure we report the redshift evolution between
$z=4.3$ (purple curves) and $z=0.07$ (red curves) of the iron
abundance (l.h.s.~panels) and O/Fe ratio distributions (r.h.s.~panels)
of these two gas subsamples.

From the evolution of the iron distribution we can infer two general
trends: the gas that is finally selected in the hot phase and in
clusters has undergone a stronger evolution in time compared to the
gas that is finally located in the warm, low-density gas, like the
WHIM or the gas selected within the pair bridge.  In fact, comparing
the two l.h.s. panels in Fig.~\ref{fig:distribN_met_evol}, we note
that the shape of the iron abundance distribution changes in time in
the GC1 case, whereas the gas finally selected in the bridge shows
little variation, especially in the low-metallicity tail.  The
mass-weighted abundance, nevertheless, increases with time in both gas
subsamples, indicating that the mass fraction of highly enriched gas
increases in both cases at low redshift.  The solar-abundance
component noted in Fig.~\ref{fig:distrib_met} seems to be present
already from $z\lesssim 2$--$3$ (blue and cyan lines).  This can be
observed also from the evolution of the O/Fe abundance ratio
distribution, in the r.h.s.\ panels of
Fig.~\ref{fig:distribN_met_evol}, where a peak around solar values is
always present below $z\lesssim 3$ and is more prominent at high
redshift, where it is driven by SNII oxygen enrichment.  {The
  latter} is also responsible for the higher values of the O/Fe
abundance ratio reached at high redshifts, where the right-hand-side
tail is higher in normalization and more extended.  This confirms that
part of the gas has been enriched already at early times (especially
with oxygen) and survives in the diffuse form till $z=0.07$~\cite[see
  also][]{biffi2018}.  Residual, late-time enrichment {is driving
  the change in the O/Fe distribution at lower values,} where a
central peak around $Z_{\rm OFe}\sim 0.2\,Z_{\rm OFe,\odot}$ is
building up at low redshift, more prominently in the hot ICM within
GC1 than in the gas ending up in the bridge.  The gas selected within
the bridge, differently, shows a shallower distribution at central
O/Fe values, although some increase is also visible and the peak at
solar values still decreases in time relative to the lowest abundance
ratios.  This can still be attributed to late iron pollution due to
long-lived SNIa.

\begin{figure}
\centering
	\includegraphics[width=.95\columnwidth]{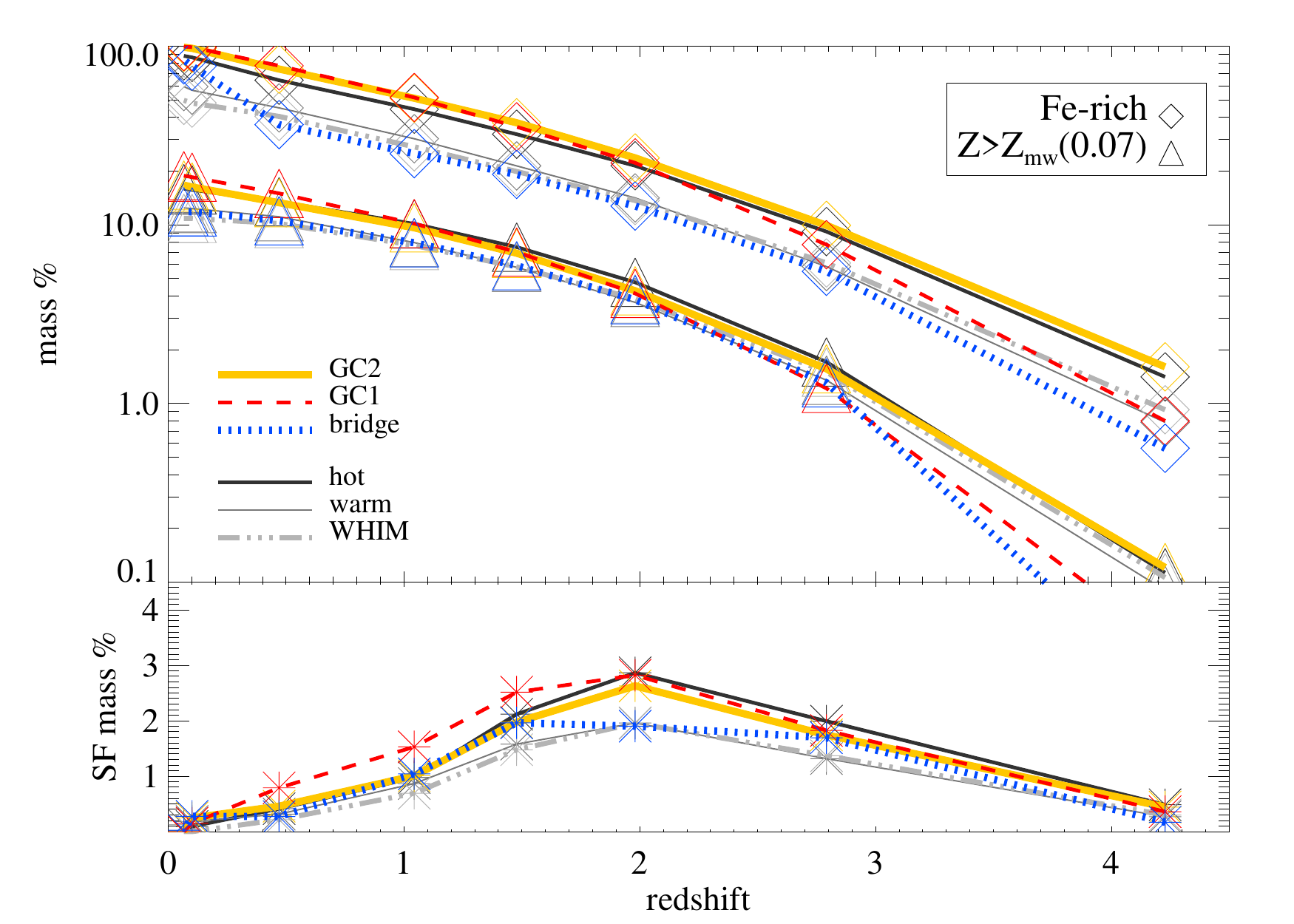}
	\caption{\label{fig:evol_SF_met}Evolution with redshift of
          {enriched} and star-forming mass fractions for the gas in
          different spatial regions and thermal phases, {selected
            at $z=0.07$} and tracked back in time.  {\it Upper inset:}
          mass fraction of Fe-rich gas {($Z_{\rm Fe} >0$;
            diamonds)} and highly-enriched gas {($Z_{\rm Fe} >
            Z_{\rm Fe,mw}^{z=0.07}$; triangles)}.  {\it Lower inset:}
          redshift evolution of the star-forming {(${\rm SFR}>0$)}
          gas mass fraction. }
\end{figure}

While Fig.~\ref{fig:distribN_met_evol} shows the differences in the
distribution shape at different redshifts, we also verified that,
despite preserving the general shape features, the normalization of
the curves increases with time when the distributions are computed in
terms of gas mass (see~Appendix~\ref{app:chem}), and normalized at
each redshift to the total mass of the selected gas. Indeed, the mass
fraction of iron- and oxygen-rich gas increases with time.  This
picture is summarized by the trends shown in
Fig.~\ref{fig:evol_SF_met}, where we quantify the redshift evolution
of the star-forming {(i.e.\ with instantaneous ${\rm SFR} > 0$)}
and enriched {($Z_{\rm Fe} > 0$)} gas fractions with respect to the
total gas mass, for each gas selection.  There we can notice how the
enriched gas fraction increases with redshift in all cases analysed,
with $\sim 90(50)\%$ of the hot(warm) gas finally enriched by
$z=0.07$.  Over the whole redshift range, $0.07\lesssim z \lesssim 4$,
the gas in the hot phase and within clusters at $z=0.07$ was
systematically more enriched by a factor of $\sim1.7$, compared to the
gas that is finally in the warmer phases.  The gas that will end up in
the bridge shares a chemical evolution similar to the diffuse, warmer
phases down to $z\sim 0.5$. {At lower redshifts, however, the mass
  fraction of the (poorly) enriched component increases more
  significantly in the bridge gas then in the warm/WHIM case.}

In Fig.~\ref{fig:evol_SF_met} we also show the time evolution of the
highly-enriched gas mass fraction, which we define as the mass
fraction of the selected gas particles that at any given redshift have
an iron abundance larger than the mass-weighted average value computed
at $z=0.07$ (i.e. with $Z_{\rm Fe} > Z_{\rm Fe,mw}^{z=0.07}$).
Similarly, this also grows in time. Such highly-enriched component,
nevertheless, was already present at high redshift in all selections,
and typically increases by a factor of $\sim 3$ in mass between
$z\sim2$ and $z=0.07$. Only for the ICM selected at $z=0.07$ within
the two clusters it increases more prominently (by a factor of
$\gtrsim 4$). Although not shown here, an overall similar, albeit
shallower, trend is observed for oxygen, with less pronounced
differences among the various gas selections.  The curves shown in the
bottom inset of Fig.~\ref{fig:evol_SF_met} show that the gas selected
in the different cases at $z=0.07$ has always undergone a peak of
star-formation activity around $z\sim2$.  Wherever the gas is finally
residing, at the peak of star formation the mass fraction of tracked
gas that was star forming is of order of a few percent at
most. Differently, at redshifts $z\gtrsim 3$ the star forming gas adds
up to less than $1\%$ of the total mass of the gas selected, with
slightly higher values for the gas that is ending up within structures
and negligible or zero star formation in the other cases. This trend
is also observed at $z\lesssim 1$, with no star-forming gas within
clusters at $z=0.07$, by definition.

\subsection{Additional structures in the local environment} 
\label{sec:clumps}

Several other extended sources, the size of groups and clusters of
galaxies, have been detected in the \erosita PV observation of the
A3391/95 field, a number of which at the same redshift of the pair
system~\cite[][Ramos-Ceja et al.,~in prep.]{reiprich2020}.  Similarly,
in the local environment surrounding the simulated cluster pair we
find additional massive structures, all comprised within the
considered volume of $20\,\cMpc$ per side (about $26\,\Mpc$ for the
redshift and cosmology considered).  In the simulations, we identify
in particular 4 group-size haloes with masses in the range $M_{500}
\sim 1.4$--$9.6\times 10^{13}\,\msun$.  Additionally, we consider the
aforementioned group B, a structure with the size of a small galaxy
group, that has entered the atmosphere of GC1 at about $z=0.16$ and
has reached its outskirts by $z=0.07$ (when is located at $d_{\rm 3D}
\sim 1.5 \times \rvir^{1}$).  At $z\lesssim 0.16$ the halo is no more
identified as an independent group but rather as a
gravitationally-bound substructure of the GC1 cluster, by the SUBFIND
substructure-finding algorithm.  The position and $\rvir$ extent of
these five systems is also reported in the projected maps of
Fig.~\ref{fig:trck_gas_prog}, at different redshifts $z\lesssim
0.5$. For the reference projection at $z=0.07$ (bottom-left panel in
Fig.~\ref{fig:trck_gas_prog}), the haloes are all labelled.

\begin{figure}
    \centering
    \includegraphics[width=.9\columnwidth]{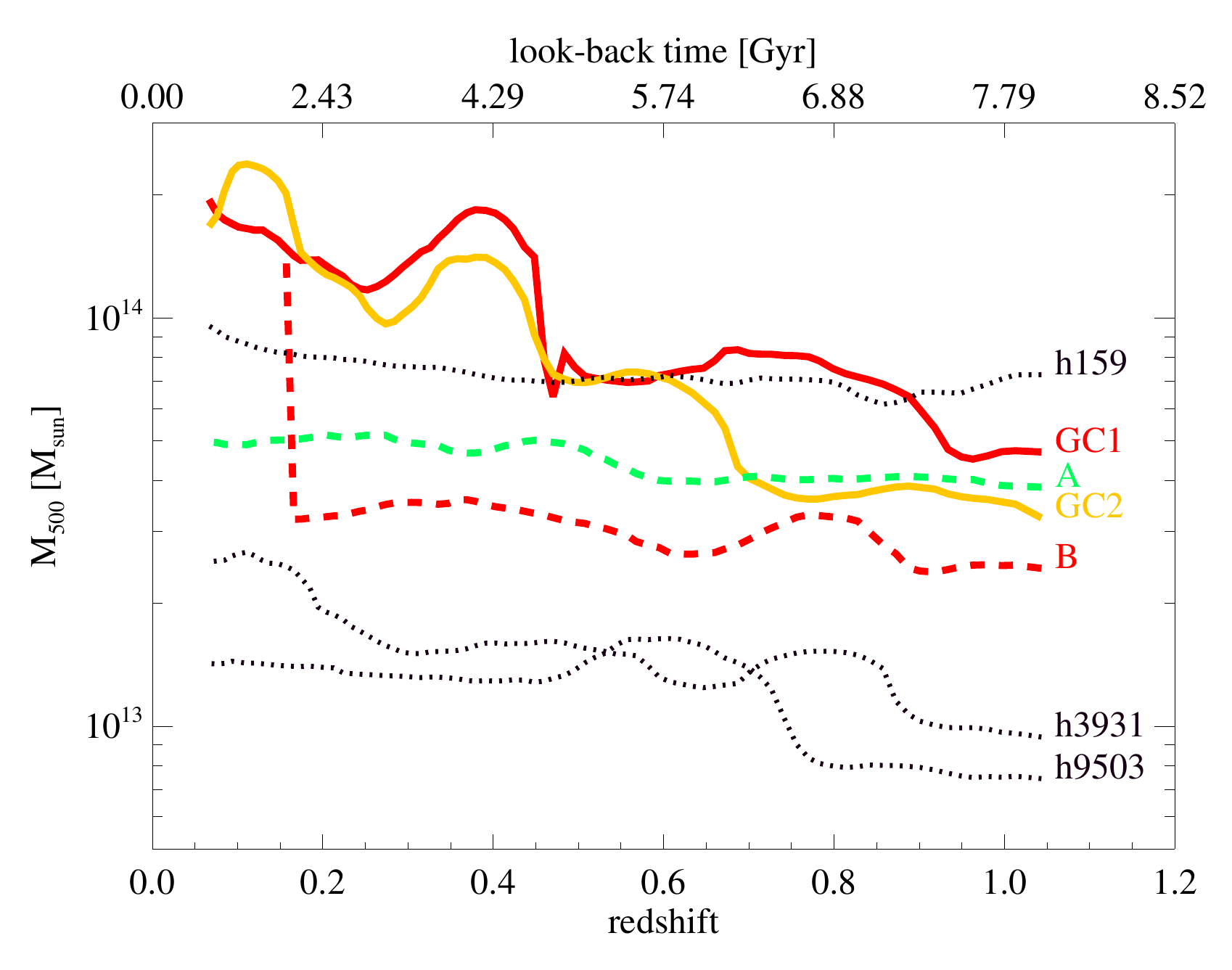}
    \caption{Redshift evolution of the total mass $M_{500}$, {for
        all the clusters and groups identified at $z=0.07$} in the
      $(20\,\cMpc)^3$ volume around the cluster pair. At each redshift
      up to $z\sim 1$, we consider the main progenitor of the given
      halo identified at $z=0.07$.}
    \label{fig:m500evol}
\end{figure}

In Fig.~\ref{fig:m500evol}, we report instead the redshift evolution
of the total mass enclosed by $\rfive$ ($M_{500}$), for the five
groups and the main clusters GC1 and GC2 as well. All the halos,
identified at $z=0.07$, are tracked back in time and the main
progenitor\footnote{For each halo, we always select the most-massive
  progenitor.} is considered at every redshift up to $z\sim 1$.  From
the Figure, we notice increases in the $ M_{500}$ masses of GC1 and
GC2, {following} merging events around redshift $z\sim 0.5$--$0.4$
and $z\sim 0.2$ {(as noticed from Fig.~\ref{fig:trck_gas_dist})}.

Motivated by the \erosita observation of the A3391/95 system, where
some clumps that appear to be infalling towards the system have been
discovered in the Northern and Southern
Filament~\cite[][]{reiprich2020}, we concentrate in particular on two
systems among those neighbour structures, {to search for signatures
  of motion towards the pair}.

Considering the most massive groups (``h159'', ``group A'' and ``group
B''; Fig.~\ref{fig:m500evol}), we focus on the northern-most group A
and the smaller group B, respectively marked with the green
{square} and red diamond in Fig.~\ref{fig:trck_gas_prog}.  {Despite
  being massive ($M_{500} \sim 6.7\times 10^{13}\,\msun$), we instead
  exclude the more complex group h159, which constitutes together with
  halo ``h9503'' another close pair, separated by $d_{\rm 3D}\sim
  3.5\,\cMpc$ and with a mass ratio of almost 7:1.  From a deeper
  inspection, we also notice that the group h159 is a multiple system
  itself, comprising three massive gravitationally-bound substructures
  of comparable size ($M_{\rm tot}\sim 1.4$--$4.9\times
  10^{13}\,\msun$), all within its boundaries ($\lesssim
  1.5\times\rvir$).}  Differently, both groups A and B are isolated
single structures and yet represent two extreme phases during the
infall towards the GC1--GC2 system, with group A still far away from
the pair location and group B already merging with one of the member
clusters.  By inspecting the evolution of group B main progenitor, we
find that its mass as a galaxy group is of order $M_{500} \sim 3
\times 10^{13}\,\msun$ (more specifically, varying between $M_{500}
\sim 2.5 \times 10^{13}\,\msun$ and $M_{500} \sim 3.6\times
10^{13}\,\msun$) at redshift $1 \gtrsim z \gtrsim 0.16$, i.e.\ before
entering the GC1 atmosphere (see Fig.~\ref{fig:m500evol}).  Group A is
an isolated, more massive group with $M_{500} \sim 5\times
10^{13}\,\msun$, and is located at a distance of $7.7\,\cMpc$ (namely
$\sim 10$ physical megaparsecs) from the pair center of mass, at
$z=0.07$.  From Fig.~\ref{fig:m500evol} we note that both groups A and
B have a rather smooth assembly history, without indications of
significant merger events.

\begin{figure*}
    \centering
    \includegraphics[width=.9\columnwidth]{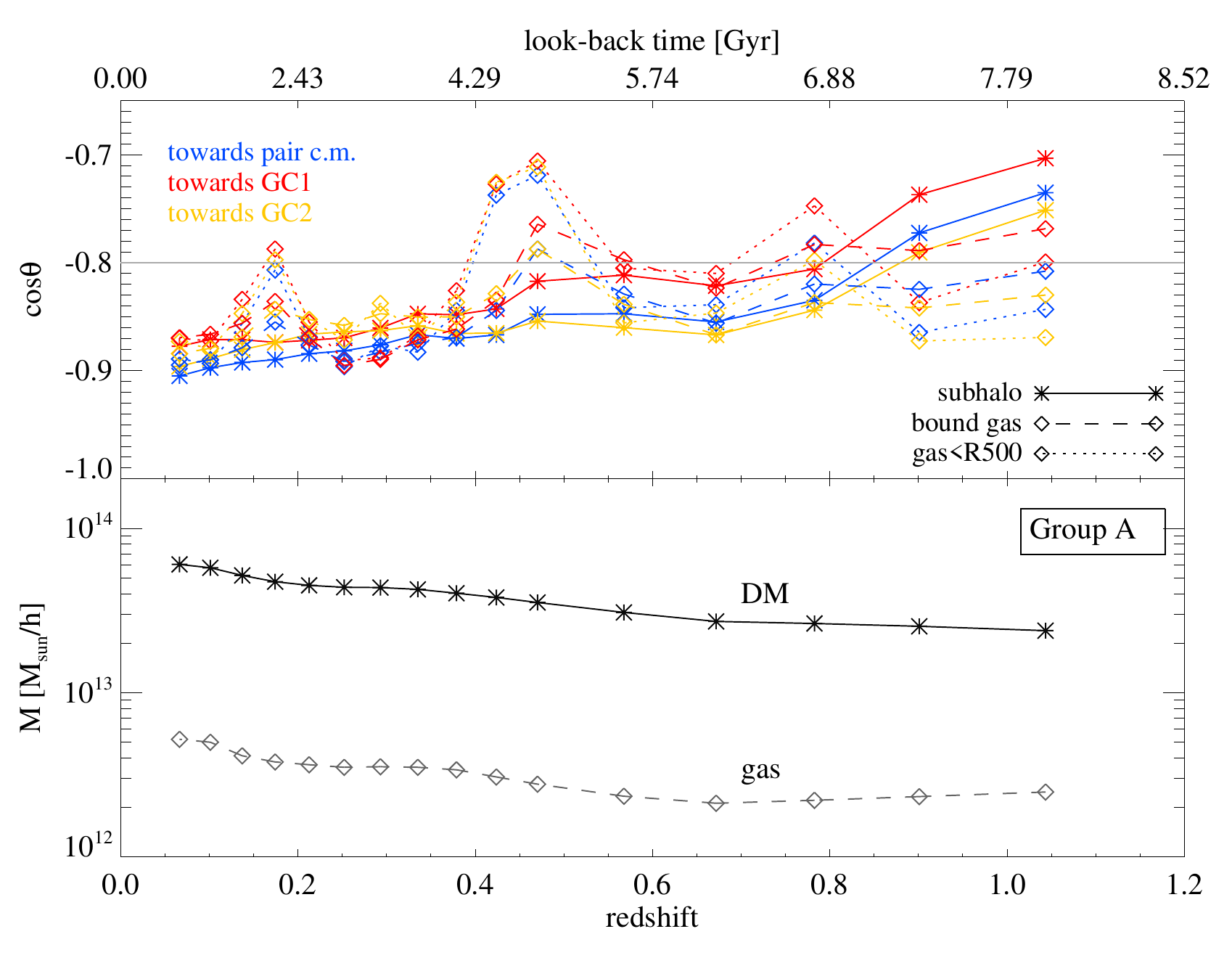}
    \includegraphics[width=.9\columnwidth]{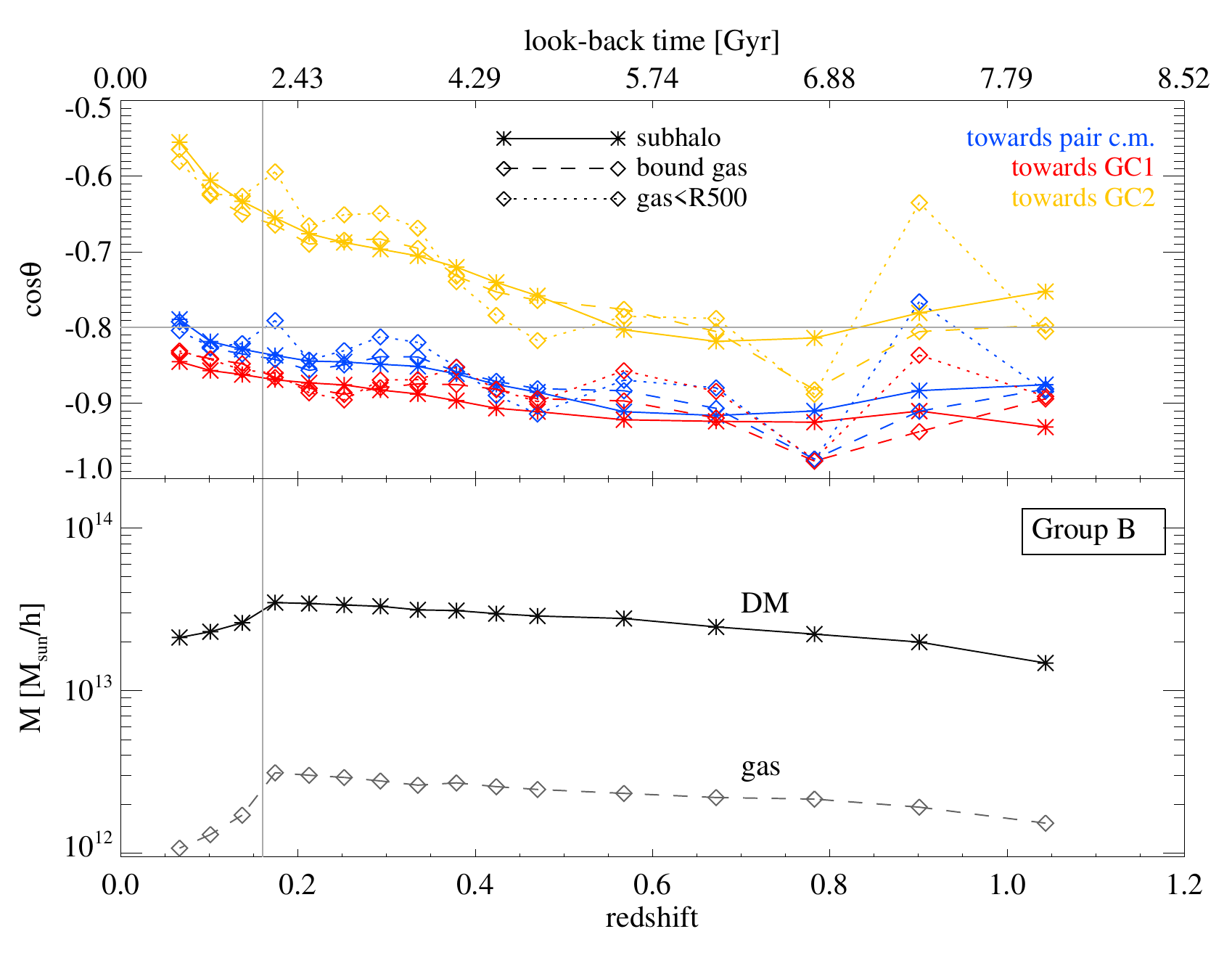}\\
    \caption{ { {\it Upper panels:} redshift evolution of the
        alignment between radial direction towards the pair (center of
        mass, blue; GC1, red; GC2; yellow) and subhalo velocity (as in
        the legend), quantified by the cosine of the angle between
        them.  For Group B (right), the gas $< \rfive$ at $z=0.16$ is
        tracked to lower redshifts, after entering the GC1 virial
        boundary.  {\it Lower panels:} redshift evolution of the
        subhalo DM (solid lines) and gas (dashed lines) bound mass.
        The two panels refer to groups A (left) and B (right).  }
    \label{fig:north-clumps-cosine}
    }
\end{figure*}

In the \erosita A3391/95 field, the Northern Clump (MCXC~J0621.7-5242)
cluster located in the Northern Filament shows in particular several
evidences suggesting a motion towards the A3391 cluster and a possible
interaction with the filament gas in which it is embedded. These
comprise, for instance, an emission enhancement towards the south, a
tail of emission extending towards the north, and a pair of bent
asymmetric radio lobes of its central Wide-Angle-Tails (WAT) radio
galaxy as well~\cite[][Veronica~et~al.~2021]{brueggen2020}.  Here, we
therefore investigate and compare the signatures of the movement of
groups A and B towards the galaxy cluster pair, at redshift $z\lesssim
1$.  This is quantified in Fig.~\ref{fig:north-clumps-cosine} (upper
insets) by the cosine of the angle ($\cos\theta$) between the group
velocity and the direction pointing towards the final position the
pair at $z=0.07$, where the Cosmic Web knot at the center of the whole
$(20\,\cMpc)^3$ region is forming.  In the Figure we report the
evolution of $\cos\theta$ with redshift, from $z\sim1$ till $z=0.07$.
For comparison, we investigate the three directions defined by the
positions at $z=0.07$ of the pair center of mass (in blue), GC1 (in
red) and GC2 (in yellow) separately.  We note from the Figure that the
cosine values, for both gas and bulk velocity, are typically comprised
between $-0.7$ and $-0.95$ below $z\lesssim1$. This indicates that the
velocity component along the infall direction is indeed dominant, and
both groups are actually moving towards the cluster pair.

Small differences are nevertheless found in the two cases, with the
smaller group B moving more specifically towards GC1 ($\cos\theta$ is
significantly lower for the red curve) and eventually merging with
it. The value of $\cos\theta$ for the direction towards the final
position of GC2 (yellow curves) starts in fact to significantly
increase at $z\lesssim 0.5$, marking the divergence of the group B
trajectory.  Instead, given the larger distance from the knot region,
for Group A we find no significant differences for the three
directions, especially for redshift $z< 0.5$.

If we consider all the gas within $R_{500}$ (dotted lines), instead of the
total bound gas only (dashed lines), then a more significant
difference with respect to the halo bulk (DM-dominated; {solid
  lines}) velocity is noticed.  In fact, the average gas velocity
within $R_{500}$ can be influenced by smooth accretion as well as by the
infall of substructures onto the group.  For group A, these accretion
events are visible for instance around redshifts $z\sim 0.75,~0.45,$
and~$0.2$, where the increase in $|\cos{\theta}|$ is accompanied by a
subsequent increase of the subhalo mass
(Fig.~\ref{fig:north-clumps-cosine}, lower inset).  In the case of
group B, the accretion of the group itself onto the GC1 cluster is
characterised by a corresponding decrease in the group total gas(DM)
bound mass by a factor of $\sim 3(1.6)$ between redshifts $z\sim 0.16$
and $z=0.07$.  The decrease of bound mass is the consequence of a
stripping phenomenon that group B undergoes while penetrating the
atmosphere of the more massive GC1 cluster.

\begin{figure*}
    \centering
    \includegraphics[width=.9\columnwidth]{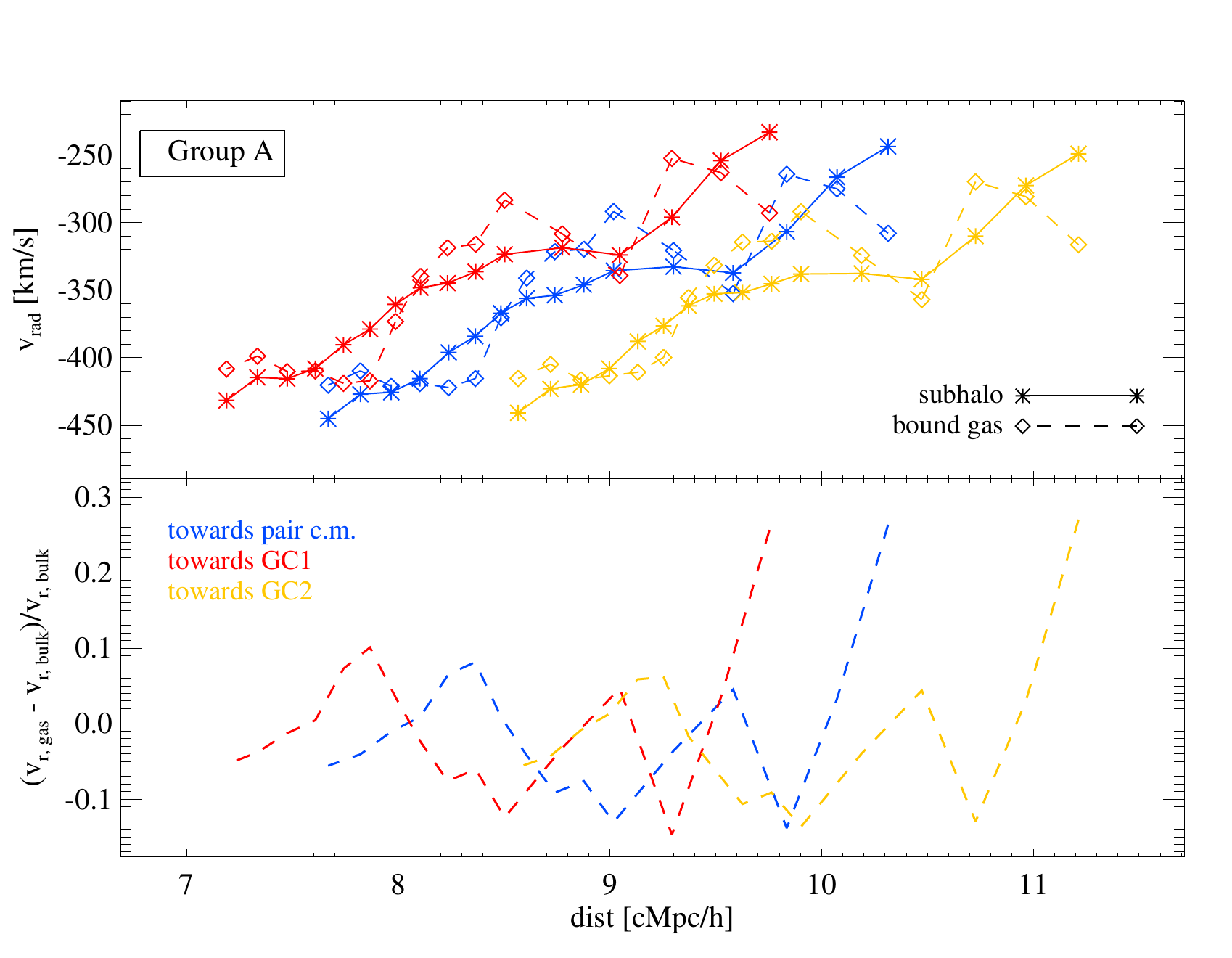}
    \includegraphics[width=.9\columnwidth]{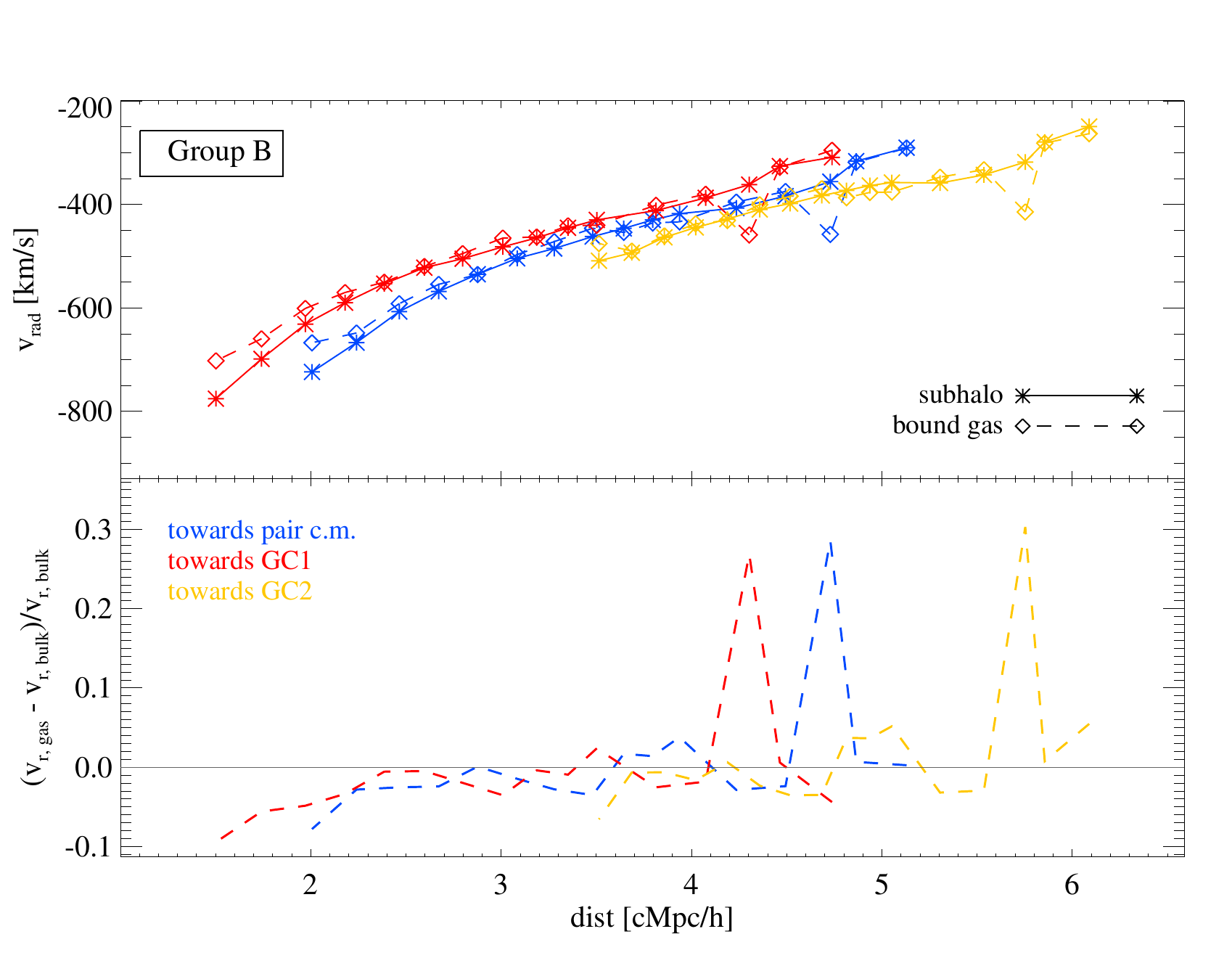}\\
    \caption{
    {Radial bulk (solid lines) and bound-gas (dashed lines) velocities (upper panels),
    and their relative absolute difference (lower panels), for groups A and B (left and right, respectively),
    as a function of the radial distance from the system (center of mass, blue; GC1, red; GC2; yellow).}
\label{fig:north-clumps-vrad}}
\end{figure*}

The radial component of the velocity, along the infall direction, is
explicitly reported in Fig.~\ref{fig:north-clumps-vrad}, across the
redshift range $0.07 < z < 1$. We show however the evolution as a
function of radial distance instead of redshift, with larger radial
distances corresponding to earlier times. As in
Fig.~\ref{fig:north-clumps-cosine}, we investigate separately the
three infall directions and we focus in particular on the velocity of
the bound components only, with the bulk velocity essentially
dominated by the dark matter.  Comparing the two panels in the
Fig.~\ref{fig:north-clumps-vrad}, we note different trends for the two
halos.  Group A shows an oscillating behaviour of both gas and bulk
velocities, with the oscillations amplified especially in the gas
component. This translates into a relative difference between gas and
bulk radial velocities swinging between $-10\%$ and $+10\%$ throughout
the redshift/distance range inspected.  Since the feature is preserved
in all the three directions considered (always defined with respect to
the final positions of GC1, GC2 and pair center of mass), it is indeed
associated to the group itself. This is reflecting a sloshing trend of
the gas component in the $xy$ plane, where the movement along the
filament mostly happens, likely due to accretion processes of smaller
substructures and diffuse gas from the filaments.  This is further
confirmed by the center-shift between gas center-of-mass and halo
center\footnote{The center of the halo is defined throughout the paper
  as the position of the minimum of the halo potential well.}, whose
$x$ and $y$ components visibly present opposite oscillating trends
(see Fig.~\ref{fig:clumps-centershift}, in Appendix~\ref{app:clumps}).
While some similar signatures for sloshing are also shown at early
times, the smaller group B presents a monotonic increase of the
relative difference between the gas and bulk radial velocities, while
getting closer than $\sim 2.6\,\cMpc$ from GC1, namely below
$z\lesssim 0.25$.  Even though the reference positions of the pair and
its members considered in Fig.~\ref{fig:north-clumps-vrad} are the
final ones at $z=0.07$, we can compute the actual distance between the
center of group B and the GC1 progenitor at $z\sim 0.25$, which
corresponds to $d_{\rm 3D}\sim 3.5\,\cMpc \sim 4 \times \rvir^{1}$. On
the $xy$ plane, the group B is found at a projected distance of $\sim
3 \times \rvir^{1}$ at $z\sim 0.25$.  { Even though group B does
  not seem to be located within a main filament, this is the typical
  distance from clusters at which simulations predict to observe
  changes in the gas properties along filaments connected to them,
  such as an increase in the gas temperature and a steeper radial gas
  density profile~\cite[namely at three to four times the cluster
    virial radius; e.g.][]{dolag2006}.}  The configuration of group B
roughly $\sim 3\,$Gyr ago and the whole picture shown since then,
i.e.\ between $z\sim 0.25$ and $z\sim 0.16$, is consistent with the
properties of the observed Northern Clump in the \erosita A3391/95
field (Veronica et al.,~submitted), which is in fact located at a
similar projected distance of about three virial radii from the A3391
cluster~(\cite{reiprich2020}; Veronica et al.,~submitted).

\section{Discussion}\label{sec:discussion}

{In this study we investigate properties of a close pair of galaxy
  clusters showing several similarities compared to the observed
  A3391/95 system.  Given that the simulated analog has been extracted
  from a cosmological hydrodynamical, not {\it ad hoc}, simulation,
  some difference with respect to the observations is naturally
  present.}

{In our investigation, we identify the A3391/95 analog by searching
  among cluster close pairs. We note, however, that the A3395 cluster
  in the observed system is rather a double merging cluster
  itself~\cite[A3395n/A3395s;][]{reiprich2002,reiprich2020}, even
  though in the literature it is sometimes treated as a whole, albeit
  double-peaked, system~\cite[][]{alvarez2018}.  The selected
  simulated pair is instead composed of two clusters, although the GC2
  member also comprises two main substructures. These are two massive
  group-size haloes that underwent a major merging around
  $3$--$3.5\,$Gyr ago ($z\sim0.3$--$0.25$, see Sec.~\ref{sec:orig})
  with a mass ratio of 1:1.6.  At the time of merging, the two systems
  had masses of $M_{500}^1=7\times10^{13}\msun$ and
  $M_{500}^2=1.1\times10^{14}\msun$, respectively.  In fact, two
  emission peaks are still visible in the innermost regions of GC2 at
  $z=0.07$, e.g.\ in the bottom-right panel of
  Fig.~\ref{fig:evol_maps}.  Compared to A3395s and A3395n, the two
  structures in the simulated GC2 cluster are at a later stage of
  interaction.  Furthermore, despite this broad similarity, we remind
  that the overall mass of the system is at least $20\%$ smaller than
  that of A3391/95, consistent with finding smaller components in GC2
  compared to A3395s and A3395n.  }

{Given the smaller masses of the simulated clusters, we can also
  expect a smaller number of prominent filaments connected to them, as
  seen in Fig.~\ref{fig:evol_maps}.  In fact, numerical studies
  predict that most of the close pairs of clusters with separation of
  $5\,\cMpc$ or less are typically connected by straight filaments,
  with less massive clusters connected typically to less filaments (1
  or 2, instead of up to 4--5)~\cite[][]{colberg2005}.  }

{In general, some of the structures in the observed A3391/95 system
  have no direct match in the simulations. This is the case, for
  instance, of the ESO\;161-IG\;006 galaxy group located in the bridge
  region. In the \erosita observations, part of the X-ray emission in
  the bridge between A3391 and A3395 is associated to the
  ESO\;161-IG\;006 group, although not dominated by it.}
{\cite{reiprich2020} show in fact that the X-ray diffuse emission
  in the bridge qualitatively spans a horizontal scale of $\sim
  3\,\Mpc$, larger than the region occupied by the ESO\;161-IG\;006
  group.  The definition of the bridge is however somewhat uncertain.
  \cite{brueggen2020} model the bridge in A3391/95 as a cylindrical
  volume with $600\,\kpc$ radius and $1.3\,\Mpc$ length, excluding the
  group emission. A similar, slightly larger, size (i.e.\ $\sim
  800\,\kpc$ and $1.6\,\Mpc$ for radius and length, respectively) was
  also assumed by~\cite{sugawara2017} in their modelling.}  {Here,
  we assume a relatively similar radius ($\sim 660\,\kpc$) and a
  length of about $2\,\Mpc$, which corresponds to the region outside
  the $\rvir$ of both GC1 and GC2, given their physical separation.}
{In the simulated analog, none of the galaxy groups in the area lie
  between GC1 and GC2. This implies that the gas selected in the
  interconnecting region is mostly warm and diffuse, except for a
  minor fraction of colder gas located in galaxy-size haloes (see
  Sec.~\ref{sec:chem}).  The transition between outer cluster halos
  and pure inter-cluster gas is however difficult to determine. In our
  definition, gas and subhaloes in the clusters' periphery ($\rvir < r
  < R_{\rm vir}$) are in fact comprised within the geometrical
  boundaries of the bridge (see Sec.~\ref{sec:bridge}).  The presence
  of clumps in the far outskirts of clusters has been investigated
  recently in cosmological simulations by~\cite{angelinelli2021},
  finding that clumps around (and beyond) $\sim 3\times \rfive$ can be
  used to locate filaments connected to clusters, and to study the
  thermodynamics of diffuse baryons before they are processed by the
  interaction with the ICM of the host cluster. A moderate ($\lesssim
  10\%$ in mass) presence of dense halo gas associated to galaxies in
  short filaments is also found by~\cite{galarraga2020b}, who
  investigated the properties of filaments in different cosmological
  simulations, including the Magneticum set.  }

{ From the analysis done on the selected pair, and given its
  geometrical configuration, we conclude that the majority of the
  bridge gas has not been stripped from the member clusters due to
  their interaction.  The two clusters GC1 and GC2 are in fact
  separated by a physical distance that is $\sim 3\times (\rfive^1 +
  \rfive^2)$ and their virial boundaries ($R_{\rm vir}$) do not
  overlap. The GC1 and GC2 spheres of influence ($\sim 3\times
  \rvir$), instead, are already overlapping and the two clusters are
  moving towards each other, as confirmed by the evolution shown in
  Fig.~\ref{fig:trck_gas_prog}. We can expect they will finally merge.
} {Observationally, it is more difficult to determine the physical
  distance of the A3391/95 components along the l.o.s. In this
  respect, the indications of some cold gas component in the
  interconnecting bridge~\cite[][]{reiprich2020} can still be
  consistent with a scenario similar to the simulated case. Namely,
  the filament has been forming in time during the LSS assembly and
  the two clusters can be in a pre-merging phase, in which no
  stripping nor direct interaction between the cluster's atmospheres
  has happened yet. A longer interconnecting bridge of pure-filament
  gas along the l.o.s., in case of a larger physical separation, could
  be observationally confirmed by the presence of colder ($T< 1\,$keV)
  WHIM gas in a detailed spectral analysis of the bridge region.}

{An early interaction phase of the two pair members was instead
  found in~\cite{planck2013}, for a similar simulated system
  resembling A399--A401, with a significant part of the bridge gas
  stripped from the two clusters. The main difference, with respect to
  the A3391/95-like system considered here, was the 3D distance of the
  two clusters relative to their size. In that case, the two clusters
  (more massive than GC1 and GC2) were separated by a smaller physical
  distance relative to their size, with the two virial radii
  overlapping in all projections.}

{ Despite the several similarities between the simulated system and
  the observed A3391/95, we finally note that none of the groups found
  in the surroundings of the selected pair at $z=0.07$ matches {\it
    exactly} the properties of the observed Northern Clump.
  Nonetheless, by selecting two interesting candidates we could
  investigate the expected signatures of infall towards the overdense
  Cosmic Web knot where the pair is located. The groups A and B serve
  as extreme limits, both with similarities and differences with
  respect to the real Northern Clump. Group A is an isolated massive
  group moving along with the main northern filament towards the pair
  system, like the Northern Clump, but is located at a too large
  distance to experience the direct influence of any of the main
  clusters. Group B, instead, is a smaller group outside of the main
  north-south filament and is already merging with one of the
  clusters.  Group B, however, showed a configuration similar to the
  Northern Clump about $\sim 3\,$Gyr ago ($z\sim 0.25$), when it was
  at a distance of $\sim 3\times \rvir$ from GC1. We use this example
  to speculate that the features presented by the Northern Clump in
  the \erosita observations are consistent with a radial motion of the
  system towards A3391 and the pair (Veronica et al.,
  submitted). Indeed, from the evolution of the group B between $z\sim
  0.25$ and the final snapshot ($z=0.07$), we observe slowing down of
  the gas compared to DM, and stripping of the gas in the later
  merging stages ($z\lesssim 0.16$), when the group B finally enters
  the virial boundary of GC1.  }

\section{Summary and conclusions}\label{sec:conclusion}

{We present results from the Magneticum cosmological hydrodynamical
  simulations on a case study resembling the binary cluster system in
  the observed \erosita A3391/95 field.  By investigating a
  cosmological comoving volume of $(352\,\cMpc)^3$ at high resolution,
  we consider all possible close pairs of cluster-size haloes and
  select a candidate at $z=0.07$ presenting several global
  similarities with the observed A3391/95 system.  Specifically, by
  restricting to all clusters with $M_{500}>10^{14}\msun$ in the local
  Universe, we expect to find $300$--$400$ pairs with physical
  separation smaller than $10\,\Mpc$ per $(\Gpc)^3$. This is confirmed
  when larger cosmological volumes of the Magneticum suite are
  inspected.  For our study, we focus in particular on the large-scale
  region of $20\,\cMpc$ per side around the pair, approximating a
  physical size of $\sim 26\,$Mpc for the given redshift and
  cosmology.  The selected simulated system is located in a node of
  the cosmic web which is assembling, with filaments building up in
  time and substructures moving along with them towards the cluster
  pair.  The main clusters in the pair are slightly less massive than
  the members of the observed A3391/95 system, despite a roughly
  similar mass ratio (${\rm M1/M2}\sim 1.2$) and configuration. The
  projected distance between the pair clusters is indeed similar
  ($2.6\,\Mpc$ in the reference projection), with the regions enclosed
  by $3\times\rvir$ overlapping.  The simulated system is in fact one
  of the $\sim 7$ close pairs found at $z=0.07$, with member masses of
  $1.5< M_{500}^{1,2} [10^{14}\msun ]< 3.5$ and a three-dimensional
  separation of $d_{\rm 3D}\sim 4.5\,\Mpc$.  Typically, our
  simulations predict in fact to find roughly $\sim 25$ close ($d_{\rm
    3D}\lesssim10\,\Mpc$) pairs per $(\Gpc)^3$, whose members have
  masses in that range and mass ratio of ${\rm M1/M2}\lesssim 1.2$.}

{Given the encouraging similarity to the A3391/95 system in the
  \erosita field, we then explored the properties and origin of the
  pair, and of the diffuse gas selected at $z=0.07$ in different
  thermal phases and spatial regions. In the following we summarize
  the main results obtained from this study.}

\begin{itemize}

    \item The gas located in the pair bridge originates from distances
      of few-to-ten comoving megaparsecs away with respect to its
      final position. Furthermore, its accretion trajectories are
      almost perpendicular to the filamentary directions through which
      the gas found within the main clusters has been accreted,
      {marking a distinctly different origin};
    
    \item Most ($\sim 90\%$) of the gas in the bridge was never inside
      the $\rvir$ radius of either cluster progenitor, and a
      substantial fraction ($\gtrsim 80\%$) was always beyond
      $2\times\rvir$ at $z\gtrsim 0.25$. The simulation study thus
      predicts that {\it the majority} of the bridge gas has not been
      stripped from the member clusters despite their closeness;
    
    \item 
    {The bridge gas is characterised by a typical temperature of
      $\sim 1\,$keV and overdensity around $\delta \lesssim
      100$. Lower densities are expected at larger distances from the
      bridge spine ($\gtrsim 1\,\Mpc$), or in longer filaments, where
      the colder WHIM phase is expected to become more
      significant~\cite[][]{galarraga2020b};}
    
    \item 
    {In the whole pair region, }the enrichment level of the warm
    diffuse gas and WHIM is fairly homogeneous and lower than for the
    hot ICM in the clusters (by a factor of $\sim 2$, for the iron
    abundance) or the cold-dense gas in galaxies and star-forming
    regions.  Nonetheless, a highly-enriched component is present at
    $z\lesssim 2$ in all gas selections, and is mostly related to
    early enrichment within galaxies and star-formation sites.  This
    component, roughly $\sim10$--$20\%$ of the gas mass depending on
    the specific selection, persists {also in the bridge,} within
    small clumps (galaxies) or in the diffuse form after being
    stripped from those haloes;
 
    \item 
    Similarly to the A3391/95 field observed by \erosita, in the
    simulation we also find additional group-size objects in the
    surroundings of the pair (see also Ramos-Ceja et al., in prep.),
    {mostly aligned along a prominent north-south filament
      structure spanning about $\sim 26\,\Mpc$ ($20\,\cMpc$) in
      projection (Fig.~\ref{fig:evol_maps}--\ref{fig:trck_gas_prog})}.
    We identify in particular 5 groups in the mass range
    $M_{500}\sim1.4$--$9.6\times 10^{13}\,\msun$ at $z=0.07$.
    
    \item 
    The additional systems, along with filaments, trace the assembly
    of the LSS around the overdense node where the pair is finally
    located. In fact, by inspecting the properties and trajectories of
    two group-size haloes in particular, groups A and B, we find clear
    signatures of their motion along with filaments towards the
    pair. In both cases we find a strong alignment of the halo bulk
    (gas and dark matter) velocity with the infall direction towards
    the pair.  The farther and isolated group A is characterised by a
    visible sloshing of the gas component with respect to the DM one,
    while no significant mergers characterize its mass accretion
    history, except for accretion of gas and smaller clumps (galaxies)
    through the filament. This is marked by the oscillating evolution
    of gas and DM infall velocity components as well as by the center
    shift between halo center and gas center of mass.
    
    \item
    The smaller group B is identified as a substructure
    gravitationally bound to the pair cluster GC1 at redshift
    $z\lesssim 0.16$ and is located in its outskirts at $z=0.07$.
    {Tracing the trajectory of the group} towards the pair we find
    evidences of the influence of the GC1 cluster. {At} a projected
    distance of about $3\times \rvir^1$ from GC1, around $z\sim0.25$,
    the gas radial velocity starts to systematically decrease compared
    to the DM one. While entering the outer atmosphere of GC1
    ($z\lesssim0.16$) it undergoes gas stripping, as marked by the
    decrease of bound gas mass by a factor $\sim 3$.  {The
      configuration corresponding to $3\,$Gyr ago ($z\sim0.25$) shares
      several similarities with the case of the real Northern Clump
      relative to A3391, supporting the picture of its actual
      infalling motion towards the A3391/95 system (see Veronica et
      al., submitted).}

\end{itemize}

The \erosita superior soft response and large FoV has allowed for the
first direct detection of the faint X-ray emission from the diffuse
gas in the bridge and filaments connected to A3391/95, drawing a
consistent picture of the LSS that well compares to theoretical
predictions.  Our conclusions from simulations further support that
the geometrical configuration of the A3391/95 system, its pre-merger
phase and physical separation, together with the large-scale structure
of the field in which it is located, provide an optimal target for
unveiling the elusive warm-hot gas that populates the Cosmic Web
filaments and for characterising its physical properties.

The \erosita all-sky survey mode will finally enable for a more
statistical approach, by providing a significantly larger sample of
candidate multiple cluster systems and gas emission filaments.
Combined and compared to theoretical predictions from simulations,
where samples of binary or multiple cluster systems in various
configurations can be statistically investigated as well, this will
greatly benefit the baryon census in the Universe, especially through
the detailed characterization of the chemo-energetic properties of the
diffuse pristine gas.

\begin{acknowledgements}
The authors would like to thank the anonymous referee for constructive
comments and suggestions that contributed to improve and clarify the
presentation of this work.  VB kindly acknowledges A.~Merloni for
providing comments on the manuscript and inspiring discussions with
U.~Maio on cosmic gas chemistry. This research was funded by the
Deutsche Forschungsgemeinschaft (DFG, German Research Foundation) ---
415510302, and also partially supported by the Excellence Cluster
ORIGINS, which is funded by the DFG under Germany's Excellence
Strategy --- EXC-2094-390783311.  Part of this work has been funded by
the Deutsche Forschungsgemeinschaft (DFG, German Research Foundation)
– 450861021.  The Magneticum Pathfinder simulations have been
performed at the Leibniz-Rechenzentrum with CPU time assigned to the
projects pr86re and pr83li.  KD acknowledges support through the
COMPLEX project from the European Research Council (ERC) under the
European Union’s Horizon 2020 research and innovation program grant
agreement ERC-2019-AdG 860744.  This work was supported in part by the
Fund for the Promotion of Joint International Research, JSPS KAKENHI
Grant Number 16KK0101.  This work is based on data from eROSITA, the
soft X-ray instrument aboard SRG, a joint Russian-German science
mission supported by the Russian Space Agency (Roskosmos), in the
interests of the Russian Academy of Sciences represented by its Space
Research Institute (IKI), and the Deutsches Zentrum für Luft- und
Raumfahrt (DLR). The SRG spacecraft was built by Lavochkin Association
(NPOL) and its subcontractors, and is operated by NPOL with support
from the Max Planck Institute for Extraterrestrial Physics (MPE).  The
development and construction of the eROSITA X-ray instrument was led
by MPE, with contributions from the Dr. Karl Remeis Observatory
Bamberg \& ECAP (FAU Erlangen-Nuernberg), the University of Hamburg
Observatory, the Leibniz Institute for Astrophysics Potsdam (AIP), and
the Institute for Astronomy and Astrophysics of the University of
T\"ubingen, with the support of DLR and the Max Planck Society. The
Argelander Institute for Astronomy of the University of Bonn and the
Ludwig Maximilians Universit\"at Munich also participated in the
science preparation for eROSITA.
\end{acknowledgements}

\bibliographystyle{aa}
\bibliography{mybib}

%-------------------------------------------------------------------
\begin{appendix}
\onecolumn

\section{Appendix -- Cluster pairs in the Magneticum Simulations}
\label{app:pairs}

{In Table~\ref{tab:pairs_stat}, we report the number of cluster
  pairs found in the Magneticum simulation (Box2/hr) at $z=0.07$, for
  different separations and mass ratios\footnote{In order to compute
    the mass ratio M1/M2, we consider $M_{500}$.}.  We consider unique
  pairs of clusters with $M_{500}^{1,2} > 10^{14}\,\msun$ ($448$
  haloes, uper half in the Table), as well as pairs where both haloes
  have masses in the range $1.5 < M_{500}^{1,2}\,[10^{14}\,\msun] <
  3.5$ ($165$ haloes; lower half of the Table).}

\begin{table*}
\centering
\caption{\label{tab:pairs_stat}%
    Number of pairs in the Magneticum Simulation cosmological volume Box2/hr at $z=0.07$, for different selection criteria based on the pair members three-dimensional (3D) and projected (2D) 
    distance and mass-ratios.}
\setlength{\tabcolsep}{10pt}\renewcommand{\arraystretch}{1.2}
    \begin{tabular}{l ccc cc}
    \hline
    & $d_{3\rm D} < 10\,{\rm Mpc}$ 
    & $d_{3\rm D} < 15\,{\rm Mpc}$ 
    & $d_{3\rm D} < 20\,{\rm Mpc}$
    & $d_{2\rm D} < 5\,{\rm Mpc}$ 
    & $d_{2\rm D} < 10\,{\rm Mpc}$
    \\
    \hline\hline
    \multicolumn{1}{l}{$M_{500}^{1,2} > 10^{14}\,\msun$} \\
    \hline
    any M1/M2     & 40 & 88 & 146   %& 265 
                  & 135 & 535 \\  
    M1/M2 $< 2$   & 26 & 55 & 97  & & \\  
    M1/M2 $< 1.5$ & 19 & 42 & 78  & & \\  
    M1/M2 $< 1.2$ & 13 & 27 & 51  & & \\  
    \hline\hline
    \multicolumn{1}{l}{$1.5 < M_{500}^{1,2}\,[10^{14}\,\msun] < 3.5$} \\
    \hline
    any M1/M2     & 8 & 15 & 25 
                  & 27 & 82\\  
    M1/M2 $< 2$   & 8 & 15 & 24 & & \\  
    M1/M2 $< 1.5$ & 8 & 15 & 24 & & \\  
    M1/M2 $< 1.2$ & 7 & 11 & 19  & & \\  
\hline
    \end{tabular}
    \tablefoot{The 2D distance $d_{2\rm D}$ is considering any projection out of the three main Cartesian directions, namely we report the number of pairs for which the projected distance on at least one of the projection planes is $<5(10)$\,Mpc (i.e. $d_{xy}< 5(10)$\,Mpc~$\lor\ d_{xz}< 5(10)$\,Mpc~$\lor\ d_{yz}< 5(10)$\,Mpc).}
\end{table*}

\section{Appendix -- Gas sloshing in groups A and B}
\label{app:clumps}

\begin{figure}
    \centering
    \includegraphics[width=.38\columnwidth,trim=0 10 0 10,clip]{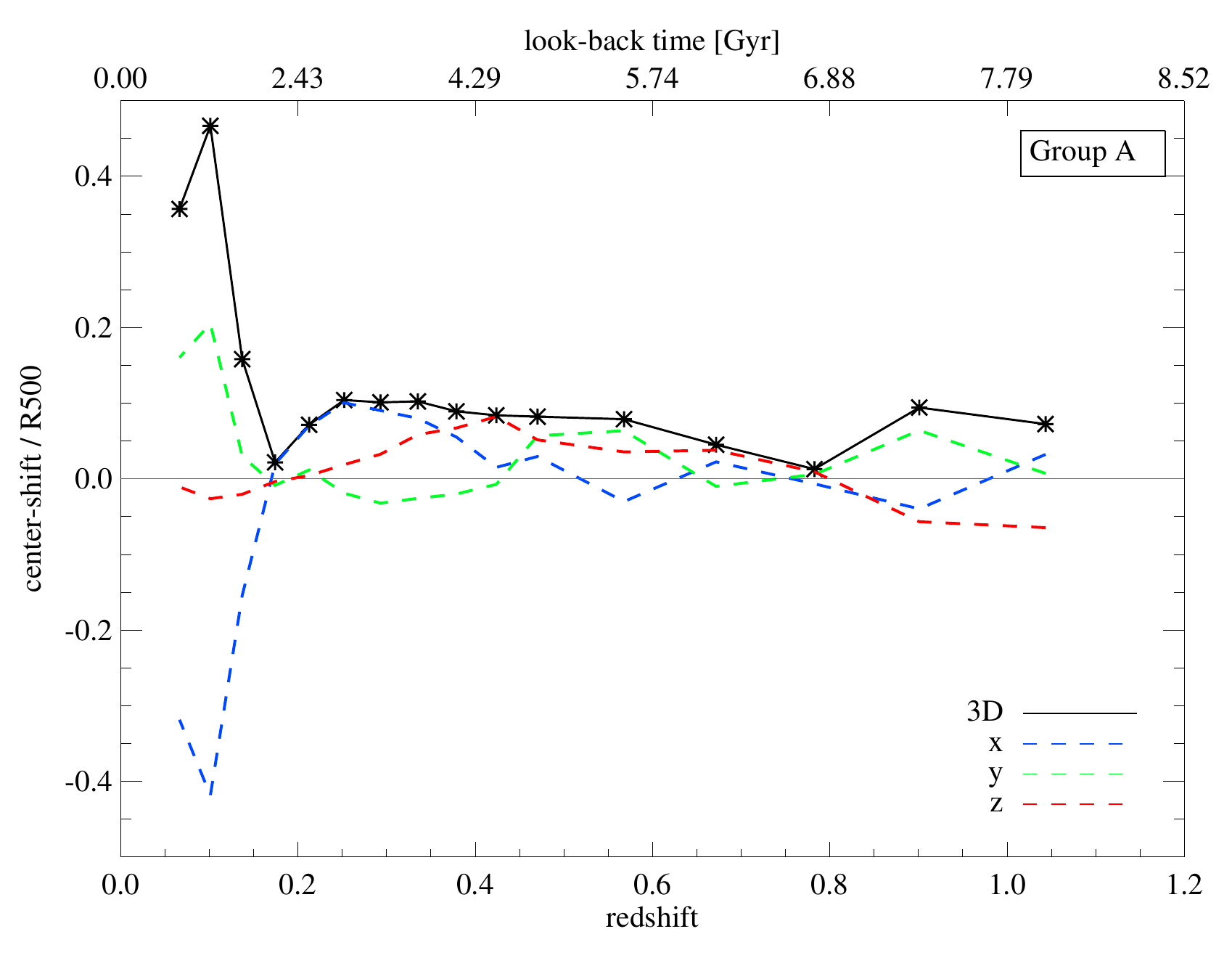}\qquad
    \includegraphics[width=.38\columnwidth,trim=0 10 0 10,clip]{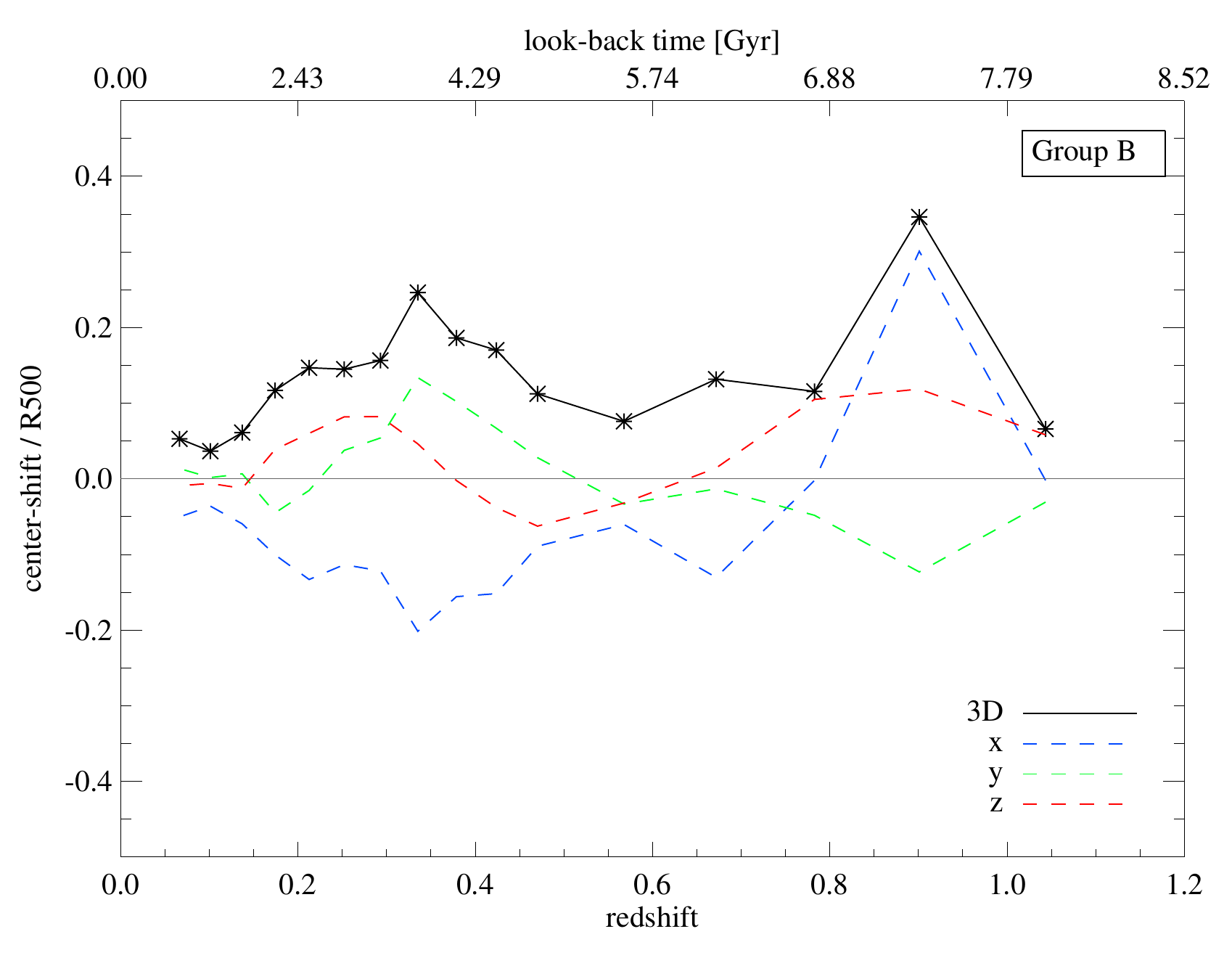}
    \caption{Redshift evolution of the center shift between the (bound) gas center of mass and the halo center, in units of $\rfive$, for the groups A (left) and B (right) analysed in Sec.~\protect\ref{sec:clumps}. 
    We report the modulus and each separate component, as in the legend.
    }
    \label{fig:clumps-centershift}
\end{figure}

In Fig.~\ref{fig:clumps-centershift} we report the center-shift
evolution for the groups A (left) and B (right) discussed in
Sec.~\ref{sec:clumps}.  The center-shift is computed as the
three-dimensional difference between the center of the halo potential
well (defined as the position of the most-bound particle) and the
bound-gas center of mass, and is given in units of the halo $\rfive$
radius, at any given redshift between $z\sim1$ and $z\sim0.07$.  In
addition to the modulus of the shift, we also show the three
components along the major simulation axes, separately (as in the
legend).  Comparing the two panels, we note that the amplitude of the
shift is overall larger in the smaller infalling group B, relative to
its size.  Considering the separate shift components, an oscillating
opposite trend in the $xy$ plane is observed in both cases, which
indicates a sloshing feature of the gas component relative to the DM.

\section{Appendix -- Metallicity distribution}
\label{app:chem}

\begin{figure*}
\centering
	\includegraphics[width=.3\textwidth,trim=0 5 0 5,clip]{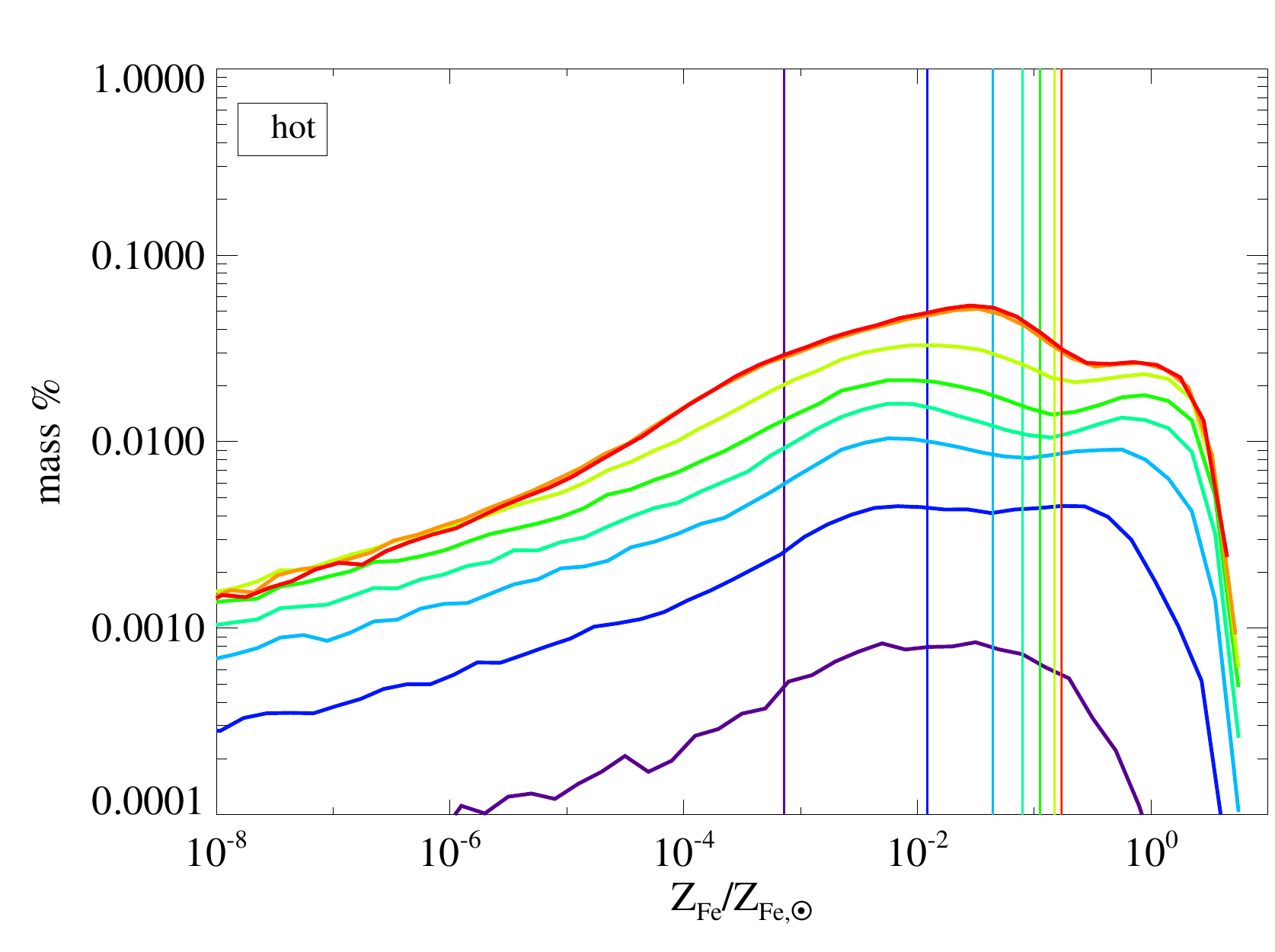}
	\includegraphics[width=.3\textwidth,trim=0 5 0 5,clip]{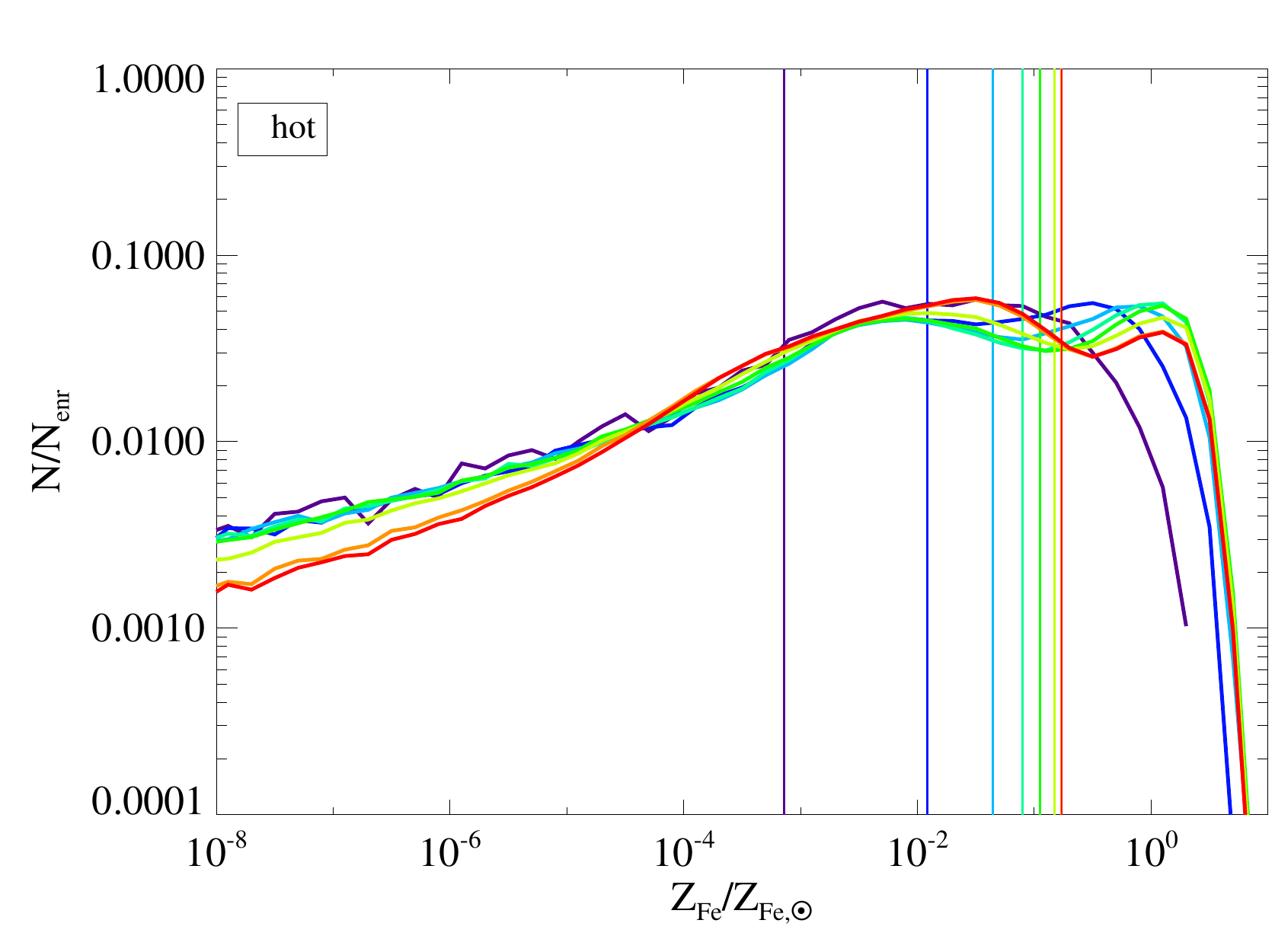}
	\includegraphics[width=.38\textwidth,trim=0 5 0 5,clip]{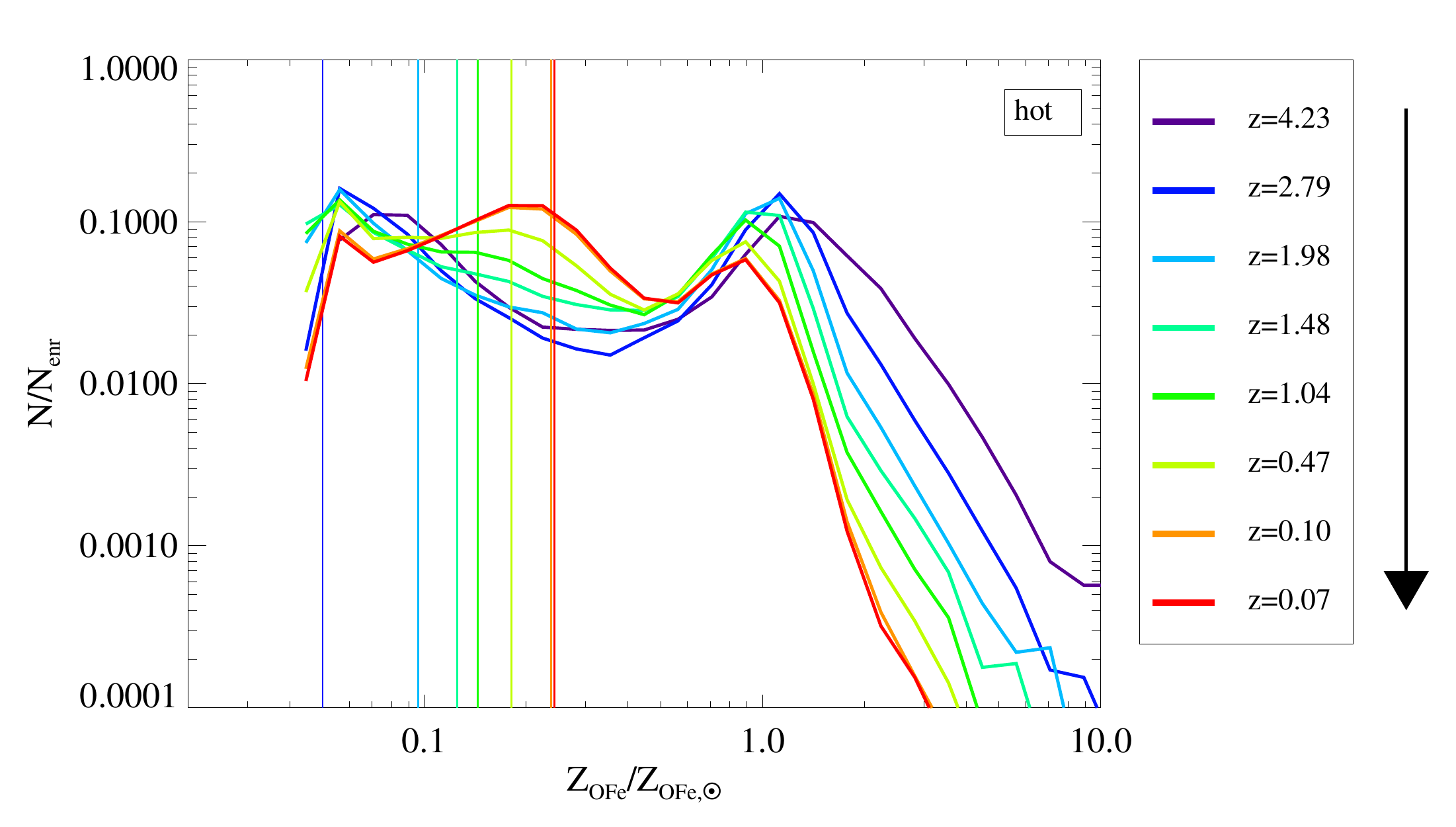}\\
	\includegraphics[width=.3\textwidth,trim=0 5 0 5,clip]{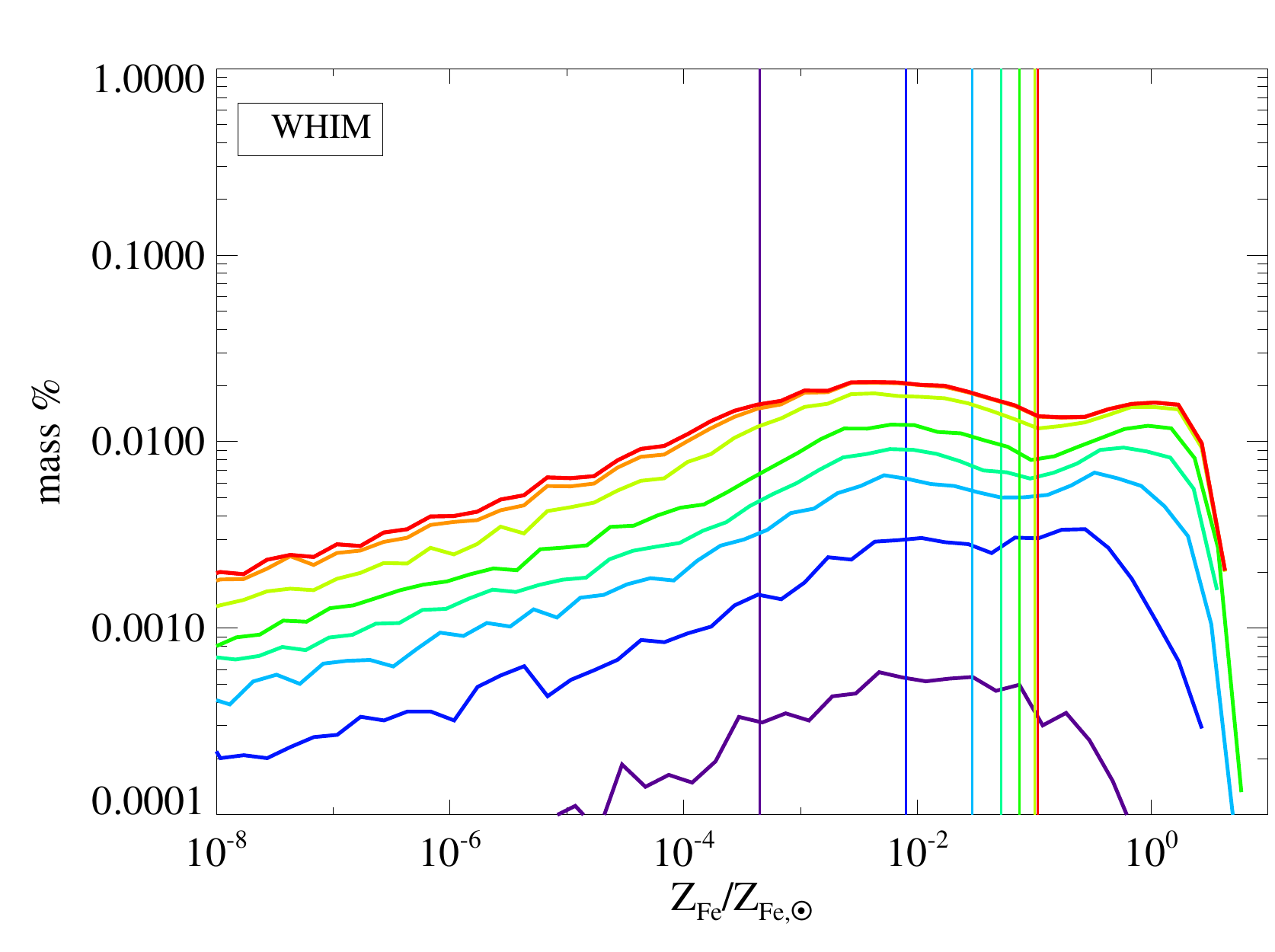}
	\includegraphics[width=.3\textwidth,trim=0 5 0 5,clip]{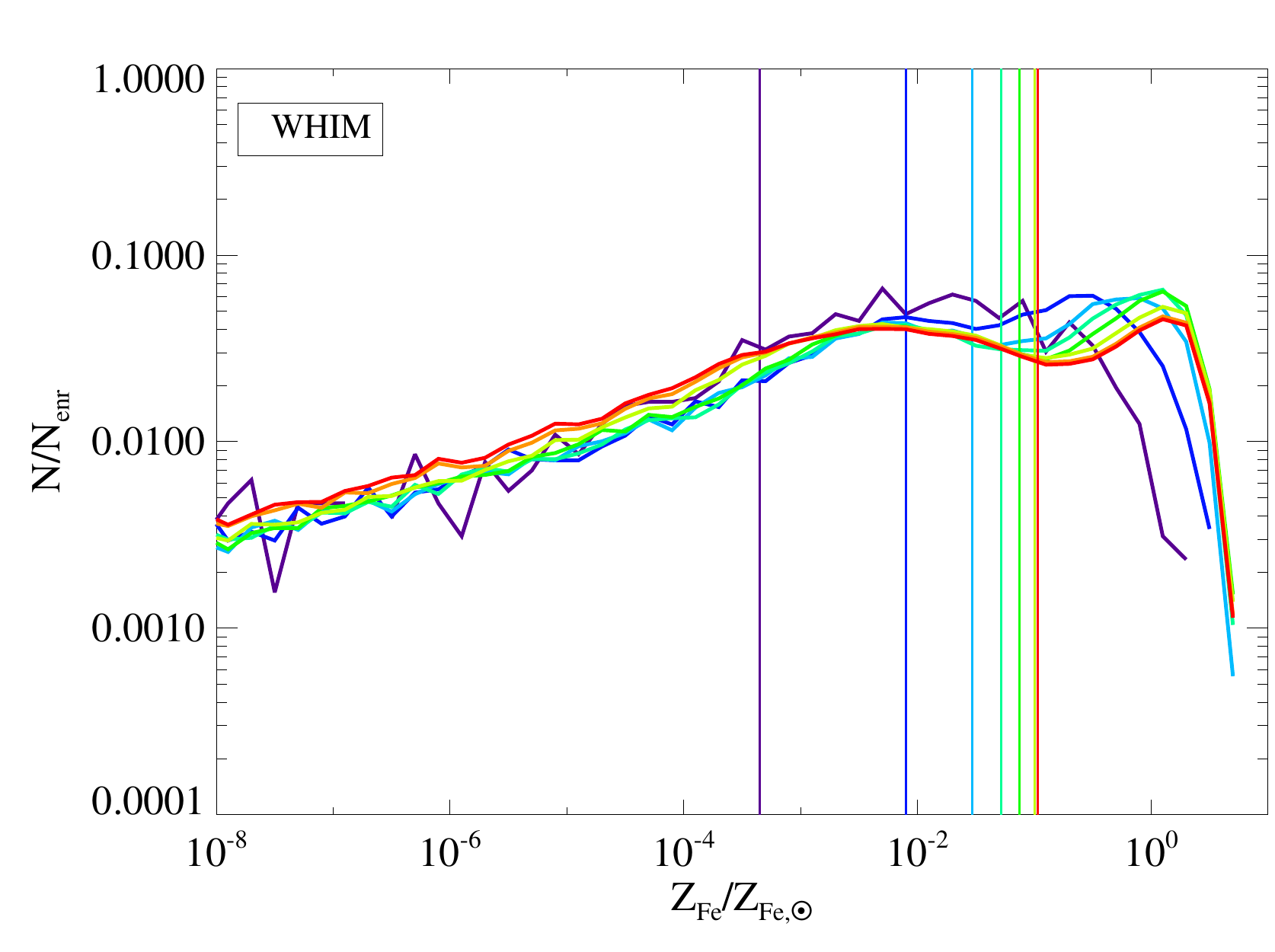}
	\includegraphics[width=.38\textwidth,trim=0 5 0 5,clip]{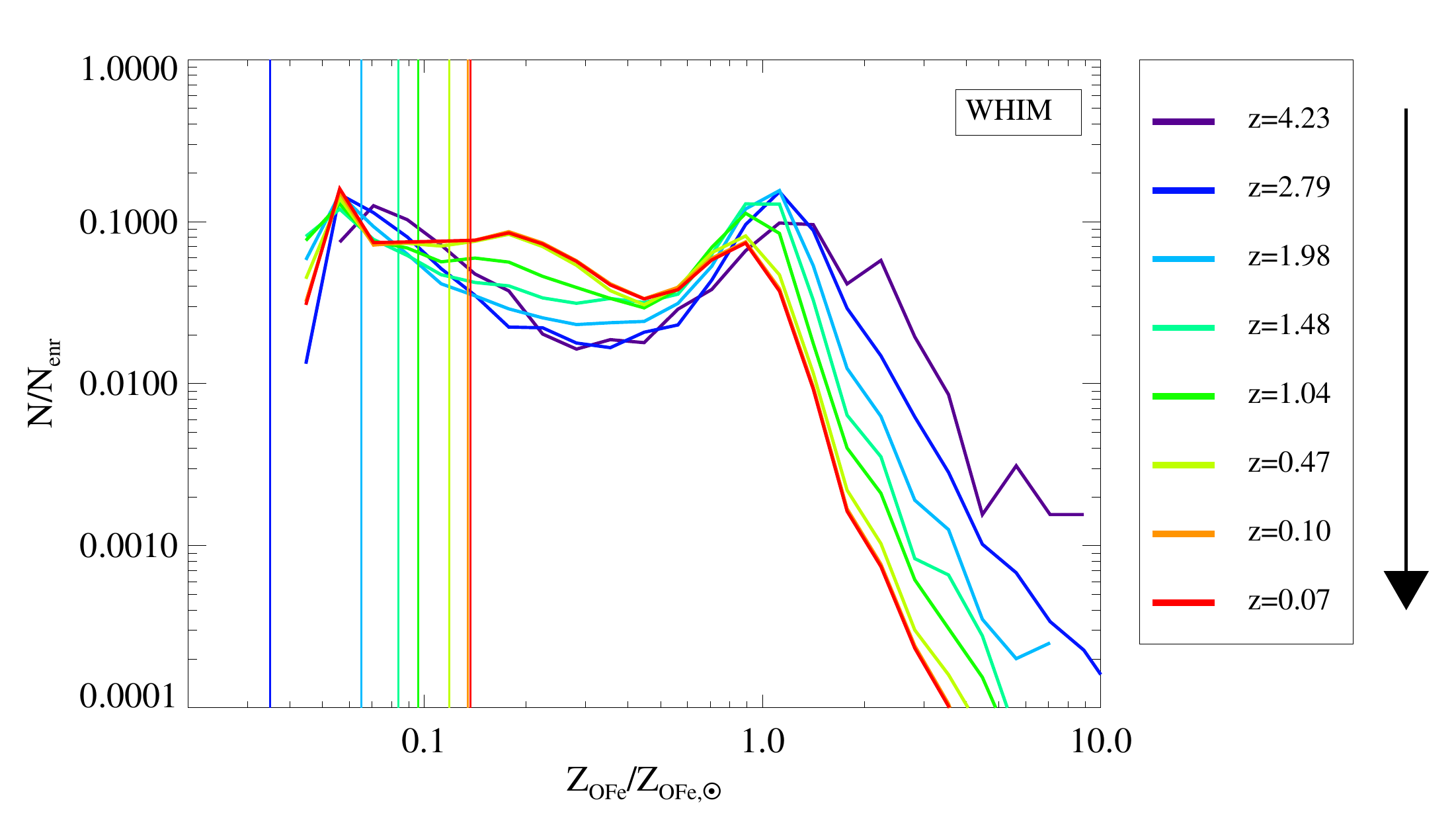}
	\caption{\label{fig:distribMass_met_evol} Iron abundance and
          O/Fe ratio distribution for the hot gas and WHIM (upper and
          lower rows, respectively), for different redshifts as in the
          legend.  Displayed in the left-hand-side panels for
          comparison, the mass fraction distribution of the hot-gas
          iron abundance. At each redshift, the mass fractions are
          normalized to the total gas mass of the selected subsample.
        }
\end{figure*}

In Fig.~\ref{fig:distribMass_met_evol} we show the distributions of
iron abundance and O/Fe abundance ratio, for the hot gas ($T > 10^7
{\rm K}$) and WHIM ($10^5 < T[{\rm K}] < 10^7$ and $\delta < 100$) in
the $(7\,\cMpc)^3$ region around the selected pair (see
Sec.~\ref{sec:chem}).  Gas particles in the two cases are selected at
$z=0.07$ and tracked back in time.  By comparison with
Fig.~\ref{fig:distribN_met_evol}, we notice that the hot and WHIM gas
present overall similar trends to the hot ICM in clusters and the
bridge gas, respectively --- as discussed in
Secs.~\ref{sec:chem}--\ref{sec:chem_evol}.  Furthermore, the mass
fraction distribution of the iron abundance is reported (left-most
panels) to demonstrate that the normalization of the distributions for
a given gas selection indeed increases with time, indicating as
expected that the mass fraction of the enriched gas grows from high to
low redshift (as in Fig.~\ref{fig:evol_SF_met}).

\end{appendix}

\end{document}